\documentstyle[12pt,epsf]{article}
\newcommand{\la}[1]{\label{#1}}

\newcommand{\pp}[1]{\langle\phi^\dagger\phi(#1)\rangle}

\newlength{\numlen}

\newlength{\indexlength}

\newcommand{\be}{\begin{equation}}
\newcommand{\ee}{\end{equation}}
\newcommand{\ba}{\begin{eqnarray}}
\newcommand{\ea}{\end{eqnarray}}

\newcommand{\rmi}[1]{{\mbox{\scriptsize #1}}}
\newcommand{\pdp}{\langle\phi^\dagger\phi\rangle}

\newcommand{\eq}{eq.~}

\newcommand{\nr}[1]{(\ref{#1})}

\newcommand{\tr}{{\rm Tr\,}}

\newcommand{\bfx}{\mbox{\bf x}}

\newcommand{\fr}[2]{{\frac{#1}{#2}}}

\newcommand{\msbar}{\overline{\mbox{\rm MS}}}

\def\lsim{\raise0.3ex\hbox{$<$\kern-0.75em\raise-1.1ex\hbox{$\sim$}}}
\def\gsim{\raise0.3ex\hbox{$>$\kern-0.75em\raise-1.1ex\hbox{$\sim$}}}

\makeatletter \@addtoreset{equation}{section} \makeatother

\newcommand{\bra}[1]{\langle #1 |}
\newcommand{\ket}[1]{| #1 \rangle}

\def\ltap{\raisebox{-.55ex}{\rlap{$\sim$}} \raisebox{.4ex}{$<$}}
\def\gtap{\raisebox{-.55ex}{\rlap{$\sim$}} \raisebox{.4ex}{$>$}}
\def\gsim{\mathrel{\gtap}}
\def\lsim{\mathrel{\ltap}}

\def\const{\mbox{const}}
\def\e{\mbox{e}}

\begin{document}
\begin{titlepage}
\begin{flushright}
CERN-TH/96-13\\
INR-913/96\\
hep-ph/9603208\\ 
March 1, 1996
\end{flushright}

\begin{centering}
\vfill
{\bf ELECTROWEAK BARYON NUMBER NON-CONSERVATION IN THE EARLY
UNIVERSE AND IN HIGH ENERGY COLLISIONS}

\vspace{1cm}

V. A. Rubakov$^{\rm a,}$\footnote{rubakov@ms2.inr.ac.ru} and
M. E. Shaposhnikov$^{\rm b,a,}$\footnote{mshaposh@nxth04.cern.ch} \\
\vspace{0.3cm}
{\em $^{\rm a}$ Institute for Nuclear Research of the Russian Academy
of Sciences,\\ 60-th October Anniversary Prospect 7a, Moscow 117312,
Russia\\}
\vspace{0.3cm}
{\em $^{\rm b}$Theory Division, CERN, CH-1211 Geneva 23, Switzerland}

\vspace{0.7cm}

{\bf Abstract}

\end{centering}
\vspace{0.7cm}

We review  recent progress in the study  of the anomalous
baryon number
non-conservation at high temperatures and in high energy collisions.
Recent results on high temperature phase transitions are described,
and
applications to  electroweak baryogenesis are considered.
The current status of the problem of electroweak instanton-like
processes at high energies is outlined. This paper is written on
the occasion of Sakharov's
75th anniversary and will appear in the memorial volume of Uspekhi
(Usp. Fiz. Nauk, volume 166, No 5, May 1996).

\vspace{2.5cm}
\noindent
CERN-TH/96-13\\
INR-913/96\\
March 1, 1996

\vfill \vfill
\noindent

\end{titlepage}
\tableofcontents
\newpage
\section{Introduction}
In his famous paper \cite{sakharov} Sakharov discussed for the first
time the
possibility of explaining  the charge asymmetry of the Universe in
terms of particle theory. The paper was submitted to JETP Letters in
September 1966, two years after the discovery of CP-violation in
$K^0$
decays \cite{Christen} and one year after the microwave black-body
radiation, predicted by the Big Bang theory \cite{gamow}, was found
experimentally \cite{penzias}. To
explain baryon asymmetry, Sakharov proposed an approximate character
for the baryon conservation law, i.e., baryon number
non-conservation and proton decay. Three years later Kuzmin published
a paper
\cite{kuzmin} where a different model leading to the baryon
asymmetry was constructed. One of its consequences was another
process with B non-conservation, namely, neutron-antineutron
oscillations.
Since that time the idea that baryon number may not be exactly
conserved in Nature has been elaborated upon considerably,
both in the context of the generation of the baryon asymmetry
of the Universe \cite{ikkt:bau,yosh,dimsus,Weinberg,ellis,iks:bau}
(for reviews see refs. \cite{dz:rev,dolgov:rev,koltur:rev})
and because of theoretical developments that have lead to a unified
picture
of fundamental interactions. In the mid--70's, grand unified theories
with inherent violation of baryon number
were put forward \cite{Pati1,Pati2,Georgi1,Georgi2,fritsch}.
Almost at the same time it was realized \cite{hooft:prl,hooft:pr}
that non-perturbative effects related to instantons
\cite{BRST} and the complex structure of gauge theory vacuum
\cite{hooft:prl,callan,jackiw}
lead to the non-conservation of baryon number even in the
electroweak theory; it has been understood later \cite{krs85}
that similar effects are relevant for the baryon asymmetry.

In his paper \cite{sakharov} Sakharov writes: ``According to our
hypothesis, the
occurrence of C asymmetry is the consequence of violation of CP
invariance in the nonstationary expansion of the hot Universe during
the superdense stage, as manifest in the difference between the
partial probabilities of the charge-conjugate reactions.'' Today,
this
short extract  is usually dubbed as three necessary
Sakharov's conditions for baryon asymmetry generation from the
initial charge symmetric state in the hot Universe, namely:\\
(i) Baryon number non-conservation.\\
(ii) C and CP violation.\\
(iii) Deviations from thermal equilibrium.\\
All three conditions are easily understood.

(i) If baryon number were conserved, and the initial baryonic charge
of
the Universe were zero, the Universe today
would be symmetric, rather
than asymmetric \footnote{Of course, there is a loop-hole in this
argument, which Sakharov knew. The Universe may be globally
symmetric, but locally asymmetric, with the size of the baryonic
cluster of matter  large enough (say, of the order of the present
horizon
 size). The inflationary models of the Universe expansion,
together with specific models of particle interactions may provide a
mechanism of the local asymmetry generation, keeping the conservation
of the baryon number intact \cite{dolgov:rev}.}. The statement of the
necessity of the baryon number non-conservation was quite
revolutionary at that time. Today it is very natural theoretically;
still, lacking positive
results from  experiments searching for B non-conservation, the
baryon asymmetry of the Universe is a unique observational evidence
in favour of it.

(ii) If C or CP were conserved, then the rate of reactions with
particles would be the same as the rate of reactions with
antiparticles.
If the initial state of the Universe was C- or CP-
symmetric, then no charge asymmetry could develop from it
\footnote{Again, there  are exotic mechanisms making use of the
inflationary stage of the Universe expansion, in which the underlying
theory conserves C or CP, the
Universe as a whole is charge symmetric, but the visible part is not,
see review \cite{dolgov:rev}.}. In  more
formal language, this follows from the fact that if the initial
density matrix of the system $\rho_0$ commutes with C- or
CP-operations,
and the
Hamiltonian of the system is C- or CP-invariant, then at any time the
density matrix $\rho(t)$ is C- or CP-invariant, so that the average
of
any C- or CP-odd operator is zero.

(iii) Thermal equilibrium means that the system is stationary (no
time dependence at all). Hence, if the initial baryon number is zero,
then it is zero forever.

Clearly, the issue of the baryon asymmetry generation requires
the development of many different areas of theoretical physics, such
as
model building, study of perturbative and
non-perturbative effects leading to
B-violation, finite
temperature field theory and non-equilibrium statistical mechanics,
theory of phase transitions.

In this paper we do not aim to give a complete review of
various theories of baryogenesis proposed so far.
The reader may
consult with  a number of reviews  on this subject
\cite{dz:rev,dolgov:rev,koltur:rev,s:nobel,mrts,s:rev,turrev,ckn:1993}.
Instead, we pick up a
specific non-perturbative
mechanism of the baryon number non-conservation,
associated with triangle anomaly. The choice of this
mechanism is explained, partially, by the authors' personal taste. In
addition, anomalous fermion number non-conservation is  a
general phenomenon for theories with chiral fermions, and is
present, e.g., in the standard model of  electroweak interactions.
This mechanism, being operative at high temperatures,
 may lead to the baryogenesis at the electroweak scale.

The possibility that baryon asymmetry may be
due to physics which is probed at accessible energies
has attracted a lot of attention
recently and serves as a powerful motivation for the development of
 high temperature field theory, theory of  phase transitions,
non-equilibrium statistical mechanics.

The fact that baryon number is rapidly violated at high
temperatures \cite{krs85} (for earlir discussion
see \cite{linde:pl77,dimsus,kman})
 and under other extremal conditions
\cite{Rubakov:1986,Ambjorn:1985,Matveev:1986,Matveev:1987,Mityaa1,Mityaa2}
naturally leads one to enquire whether electroweak baryon
number non-conservations occurs at high enough rate in collisions of
energetic particles. This problem has attracted considerable
interest
in recent years, after the first --- and encouraging at the time ---
quantitative results were obtained
\cite{Ringwald:1990,Espinosa:1989}. In spite of
remarkable theoretical developments, this problem is still not
completely solved;
existing results indicate that the electroweak baryon number
violating
processes occur at unobservable rates even at very high energies.

The paper is organized as follows. In Section 2 we provide
the necessary background and discuss the mechanism of anomalous
non-conservation of fermionic quantum numbers together with relevant
bosonic classical solutions (instantons and sphalerons).
Section 3 contains preliminary discussion of the role of
baryon number violating
electroweak processes in early Universe.
The fermion number non-conservation at high
temperatures  is considered in Section 4.  In Section 5 we present
recent developments in the theory of high temperature phase
transitions. In Section 6 we briefly
address the question of survival of the
primordial baryon asymmetry. The discussion of various
electroweak baryogenesis
mechanisms is contained in Section 7.
We turn to electroweak baryon number non-conservation
in particle collisions in Section 8. Section 9 contains concluding
remarks.
\section{Basics of anomalous non-conservation of fermion quantum
numbers}
Let us discuss non-perturbative non-conservation of fermion quantum
numbers
in the context of a model with the gauge group $SU(2)$ and the
massless
left-handed fermionic doublets $\psi_{L}^{(i)}$, $i=1,\dots, n_{L}$.
The
absence of global anomaly \cite{Wittenglobal} requires that $n_{L}$
is
even. We also add a Higgs doublet $\phi$ that breaks the $SU(2)$
symmetry
completely. Then this theory is a simplified version of the
electroweak
sector of the minimal standard model. All relevant features of the
standard
model are present in this simplified theory; later on we shall
comment on
minor complications due to $U(1)_{Y}$ gauge symmetry, right-handed
fermions
and Yukawa interactions leading to fermion masses. One may regard the
simplified theory as the standard model in the approximation where
$\sin\theta_{W}$ and all fermion masses are set to zero; for three
families of quarks and leptons one has $n_{L}=12$ and
\begin{equation}
   \psi_{L}^{(i)} =
	       \{ q_{L}^{f,\alpha}, l_{L}^{f} \},
\label{I1a*}
\end{equation}
where $f=1,2,3$ is the family index and $\alpha=1,2,3$ labels the
colour of
quarks.

At the classical level, there exist $n_{L}$ conserved global $U(1)$
currents,
\[
    J^{(i)}_{\mu} =
	\bar{\psi}^{(i)}_{L}\gamma^{\mu} \psi^{(i)}_{L},
\]
which correspond to the conservation of the number of each fermionic
species. At the quantum level these currents are no longer conserved
due to
the triangle anomaly \cite{Adler,BellJackiw,Bardeen}
\begin{equation}
      \partial_{\mu} J_{\mu}^{(i)} =
		 \frac{1}{32\pi^{2}}
		 \mbox{Tr} \left(F_{\mu\nu}
		 \tilde{F}_{\mu\nu}\right).
\label{I2+}
\end{equation}
Therefore, one expects that fermion numbers
$N_{F}^{(i)}=\int~d^{3}x~J^{(i)}_{0}$ are not conserved in any
process
where the gauge field evolves in such a way that
\begin{equation}
     N[A] = \frac{1}{32\pi^{2}}
                  \int~d^{4}x~
		 \mbox{Tr} F_{\mu\nu}
		 \tilde{F}_{\mu\nu}
		  \neq 0.
\label{I2*}
\end{equation}
Namely,
\begin{equation}
       \Delta N_{F}^{(i)} = N[A], ~~~~i=1,\dots,n_{L}.
\label{I2**}
\end{equation}
It is clear from eq. (\ref{I2*}) that in weakly coupled theories,
one has to
deal with strong fields: the field
$F_{\mu\nu} = \frac{g}{2i} \tau^{a} F^{a}_{\mu\nu}$
should be of order $1$, and $A_{\mu}^{a}=O(g^{-1})$. So, it is
natural
that the (semi)classical treatment of bosonic fields is often
reliable.

Equation (\ref{I2**}) may be viewed as the selection rule: the number
of fermions changes by the same amount for every species. In terms of
the
assignment (\ref{I1a*}) it implies, in particular,
\[
         \Delta N_{e}=
             \Delta N_{\mu}=
                  \Delta N_{\tau}= N[A],
\]
\begin{equation}
	\Delta B = \frac{1}{3}\cdot 3\cdot 3\cdot N[A],
\label{I3*}
\end{equation}
where the factor $1/3$ comes from the baryon number of a quark, while
the factor $3 \cdot 3$ is due to colour and number of generations.
So,
the amounts of non-conservation of baryon and lepton numbers are
related:
\[
                  \Delta N_{e}=
                         \Delta N_{\mu}=
                             \Delta N_{\tau}=
	                     \frac{1}{3}\Delta B,
\]
$(B -L)$ is conserved while $(B + L)$ is violated.

\begin{figure}
  \begin{center}
    \mbox{\epsffile{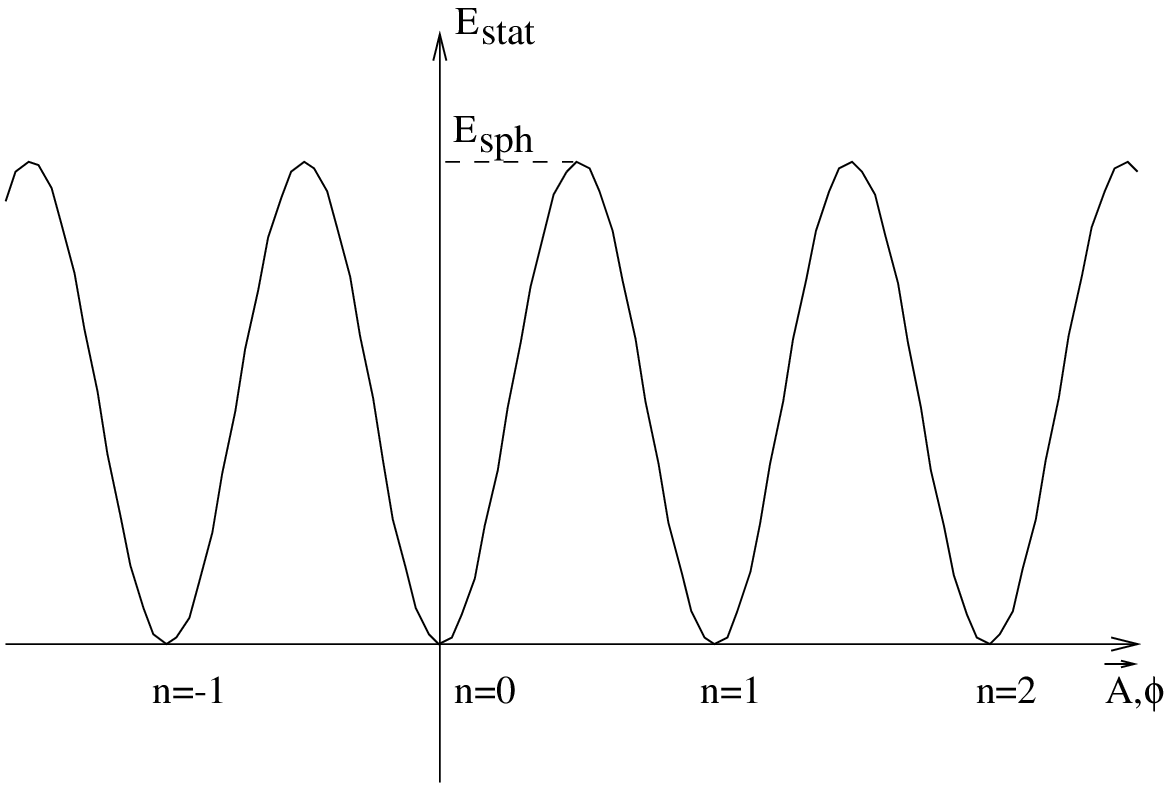}}
  \end{center}
\caption{\protect
Schematic plot of the static energy as function of gauge and
Higgs fields. The minima correspond to the classical vacua.}
\label{periodic}
\end{figure}

The analysis of gauge field configurations with the non-zero
topological number
(\ref{I2*}) is conveniently performed in the gauge
\[
	       A_{0} = 0.
\]
In this gauge, there exists a discrete set of classical vacua, i.e.
pure
gauge configurations
\[
       A_{i} = \omega \partial_{i} \omega^{-1},
\]
\[
	 \phi = \omega \phi_{0},
\]
where
        $\phi_{0} = (0, v/\sqrt{2})$
is the Higgs field in the trivial vacuum.
The gauge functions $\omega$ depend only on spatial coordinates
$\omega =\omega({\bf x})$ and are characterized by an integer:
\[
     n[\omega] =
       -\frac{1}{24\pi^{2}} \int ~d^{3}x~\epsilon^{ijk}
       \mbox{Tr} ( \omega\partial_{i}\omega^{-1}\cdot
       \omega\partial_{j}\omega^{-1}\cdot
       \omega\partial_{k}\omega^{-1}).
\]
The vacua with different $n[\omega]$ cannot be continuously deformed
into
each other without generating non-vacuum gauge fields,
so these vacua are separated by a potential barrier.
Therefore, the gauge--Higgs system is similar to a particle in
periodic
potential, as shown in fig. \ref{periodic}. An explicit construction of
the minimum energy path connecting the neighboring vacua was carried out
in ref. \cite{akiba}, and the fermion sea contribution to this path was
evaluated in ref. \cite{goeke}.

The topological number density entering eq. (\ref{I2*}) is a total
derivative,
\[
 \frac{1}{32\pi^{2}} \mbox{Tr} (F_{\mu\nu}\tilde{F}_{\mu\nu}) =
	    \partial_{\mu} K_{\mu},
\]
where
\[
   K_{\mu} = \epsilon^{\mu\nu\lambda\rho}
	       \mbox{Tr} \left(F_{\nu\lambda} A_{\rho} -
	       \frac{2}{3} A_{\nu} A_{\lambda} A_{\rho}\right).
\]
If one is interested in vacuum--vacuum transitions, then
\[
 N[A] = \int~d^{3}xdt~
	    \partial_{\mu} K_{\mu}=
	    \left[\int~d^{3}x~ K_{0}\right]^{t=+\infty}_{t=-\infty}
\]
\[
{}~~~~~~~~~~~~~~~~~~=n[\omega_{t=+\infty}]-
                         n[\omega_{t=-\infty}].
\]
So, the topological number of the gauge field is non-zero for
transitions
between the distinct vacua.

At zero energies and temperatures, the transition between vacua with
different $n[\omega]$ is a tunnelling event which
is described by instantons
\cite{BRST} (constrained instantons in theories with the Higgs
mechanism
\cite{Affleck1}). In pure Yang--Mills theory an instanton is the
solution to
the Euclidean field equations which is an absolute minimum of the
Euclidean
action in the sector $N[A]=1$. Properties of instantons are reviewed
in
ref. \cite{SVZinst}. The instanton field, up to gauge
transformations, is
\begin{equation}
       A^{a}_{\mu} = \frac{1}{g}\eta_{\mu\nu a}
		      \frac{2 x^{\nu}}{x^{2} + \rho^{2}},
\label{I5*}
\end{equation}
where $\eta_{\mu\nu a}$ are the 't Hooft symbols, and $\rho$ is an
arbitrary scale to be integrated over. The instanton action is
\[
	S_{inst} = \frac{8\pi^{2}}{g^{2}}
\]
and the tunnelling amplitude is proportional to
\begin{equation}
     A_{inst} \propto \e^{- S_{inst}}.
\label{I7**}
\end{equation}
In the electroweak theory, the tunneling probability is unobservably
small,
\begin{equation}
     \sigma_{inst} \propto \exp\left(-\frac{4\pi}{\alpha_{W}}\right)
     \sim 10^{-170},
\label{I6*}
\end{equation}
where
\[
        \alpha_{W} = \frac{g^{2}}{4\pi} =
	\frac{\alpha}{\sin^{2}\theta_{W}}=\frac{1}{29}.
\]

In theories with the Higgs mechanism there appears a slight
complication.
There are no solutions to Euclidean field equations, i.e. no exact
minima
of the Euclidean action in sectors with $N[A]\neq 0$. The reason is
that
the action for configurations like (\ref{I5*}), with
appropriate Higgs field,
depends on the instanton size $\rho$ and decreases as $\rho$ tends to
zero.
To evaluate the functional integral in that case one introduces a
constraint that fixes the size of the configuration \cite{Affleck1},
then
minimizes the action under this constraint and finally integrates
over
$\rho$. The outcome of this procedure is as follows. The instanton
contribution into the functional integral becomes
\cite{hooft:pr,Affleck1}
\begin{equation}
  \int~d^{4}x_{0}~ \frac{d\rho}{\rho^{5}}~ \mu(\rho)
  \exp\left(-\frac{8\pi^{2}}{g^{2}} - \pi^{2}v^{2}\rho^{2}\right),
\label{I7*}
\end{equation}
where $x_{0}$ is the instanton position and $\mu(\rho)$ is
a function of $g$ and $\rho$ that varies relatively
slowly. The integral (\ref{I7*})
is saturated at $\rho \lsim v^{-1}$, so that the size of the
constrained instantons is smaller than the inverse $W$-boson mass
$m_{W} = gv/2$. The constrained instanton configuration
is conveniently described in the singular gauge where the original
pure Yang--Mills instanton has the form
\begin{equation}
       A^{a}_{\mu} = \frac{1}{g}\bar{\eta}_{\mu\nu a}
		      \frac{2 \rho^{2}x^{\nu}}{x^{2}(x^{2} +
\rho^{2})}.
\label{I7a*}
\end{equation}
The  constrained instanton
 is given by
eq. (\ref{I7a*}) at $x\ll m_{W}^{-1}$, and exponentially decays at
large $x$,
\[
     A^{inst} \propto \e^{-m_{W}|x|}.
\]
Clearly, the tunnelling rate is still suppressed by the exponential
factor
(\ref{I6*}).

In this paper we discuss processes at high temperatures or energies.
The relevant energy scale is set by the height of the barrier between
different vacua as sketched in fig. \ref{periodic}. This height is
determined by the static saddle point solution to the
Yang--Mills--Higgs equations, the sphaleron \cite{Manton,kman}. This
solution was found previously in refs.
\cite{Dashen,Soni,Boguta,Forgacs}, but its relevance to topology was
realized only in ref. \cite{kman}. 
By simple scaling one obtains that the
static energy of the sphaleron solution in our simplified model,
which is equal to the height of the barrier at zero temperature, is
\begin{equation}
E_{sph} = \frac{2m_{W}}{\alpha_{W}}
B\left(\frac{m_{H}}{m_{W}}\right),
\label{I8*}
\end{equation}
where  $m_{H}$ is the mass of the Higgs boson. The function
$B(m_{H}/m_{W})$ has been evaluated numerically \cite{kman};
it varies from 1.56 to 2.72 as $m_{H}/m_{W}$ varies from zero to
infinity \footnote{At very large $m_{H}/m_{W}$ the situation is more
complicated \cite{YaffeSphaleron,Brihaye1989}, but the estimate
(\ref{I8*}) remains valid.}. So, the height of the barrier in the
electroweak theory is of order 10 TeV.

At energies above $E_{sph}$ the
system can in principle evolve from a neighbourhood of one vacuum
to another
in a classical way, without tunnelling
\footnote{Of course, this travel does not typically proceed exactly
through the sphaleron configuration. Some (not necessarily small)
deformations of sphalerons are considered in refs.
\cite{Minos1,Minos2,Minos3}
where topological properties of these configurations are investigated.};
as outlined above, this
classical process will lead to non-conservation of baryon and
lepton numbers. Clearly, having enough energy is a necessary, but not
a
sufficient condition for the absence of exponential suppression of
the
baryon and lepton number violation rates. Whether the exponential
suppression actually disappears or not is a matter of complicated
dynamics
which is one of the main subjects of this paper.

There
are at least two possible ways to see that fermion quantum numbers
are
indeed violated in instanton-like processes. One of them
\cite{hooft:prl,hooft:pr} makes use of zero fermion modes in
Euclidean
background fields with $N[A]\neq 0$. This approach is reviewed in
ref. \cite{SVZinst} and its Minkowskian counterpart is considered in
ref. \cite{Christ}. A more intuitive way \cite{CDGCrossing,Christ} is
related
to the phenomenon of level crossing, which is as follows. Consider
left-handed fermions in the background field
${\bf A}({\bf x},t)$
which changes
in time from one vacuum at $ t=-\infty$ to another vacuum at
$t=+\infty$
(we again use the gauge $A_{0} = 0$). At each time $t$ one can
evaluate the
fermionic spectrum, i.e. the set of eigenvalues of the Dirac
Hamiltonian
in the static background
${\bf A}({\bf x},t)$ where $t$ is viewed as a parameter. The spectrum
varies with $t$; some levels cross zero from below and some cross
zero from
above. The relevant quantity is the net change of the number of
positive
energy levels, which is the difference between the total number of
levels
that cross zero from above and from below in the course of
the entire evolution
from $t = -\infty$ to $t=+\infty$. A general
mathematical theorem \cite{APZ} says that this difference is related
to the topological number of the gauge field,
\begin{equation}
	N_{+} - N_{-}
               =n[\omega_{t=+\infty}]-
                         n[\omega_{t=-\infty}]=
			 N[A].
\label{I10*}
\end{equation}
Recall now that at vacuum values of ${\bf A}$, the ground state of
the
fermionic system has all negative energy levels filled and all
positive
energy levels empty. A real fermion corresponds to filled positive
energy
level and antifermion is an unoccupied negative energy level. As the
energy levels cross zero, the number of real fermions changes, and
the net
change in the fermion number of each left-handed doublet is
\[
    \Delta N_{F}^{(i)} = N_{+} - N_{-}.
\]
Combining this relation with eq.(\ref{I10*}) we see that the fermion
number
is not conserved indeed, and the amount of non-conservation is in
perfect
agreement with the anomaly relation, eq. (\ref{I2**}).

Although the above discussion was for massless fermions, the results
remain
valid for the standard electroweak theory where fermions
acquire masses via
the Yukawa coupling to the Higgs field
\cite{KrasnikovRT,Ringwald:1988,AnselmIohansen}.
Indeed, the triangle anomaly for baryon and lepton currents remains
valid
in the standard model, so that relation (\ref{I3*}) must hold. The
counting
of fermion zero modes in the instanton background confirms this
expectation
\cite{KrasnikovRT,AnselmIohansen,Minos4}. Also, the level crossing
phenomenon has
been explicitly found in theories of
this type \cite{VRHeavy,Kunz:1994,Minos4}. So, the
complications due to right-handed fermions and fermion masses do not
change
the picture of baryon and lepton number non-conservation.

Finally, the presence of $U(1)_{Y}$ gauge symmetry in the standard
model
does not modify the analysis to any considerable extent either. There
are
no instantons of the $U(1)_{Y}$ gauge field, while the effect of the
$U(1)_{Y}$ interactions on the  measure $\mu(\rho)$ for $SU(2)$
instantons
in eq.(\ref{I7*}) is tiny. Also, the energy of the $SU(2)$ sphaleron
is
still given by eq.(\ref{I8*}) where the factor $B$ depends also on
$\sin^{2}\theta_{W}$.
For the actual value
$\sin^{2}\theta_{W}=0.23$, the deviation of $B$ from its $SU(2)$
values
is numerically small \cite{kman,KlinkhamerU1,Brihaye:1993}.

\section{Baryon asymmetry: preliminaries}

In this section we  qualitatively discuss the issues relevant
to the main topic of this review -- electroweak baryon
number non-conservation at high temperatures and generation of the
baryon asymmetry of the Universe. These issues will be considered in
much more detail in the following sections, so this section may be
regarded as a guide for a reader not familiar with the subject.
Most of what is said in this section should not be taken too
literally:
we will somewhat oversimplify the picture of the electroweak
physics in the early Universe and hence will use fairly loose terms.

In hot Big Bang cosmology, there is an epoch of particular interest
from the point of view of the electroweak physics. This is the
epoch of the electroweak phase transition, the relevant temperatures
being of the order of a few hundred GeV
\cite{kir,kirlin,doljack,weitemp}.
Before the phase transition (high temperatures),
the Higgs expectation value is zero, while after the phase
transition the Higgs field
develops a non-vanishing expectation value. The critical
temperature $T_c$ depends on the parameters of the
electroweak theory; in the Minimal Standard Model (MSM) the only
grossly
unknown parameter is the mass of the Higgs boson,
$m_{H}$. In extensions of the MSM, there are more parameters that
determine
$T_c$.

At sufficiently small $m_H$ in MSM, the phase transition is
of the first order, while at large $m_H$ the exact nature of the
phase
transition is still not clear:  it may be weakly first order, second
order or smooth cross-over. It is important that the masses of $W$-
and
$Z$-bosons immediately after the phase transition, $m_W(T_c)$ and
$m_Z(T_c)$, are smaller than their zero temperature values; the
precise
behaviour of $m_W(T)$ and $m_Z(T)$ again depends on the parameters of
the
model (on $m_H$ in the MSM). Generally speaking, the stronger
the first order
phase transition, the larger $m_W(T_c)$ and $m_Z(T_c)$. The
electroweak
phase transition is considered in more detail  in section 5.

Let us now turn to the rate of the electroweak baryon number
non-conservation at high temperatures. While at zero temperatures
the $B$ non-conservation comes from tunnelling and is unobservably
small because of the tunnelling exponent, it may proceed at high
temperatures via thermal jumps over the barrier shown
in fig. \ref{periodic}
\cite{krs85}. At temperatures below the critical one, $T<T_c$,
the probability to find the system at the saddle point separating
the topologically distinct vacua is still suppressed, but now by
the Boltzmann factor,
\begin{equation}
 \Gamma \propto \exp \left( - \frac{E_{sph}(T)}{T}\right),
\label{p4*}
\end{equation}
where
\[
      E_{sph}(T)=
	   \frac{2m_{W}(T)}{\alpha_W} B\left(\frac{m_H}{m_W}\right)
\]
is the free energy of the sphaleron. Once the system jumps up to the
saddle point (i.e. once the sphaleron is thermally created), the
system
may roll down to the neighbouring vacuum, and the baryon and lepton
numbers may be violated. Therefore, the factor (\ref{p4*}) is also
the suppression factor for the rate of the electroweak baryon number
non-conservation at $T<T_c$.

At $T>T_c$, the exponential suppression
of the baryon number non-conserving transitions is absent. The
power-counting estimate of the rate per unit time per unit volume in
the unbroken phase is then \cite{armc:zero,ks}
\begin{equation}
	  \Gamma = \const \cdot \left(\alpha_W T \right)^4,
\label{p5*}
\end{equation}
where the constant is of order $1$. The rate of the electroweak
$B$ non-conservation is considered in detail in section 4.

The rates (\ref{p4*}) and (\ref{p5*}) are to be compared with the
rate of expansion of the Universe,
\[
	\frac{1}{t_U} = \const \cdot \left(\frac{T}{M_{Pl}}\right) T,
\]
where the constant is of order $10^{-1}$. Clearly, in the unbroken
phase the $B$ non-conservation rate is much higher than the
expansion rate in a wide interval of temperatures,
$T_c <T < 0.1\cdot M_{Pl}\alpha_{W}^{4} \sim 10^{12}$ GeV.
Therefore, the electroweak $B$ non-conserving reactions are fast
at these temperatures. After the phase transition the situation
is more subtle: the rate of $B$ non-conservation exceeds the
expansion
rate if the phase transition is weakly first order ($m_W(T_c)$ is
small) or second order or of the cross-over type;
on the other hand, the rate of $B$-violating processes
is much lower than the
expansion rate if the phase transition is strongly first order
($m_{W}(T_c)$ is large enough). The electroweak $B$
non-conservation switches off immediately after the
phase transition if $\frac{E_{sph}(T_c)}{T_c} > 45$ (see sections 6
and 7)
and operates after the phase transition in the opposite case.
This inequality is not satisfied in the MSM (section 7)
with $m_{top}=175$ GeV and experimentally allowed Higgs
mass $m_{H} > 65$ GeV.
So, the $B$-violating reactions are fast after the phase
transition in the MSM.

In the extensions of the MSM, the properties of the
phase transition are determined by more parameters than just the zero
temperature Higgs boson mass. So, for some region of the parameter
space,
the electroweak $B$ non-conservation is negligible after the phase
transition.

Clearly, the above observations are directly relevant to the problem
of the
generation of the baryon asymmetry of the Universe whose quantitative
measure is the dimensionless ratio of the baryon number density to
entropy density,
\[
     \Delta_B = \frac{B}{s}.
\]
This quantity is almost constant during the expansion of
the Universe at the stages when baryon number is conserved,
and its present value is
\[
      \Delta_B = (4-6)\times 10^{-11}.
\]
Several possibilities
to generate the baryon asymmetry are  discussed in the literature,
which differ by
the characteristic temperature at which the asymmetry is produced.

(i) {\em Temperature of grand unification}, $T \sim 10^{15}-10^{16}$
GeV.

A viable possibility is that the observed baryon asymmetry is
generated by baryon number violating interactions of grand unified
theories. The effect of the electroweak processes is basically that
$(B+L)$, generated at grand unified temperatures,
is washed out at some later
time
(recall that $(B-L)$ is conserved by anomalous electroweak
processes).
The asymmetry may survive from the grand unification epoch only if a
large
$(B-L)$ asymmetry is generated at $T\sim 10^{15}-10^{16}$ GeV, and
there are no strong lepton number violating interactions at
intermediate
temperatures, $100$ GeV $< T < 10^{12}$ GeV (otherwise all fermionic
quantum numbers are violated at these temperatures, and the
baryon asymmetry is washed out). The first requirement
points to non-standard,
$(B-L)$ violating modes of proton decay, though this indication is
not
strong. We discuss in section 6 some issues related to this scenario
of
baryogenesis.

(ii) {\em Intermediate temperatures}, $1$ TeV $\ll T \ll 10^{15}$
GeV.

An interesting possibility is that there exist {\em lepton} number
violating interactions at intermediate scales, and these interactions
generate a lepton asymmetry of the Universe at intermediate
temperatures.
Then this lepton asymmetry is partially reprocessed into baryon
asymmetry by anomalous electroweak interactions \cite{Fuk86}.
Possible
manifestations of this scenario are Majorana neutrino masses (which
actually may be helpful from the point of view of solar neutrino
experiments, for a review see, e.g., ref. \cite{Smir95})
and/or lepton number violating processes
like $\mu \to e\gamma$, $\mu \to eee$ and $\mu$-$e$ conversion. A
more
detailed discussion of this possibility, together with the analysis
of
concrete models, can be found
in refs. \cite{Lang86,Lu92,Ack93,Mur94,Gan94}.

Another mechanism able to generate the baryon asymmetry
at intermediate temperatures \cite{affldine}
deals with  coherent
production of  scalar fields carrying baryon number. At a
later stage the ``scalar'' baryon number stored in scalar fields is
transferred into an ordinary one. The most recent consideration of
this interesting possibility in the framework of the sypersymmetric
standard model can be found in ref. \cite{Di95}.

(iii) {\em Electroweak temperatures},
$T \sim \mbox{(a~few)}\times 100$ GeV.

The remaining possibility is that the observed baryon asymmetry is
generated by anomalous electroweak interactions themselves. Since the
Universe expands slowly during the electroweak epoch, a considerable
departure from equilibrium (the third Sakharov condition) is
possible only from the first order phase transition.
Indeed, this transition, which proceeds through the
nucleation, expansion and collisions of the bubbles of the new
phase, is  quite a violent phenomenon. The dynamical aspects of the
first  order phase transitions
in the Universe are
considered in section 5.

A necessary condition for the electroweak baryogenesis is that
the baryon asymmetry created during the electroweak phase transition
should not be washed out after the phase transition completes. In
other words, the rate of the electroweak $B$-violating transitions
has to be negligible immediately after the phase transition. As
outlined
above, and discussed in detail in section 7, the latter requirement
{\em is not fulfilled} in the Minimal Standard Model, so the
electroweak
baryogenesis is only possible in extensions of the MSM. Extending the
minimal model is useful in yet another respect: it generally provides
extra sources of CP violation beyond the Kobayashi--Maskawa
mechanism, so that the second Sakharov condition is satisfied
more easily \footnote{Though it is still not excluded that the KM
mechanism alone is sufficient for baryogenesis, see section 7.}. The
phenomenological consequences of these extra sources of CP
violation are electric dipole moments of neutron and
electron \cite{Barr88,Barr93},
whose values are expected, on the basis of the considerations
of baryogenesis \cite{Kaz92}, to be close to existing experimental
limits.

Several specific mechanisms of electroweak baryogenesis are outlined
in section 7. The outcome is that the observed baryon asymmetry may
naturally be explained within extended versions of the Standard
Model.
This result is particularly fascinating as the physics involved
will be probed at LEP-II and the LHC relatively soon. Naturally, most
of our review is devoted to the topics related to the electroweak
baryogenesis.
\section{Sphaleron rate at finite temperatures}
In this section
we attempt to describe the present situation with the computation
of the rate of  fermion number non-conservation at high
temperatures. We shall try to separate the exact results from
(plausible) assumptions. We begin with the qualitative discussion of
the rate and derive  the Van't Hoff--Arrhenius type formulae for the
rate, valid at sufficiently low temperatures. Then we derive an
exact real-time Green function representation for the rate and show
how it can be related to the more qualitative discussion. The
quantum corrections to the rate are discussed.
At the end, we present some numerical
results for the sphaleron rate.

Of course, there is much similarity between the description
of sphaleron processes and reaction-rate theory in condenced matter
physics. The latter is reviewed, e.g., in ref. \cite{ratetheor}.

\subsection{Qualitative discussion}
As outlined in Section 2,
the anomalous
fermion number
non-conservation is associated with the
transitions of the bosonic fields from
the classical vacuum of fig. \ref{periodic} to the
topologically distinct one. For the case of zero temperatures,
low
fermion densities and
low
energies of colliding particles, the
initial state of the system as well as the final state are close to
the vacuum configurations. So, to
experience
fermion number
non-conservation,
the system has to tunnel through the barrier. This
process can be described by instantons and is
strongly
suppressed by the semiclassical exponent  $\exp
(-\frac{4\pi}{\alpha_W})$.

In order to deal with topological transitions at non-zero
temperatures let us consider first a simple example of the system
with one particle in the double-well potential with
the Lagrangian:
\be
L = \fr12 {\dot x}^2 - U(x),
\ee
\be
U(x)=\fr14\lambda (x^2-c^2)^2.
\ee
 The corresponding Hamiltonian is
\be
H=\fr12 p^2 +U(x),~p={\dot x}
\ee
and the curvature of the potential at its minimun,
 $x=c$, is
\[
   U''(x=c)= m^2 = 2 \lambda c^2
\]

Suppose that a particle is initially in the left well, and we want to
calculate the probability of finding this particle in the other well.
Let us take first the case of zero temperatures and consider the
transition from the classical
ground state. The probability of tunnelling can
be found in the  WKB approximation and is of
the
order of
\be
P \sim \exp(-2S_0),~~ S_0 = \int_{-c}^{c}\sqrt{2U(x)}~dx.
\ee
It is exponentially suppressed provided that the energy barrier
separating different classical ground states is sufficiently high.

At finite temperatures, in addition to the ground state in the left
well there are excited states with non-zero energy $E$. The
probability of  the state with the energy $E$ is given by the
Boltzmann distribution, $\exp(-E/T)$. Hence, the rate of the
transitions is proportional to the sum of the probabilities of
transitions from the excited levels with energy $E$ weighted with the
thermal distribution,
\be
P \sim \sum_n \exp(-E_n/T -2S(E_n)),
\ee
where
\be
S(E) = \int_{-x(E)}^{x(E)}dx\sqrt{2(U(x)-E)},~~~~~U(x(E)) = E.
\ee
At temperatures $T\gg m$ the sum can be approximated by the
integral over $x$,
\be
P \sim \int dx \exp\left[-\frac{U(x)}{T}
            -2 \int_{-x}^{x}dy\sqrt{2(U(y)-U(x))}\right],
\ee
with the result
\be
P\sim \exp(-U_0/T),
\label{P}
\ee
where $U_0=U(0)=\fr14\lambda c^4$ is the height of the barrier. This
result is  clear from the physical point of view.
Namely, $P$ in eq. (\ref{P}) counts the number of
states with energy higher than the
height of the  barrier. At temperatures
$T<m$ the number of these states is exponentially suppressed by the
Boltzmann exponent and their contribution is smaller than the
contribution of tunnelling from the vacuum state.
On the other hand,
at high temperatures $T\gg m$ the main contribution to the
transition rate comes from the states with energy higher than the
height of the barrier, which can overcome the barrier classically.
Hence, we can address the problem of
interest by the entirely classical
calculation of the rate, which is equal to the probability flux in
one direction (from left to right) at the
point $x=0$ (see ref. \cite{ratetheor} and references therein),
\be
\gamma = \langle\delta(x)\theta(\dot{x})\dot{x}\rangle =
\frac{\int dp dx \exp(-H/T) \delta(x)\theta(p)p}
{\int dp dx \exp(-H/T)} = \frac{m}{2\pi}\exp(-U_0/T),
\label{qm}
\ee
if $T \ll U_0$. Note that the curvature of the
potential near the saddle point at $x=0$ does not enter  the final
result; the quantum constant $\hbar$ does not appear in the answer at
all. Note also that the classical treatment of the
problem is applicable only if $U_0/T < 2S_0$. In the opposite case
the rate of the quantum tunnelling is higher than the rate of the
classical transitions. At the same time, the saddle point
approximation we used for the calculation of the integral (\ref{qm})
is valid only for $U_0/T \gg 1$. If
the latter
relation does not hold,
the calculation should go
beyond the saddle-point approximation.
This discussion can be easily generalized
to the case of the
systems with many degrees of freedom, in particular to the
field theory we are interested in \cite{langer,affleck}.  

Let us consider specifically sphaleron transitions.
As in the quantum-mechanical
example discussed previously, we would like to put our
system initially in the vicinity of one of the topological vacua, say
with
$n=0$, and determine the rate at which the system moves to
neighbouring
vacuum sectors. The sphaleron configuration, lying on a minimal
energy path connecting two close-by vacua with different topological
numbers, plays a crucial role in the computation.

The energy functional near the sphaleron configuration can be written
in the quadratic approximation as
follows,
\be
{\bf H} = E_{sph} -\frac{1}{2}\omega_-^2 x_-^2 +
\frac{1}{2} \sum \omega_i^2 x_i^2 + \fr12 \sum p_i^2,
\ee
where $x_i$ and $p_i$ are the normal coordinates and momenta,
$\omega_i > 0$ are corresponding frequencies, and the index ``-''
refers to the negative mode. Now, the surface $x_{-}=0$
in the configuration
space  is the complete analogue of the saddle point $x=0$ in
the $1$ degree
of freedom model considered above. If we put our
system on this surface, and let it evolve with time, then it will
almost definitely move to the sector $n=1$ (0) (and stay there for
long
time)
if the projection of initial momenta to the normal to the surface is
positive (negative). So, to count the number of transitions with the
topological number change we should calculate the probability flux
through the surface $x_-=0$ in one particular direction
(cf.
refs. \cite{langer,affleck}). The rate per unite time and unite
volume is
given by \cite{armc:zero,bs:two,ks}
\be
\Gamma = \frac{1}{Z}\frac{1}{V}\int \prod_i dx_i dp_i dx_-dp_-
\delta(x_-)\theta(\dot{x}_-)
\dot{x}_- \exp (- {\bf H}/T),
\label{gaussint}
\ee
where $Z$ is the statistical sum
\be
Z=\int {\cal DP}{\cal DQ}\exp(-H/T)
\ee
and $V$ is the volume of the system. As in the simple example, the
rate does not depend on the curvature along the sphaleron negative
mode. After Gaussian integration over momenta the result may be
written in a compact form
\be
\Gamma = \frac{1}{V}\frac{1}{Z}\frac{\omega_-}{2 \pi T}\mbox{Im}
F_{sph},
\ee
where $\mbox{Im} F_{sph}$ is
the
result of the {\em formal computation} of the
imarinary part of the free energy near the sphaleron in the one-loop
approximation (since the sphaleron is a saddle point of the energy
functional rather than its minimum, the functional integral around it
does not exist).

The treatment of the sphaleron zero modes is fairly standard. The
total number of sphaleron zero modes is 6. Three
translational
zero modes restore the correct volume dependence
of the rate while $SU(2)$ transformations introduce some
normalization factor. The final result for the rate reads
\cite{armc:zero}
\be
\Gamma = \kappa \frac{T^4 \omega_-}{M_W}
\left(\frac{\alpha_W}{4\pi}\right)^4
N_{tr} N_{rot} \left(\frac{2M_W}{\alpha_W T}\right)^7 \exp \left(-
\frac{E_{sph}}{T}\right) .
\label{rate}
\ee
Here the factors $N_{tr} \simeq 26,~N_{rot} \simeq 5.3 \cdot 10^3$
come from the zero mode normalization \cite{armc:zero}, $\kappa$ is
the
determinant of non-zero modes near the sphaleron. Again this result
is purely classical, and it does not contain $\hbar$. It is not
applicable at very low temperatures where
$\frac{E_{sph}}{T}>S_{inst}$
(there
 quantum tunnelling is more
important
than
classically allowed transitions), and at high
temperatures where the exponent is not large compared
to $1$.

There are several important assumptions used in
the above
derivation of the
sphaleron rate. The first one is the applicability of the classical
theory to the description of the topology change in high temperature
plasma. The second one is inherent in the classical theory itself.
The
thermodynamics of the classical field theory is, strictly speaking,
ill defined due to the Rayleigh--Jeans instability (in field theory
language there are ultraviolet divergences). Even besides this, we
{\em assumed} that if the system is initially on the surface $x_-=0$
then it will move to the sector with topological charge $n=1$
provided
that
$p_->0$ and to $n=0$ in the opposite case \footnote{This is only true
in the quadratic approximation of the energy functional near the
surface. Higher order terms in the expansion of the energy functional
will, in general, introduce recrossings of the surface
\cite{armc:zero} and modification of the rate.}. None of these
asumptions has been proved with any rigour up to now. A number of
quantum corrections to the rate, associated with the contributions of
high momentum particles ($k\sim T$) can be taken into account, the
infinities in the classical theory can be properly dealt with at the
one-loop level, but the regular procedure of evaluating higher order
(two-loop, etc.) contributions to the rate is not known. Moreover,
the
approach discussed above does not shed any light on the rate at high
temperatures, when the sphaleron approximation breaks down.

Below we will describe
the
Green's function approach to the
problem
of
$B$ non-conservation, which follows from the first principles of
statistical mechanics and allows (at least in principle) the
discussion of the $B$ non-conservation rate beyond
the semiclassical approximation \cite{ks,mms:prd,mottola}.

\subsection{The Green's function approach}
Let us consider first the behaviour of the quantity $Q(t)$
\be
Q(t)=\int_0^t q(x) d^4x
\label{Q(t)}
\ee
at high temperatures in a quantum system without fermions. Here
$q(x)$ is the topological number density
\be
q(x)=\frac{g^2}{32 \pi^2}F_{\mu\nu}^a {\tilde F}_{\mu\nu}^a.
\ee
If the system starts in the vicinity of one classical vacuum (say,
$n=0$),
then, due to the sphaleron-like transitions, it will move randomly to
the vicinity of the other vacua of fig. \ref{periodic}.
Because of the periodicity
of the static energy, these processes will not increase the free
energy.
In other words, the quantity $Q$ makes a
``random walk''
in the space of configurations, and
\be
\langle Q^2 \rangle = 2\Gamma V t, ~~~t\rightarrow \infty,
\ee
where $V$ is the volume of the system, $\langle O \rangle =
Tr(O\rho)$, and $\rho$ is the equilibrium density matrix, $\rho =
\frac{1}{Z}\exp(-\frac{H}{T})$. The quantity $\Gamma$ is nothing but
the rate of the transitions with a change of the topological number
per unit time
per
unit volume. It is given by the correlation
function
\be
\Gamma = \fr12\lim_{t\rightarrow\infty}\lim_{V\rightarrow \infty}
\int_0^t \langle(q(x)q(0)+q(0)q(x))\rangle d^4x.
\label{grn2}
\ee

Now, we can derive a fluctuation--dissipation theorem showing that
the
kinetic coefficient describing the relaxation of the fermion number
is directly related to
the
rate $\Gamma$. Let us add left-handed (and
right-handed) massless fermions to our system, switch off the Yukawa
interaction with scalar fields, and suppose that the only deviation
from the thermal equilibrium is that assosiated with the presence of
small lepton and baryon numbers. In other words, all other
interactions are supposed to be
faster
than those associated with
anomalous $B$ non-conservation\footnote{In  reality this is not
always the case. For example, the chirality breaking reactions for
massive light quarks and leptons due to Yukawa couplings may be
slower that the anomalous reactions. The
modification of the kinetic
equations in this case is discussed in refs.
\cite{krs87,bs:higgs,Harvey,Dreiner}.}.
For simplicity we assume  that the averages
of all conserved fermion
numbers are equal to zero
\footnote{A more general case is considered
in refs. \cite{krs87,ks,Dreiner}.},
so
that  the only non-vanishing global charge is $\langle (B+L)\rangle=
n V $.
Here $B$ and $L$ are
{\em left} baryon and lepton numbers (in the massless limit only
left-handed particles participate in the anomalous processes).

In principle, the time dependence of the baryon number can be found
from the solution of the Liouville equation for the density matrix
$\rho(t)$,
\be
\frac{\partial \rho(t)}{\partial t}=i[\rho,H]
\ee
with the
following
initial
condition at
$t=0$:
\be
\rho_0 = \frac{1}{Z}
         \exp\left[-\frac{1}{T}\left(H +\mu_0(B + L)\right)\right],
\label{initmat}
\ee
where $\mu_0$ is the initial value of the chemical potential and
$Z$ in the statistical sum. Then,
\be
\langle B+L \rangle(t) = Tr [\rho(t) (B+L)] = Tr [\rho_0
(B(t)+L(t))],
\ee
where $B$, $L$  ($B(t),L(t)$) are  the  operators of baryon and
lepton numbers in Schr\"{o}dinger (Heisenberg) representation.
We expect that if the density $n$ is small compared
with $T^3$, then
the time dependence of $n$ follows from the kinetic equation
\be
\frac{\partial n}{\partial t} = - \Gamma_B n,
\ee
where $\Gamma_B$ is the rate of the fermion number non-conservation
we are interested in. Probably, the
easiest
way to derive this
kinetic equation is to make use of the Zubarev formalism of the
non-equilibrium density matrix \cite{zubarev,zub:book}.
Zubarev defines two density matrices to be considered: the first one
is
the
so-called local equilibrium density matrix, which is time dependent
only through $\mu(t)$,
the
instant chemical potential
for the operator $(B+L)$:
\be
\rho_{le} = \frac{1}{Z} \exp\left(-\frac{1}{T}[H +\mu(t)(B +
L)]\right).
\label{loceq}
\ee
The magnitude of $\mu$ slowly varies in time due to $B$ and $L$
non-conservation. The average value of $n$ is related to
$\mu(t)$ as
follows,
\be
n  = \frac{1}{V} Tr[(B+L)\rho_{le}]=\fr23 \mu(t)T^2 N_f,
\ee
where $N_f$ is the number of fermion generations and we take into
account that the baryon number of a quark is $\fr13$, and the
number of colours is $3$. The change in the baryon number is
related to the change in the chemical potential, $\partial n = \fr23
N_f T^2 \partial \mu(t)$.

At the same time, due to the anomaly equation,
\be
\frac{\partial}{\partial t}n(t)= 2N_f
\mbox{Tr}(q(t)\rho_0).
\ee
The main difficulty is to compute the large time asymptotics of this
expression. Zubarev argues that at $t> \Gamma_B$ the following
so-called non-equilibrium density operator can be used instead of
$\rho_0$:
\be
\rho_{Zub} = \frac{1}{Z} \exp \left[-\frac{1}{T}\left(H +\epsilon
\int
_{-\infty}^t dt'e^{-\epsilon(t-t')}\mu(t')(B(t') +
L(t'))\right)\right],
\label{zub}
\ee
where operators $B$ and $L$ are taken in the Heisenberg
representation
and the limit $\epsilon \rightarrow +0$ is assumed. This density
matrix is static in this limit. Now, expanding the density matrix
(\ref{zub}) with respect
to
 small
$\mu$, integrating over $t'$ by
parts and neglecting the time derivative of $\mu$ one obtains
\be
Tr [q(t)\rho_{Zub}]=\fr13 T^2 \mu(t) \Gamma_B,
\ee
where the rate of the baryon number dilution is written in terms of
the retarded Green function,
\be
\Gamma_B= \frac{3}{T^2}\lim_{\epsilon\rightarrow +0}\int_{-\infty}^t
dt' d^3x
[q(t),n(t')]_-\exp(-\epsilon(t-t')).
\label{grn1}
\ee
Now, the use of the spectral decomposition for the correlation
function (\ref{grn2}) and Green's function (\ref{grn1})
\cite{mms:prd,mottola} shows
that these two functions
 are in fact equal to each other up to a coefficient containing
the
number of fermion generations. Finally, one finds
the
 desired
    relation
\be
\Gamma_B = 12N_f\frac{\Gamma_{sph}}{T^3}.
\ee
If the  Yukawa interactions with the Higgs particles
are faster than the sphaleron transitions,
then one obtains the
coefficient $\frac{13}{2}$ instead of $12$ \cite{bs:higgs}.

\subsection{The relation to the ``probability flux'' formulae}
At first sight, the correlation function describing the rate of
topological transitions (\ref{grn1}) has nothing to do with the
probability flux through the surface $x_-=0$ we found in the
subsection 4.1.
Below we will see that they are in fact the same in the Gaussian
approximation to the classical theory \cite{ks}, at temperatures
below
the sphaleron mass but above the particle masses. One may
expect that the classical approximation to the correlation function
may be good enough, since the quantum bosonic distribution functions
at low momenta are the same as the classical ones.

Consider for simplicity
 the purely
bosonic theory.
Notice that expression (\ref{grn2}) allows
for the
naive classical limit $\hbar
\rightarrow 0$ (we leave aside the question of renormalization for a
moment),
\be
\Gamma_{class}= \lim_{t\rightarrow \infty}\int {\cal DP} {\cal DQ}
\exp \left[-\frac{H({\cal P,Q})}{T}\right] Q(t)q(0),
\label{clrate}
\ee
where $H({\cal P,Q})$ is the classical Hamiltonian depending on the
generalized momenta and coordinates, $q(0)$ is the
density of the topological charge at
time
$t=0$ expressed
through canonical coordinates and momenta ${\cal P,Q}$, and $Q(t)$
is the topological charge (\ref{Q(t)}) where $A(x,t)$ is
the
solution to
the classical equations of motion with initial conditions ${\cal
P,Q}$.  Derivation of the classical limit goes precisely along the
lines of the corresponding analysis for the quantum mechanics
given in ref. \cite{Dolan}.

Intuitively, the main contribution to the path integral
(\ref{clrate})
comes from configurations that start in the vicinity of one of the
classical
vacua
of our system, evolve in time, pass near the saddle
point (sphaleron), and relax in the vicinity of the other vacuum. For
these configurations, $Q(\infty)=\pm 1$, depending on the direction
of
motion.

Consider the classical configurations crossing the surface $x_-=0$
near the sphaleron at some time $t_1$. They have the form:
\be
A({\bf x})= A_{sph}({\bf x} - {\bf x}_0,\Omega)
       +\sum x_n A_n({\bf x}-{\bf x}_0,\Omega),
\ee
\[
\phi({\bf x})= \phi_{sph}({\bf x}-{\bf x}_0,\Omega)+\sum x_n
\phi_n({\bf x} - {\bf x}_0,\Omega).
\]
Here ${\bf x}_0$ and $\Omega$ are the collective coordinates corresponding
to the sphaleron translational and rotational zero modes, and $x_i$
are small. According to the Liouville theorem, the phase space is
invariant, so that we can write
\be
\left({\cal DP} {\cal DQ}\right)(0)=
\left({\cal DP} {\cal DQ}\right)(t_1)=
\Pi~ {\cal D}x_n {\cal D}p_n d{\bf x}_0 d\Omega N_{tr} N_{rot}|p_-|
dt_1
\ee
where $p_n$ are the momenta corresponding to the coordinates $x_{n}$,
$dx_- =|p_-| dt_1$, $N_{tr}$ and $N_{rot}$ are the normalization
factors for the translational and rotational zero modes.  Now,
\be
\int d^{3}x_0 \int_0^{t_1} dt_1 q(0)=\int d^{3}x_0 \int_0^{t_1} dt_1
q(-{\bf x}_0,-t_1)
\ee
is nothing but the topological charge of the configuration which
at time $t_1$ belongs to the surface $x_-=0$.  Then, if the momentum
corresponding to the negative mode is positive,  the
configuration under consideration was evolving in time with the
increase
of the coordinate $x_-$, producing on average the topological charge
$\fr12$.
In the opposite case the average topological charge is $-\fr12$.
So, for these configurations
\be
\int d^3 x_0 \int_0^{t_1} dt_1 q(-{\bf x}_0,-t_1)=\fr12 \mbox{sign}(p_-).
\ee
and $Q(\infty)= \mbox{sign}(p_-)$.
Finally, the correlation function (\ref{clrate}) is
\be
\Gamma= \fr12 \frac{1}{Z}N_{tr}N_{rot}\int{\cal D}x_n {\cal D}p_n
|p_-|dp_-\exp[-H(x_n,p_n)]
\label{fcnint}
\ee
coinciding
with
what was found previously. One can see that the set
of assumptions used in the computation of this correlation function
is precisely the same as that for the estimate of the probability
flux.

\subsection{Quantum versus classical rate}
In spite of the fact that we were able to write an exact quantum
real-time correlation function describing the sphaleron rate, there
are no
regular methods which allow for the actual
computation. The Euclidean (Matsubara)
field theory perturbative methods are of little help here, since the
analytical continuation to the real time is necessary, which is
hardly
feasible by making use of the perturbation theory.

The following arguments suggest that the leading quantum effects may
be absorbed into the parameters of the classical theory, at least at
sufficiently high temperatures. In the consideration of the sphaleron
rate the  most important role was played by the properties of the
static configurations of the gauge and Higgs fields. They are
certainly influenced by the precence of the  high-temperature
excitations. Nevertheless, it can be shown (see the discussion in
Section 5)  that all static {\em
quantum} temperature Green's functions for bosonic fields $\phi, A_i$
coincide,
up to $O(g^3)$ terms, with the
static  temperature Green's functions for the {\em
classical} bosonic theory with the Hamiltonian
\be
H= \int d^3x ~\left[\fr12 E_i(x)^2 + P^*(x)P(x) +
\frac{1}{4}G^a_{ij}G^a_{ij}+
(D_i\Phi)^{\dagger}(D_i\Phi)+U(\phi)\right],
\label{class}
\ee
\[
U(\phi)=m^2(T)\Phi^{\dagger}\Phi+\lambda(T)
(\Phi^{\dagger}\Phi)^2.
\]
Here $E_i$ and $P$ are the momenta conjugate to the fields $A_i$ and
$\phi$ \footnote{To be more precise, the equivalence holds  for any
form of the kinetic part of the classical Hamiltonian, provided it
contains momenta only. Then for static Green's functions the
integrations over momenta and coordinates are factorized. This
ensures the equivalence of the classical high temperature equilibrium
statistics and quantum 3d zero temperature euclidean theory.}.
The coupling constants $g(T)$, $\lambda(T)$ and the mass $m^2(T)$ of
the classical field theory can be found by
well-defined perturbative prescription,
to be discussed in more detail in section 5.
The classical theory does not contain fermions, which are integrated
out by the procedure of dimensional reduction (see section 5).
Static classical Green's functions are finite, provided known one-
and
two-loop counterterms are added to the Hamiltonian (\ref{class}). The
perturbative transition to the classical theory is possible in weakly
coupled theories only. Moreover, the static correlation lengths in
the theory must be large enough ($l \gg (\pi T)^{-1},~l\gg
(gT)^{-1}$) for the approximation to be valid. The
Rayleigh--Jeans instability is nothing but the
infinite renormalization of the vacuum energy in the 3d theory,
which can be
removed by vacuum counterterms.
Therefore, we see that the static energy barriers can be found
from the analysis of the
saddle points of the energy functional defined in eq. (\ref{class}).
If $m^2(T)$ is negative, the symmetry is spontaneously
broken, and the sphaleron solution does exist. Its energy now
is temperature dependent via quantum corrections to the
parameters of the classical theory. The size of the
high-temperature sphaleron is of the order of the static
correlation length
$l$ and is much larger, according to our assumption, than
the inverse temperature (typical distance between particles). Now we
recall that in the classical computation of the probability
flux through the surface $x_-=0$ the integration over momenta
is Gaussian (see eq.(\ref{gaussint})) and the main contribution comes
from
momenta $p_n^2/2 \sim T$. Therefore, for the normal sphaleron
modes with $\omega_n \ll T$ the real time motion indeed can be
considered as  the classical one. This is not true for $\omega_n >
T$,
but high energy sphaleron modes with $\omega_n \gg l^{-1}$
are close to those around the vacuum configuration and,
therefore, cancel out in the rate computation.

In the symmetric phase,
at
$m^2(T)>0$,
the sphaleron solution does
not exist. However, the typical static correlation length in the
system,
$l \sim (g^2 T)^{-1}$,
is still much larger than the
inverse
temperature. It is natural to assume that the typical size of the
configurations contributing to the rate is of the order of the
correlation length; then the argument given above indicates the
possibility of the classical description.

So, the {\em conjecture} (which has never been proved, though) is
that the rate of the fermion number non-conservation
in
 quantum
field theory at high temperatures ($T\gg m_W(T),~gT\gg m_W(T)$) is
given, up to terms $O(g^3)$, by the classical correlation function
(\ref{grn2}) with the Hamiltonian (\ref{class}).
Since the statistical mechanics of the classical field theory does
not exist due to ultraviolet divergences, the latter statement
requires clarification. To define the classical statistics, an
introduction of some high energy cutoff $\Lambda$ is necessary,
together with 3d counterterms removing divergencies from the
potential part of the classical Hamiltonian.
For this conjecture
to be true, the necessary condition is that the classical correlation
function (\ref{grn2})  exists in the limit
\be
\lim_{\Lambda\rightarrow\infty}\lim_{t\rightarrow\infty}\lim_{V
\rightarrow\infty}G
\ee
(the order of limits is essential here, see below). Some evidence in
favour of the last statement has been found by a number of direct
real time Monte Carlo simulations,
which
will be
considered in the last subsection. If the conjecture is true,
it allows an immediate  determination of the parametric
dependence of the rate
of sphaleron transitions:
\be
\Gamma = (\alpha_W T)^4 f\left(\frac{\lambda}{g^2},\frac{m^2(T)}{g^4
T^2}\right),
\ee
since the rate has dimensionality (GeV)$^4$, and the classical
dynamics of the theory is governed by the unique dimensionful
coupling $g^2 T$ and two dimensionless ratios. Deeply in the
symmetric phase, at $T\gg T_c$, the scalar degrees of freedom
decouple, since $m^2(T)\gg g^4 T^2$. Then  $\Gamma \sim (\alpha_W
T)^4$ \cite{armc:zero,ks}.

Recently, another
conjecture was put forward in ref. \cite{Bodeker}. The authors
suggest that
instead of the classical Lagrangian (\ref{class}) one should use the
Braaten--Pisarski effective action  \cite{Braaten},
which sums up so-called hard thermal
loops in the particle amplitudes. This action is a generating
functional for the gauge and matter fields at soft momenta; it was
used in ref. \cite{Braaten1}
for the calculation of the  the damping rate of fermions in
the plasma. Yet another suggestion is to use Langevin-type equations
with a friction term together with a
random force \cite{Bochkarev}, instead of
any type of deterministic equation of motion. The random force is
served to mimic the interaction of the classical soft momentum
modes with the short wave quantum fluctuations. These conjectures
remain not proved either.

\subsection{The sphaleron rate in the broken phase}
The discussion in the previous subsections shows that in the broken
phase,
the rate of sphaleron transitions is given by expression
(\ref{fcnint}),
where the classical static energy which should be used in the
evaluation of the functional integral is given by eq. (\ref{class}),
with
temperature-dependent masses and couplings.  So, in this regime, the
rate is given by
eq.(\ref{rate})
with the replacement $M_W\rightarrow M_W(T)$
where
$M_W(T)= \fr12 g(T)v(T)$, and the expectation value of the Higgs
field
is to be determined by the minimization of the classical potential
$U(\phi)$.

To completely determine the rate, one should calculate the
3-dimensional
determinant of small fluctuations around the sphaleron solution,
which is also temperature-dependent. Formally, it diverges linearly with
ultraviolet
cutoff, but this divergence is removed by the one-loop counterterm
for the scalar mass $m^2(T)$. Recently, this computation has been
done
numerically in refs. \cite{Baacke1,Baacke2,Baacke3} where
the values of the determinant can be found
at different scalar coupling constants. For the small ratio of
$\lambda(T)/g^2(T)$ the result has a simple form. Namely, instead of
taking the tree value for the vacuum expectation value for the scalar
field, one may obtain it from the
minimization
of the one-loop
effective potential,
\begin{eqnarray}
V_1(\phi) &=& \fr12 m^2(T)\phi^2
+\fr14\lambda(T)\phi^4 \nonumber \\
&& -{1\over12\pi} \biggl(6m_T^3 + m_1^3+3m_2^3 \biggr),
\la{1looppot}
\end{eqnarray}
where the mean field-dependent masses are defined  as
\be
 m_T=\fr12 g(T)\phi, ~~~~
m_1^2=m_3^2(\mu_3)+3\lambda(T)\phi^2,~~~~
 m_2^2=m_3^2(\mu_3)+\lambda(T)\phi^2.
\la{masses}
\ee
Then the 3d determinant $\kappa\simeq 1$; the precise numbers are
given in ref. \cite{Baacke3}. This
is the most complete calculation of the
sphaleron rate in the framework of 3d approach done up to now.

Recently both the bosonic and fermionic determinants in the background of
the sphaleron have been calculated  for any temperatures in refs. 
\cite{Dyakonov1,Dyakonov2}.  Probably, this is the most
involved and non-trivial computation  done until now for the sphaleron rate.
Also, the change of the fermion number and behaviour of the 
relevant fermion
level 
during the sphaleron transition has been followed.
The authors conclude that the fermionic contribution is numerically 
important for $m_t = 175 GeV$. However, those 
calculations become unreliable close to $T_c$\footnote{We thank 
D. Diakonov and K. Goeke  for discussion of this point.}. 
 Fortunately, the full theory in the 
vicinity of $T_c$  can be reduced to a 3-dimensional problem, and one 
can show that almost entire effect of the determinants can be absorbed 
into the definition of coupling constants of the effective 3-dimensional 
theory (see section 4.4 and ref. \cite{K2}). If one uses these couplings 
to redefine the sphaleron 
solution, the remaining  contributions are shown to be small \cite{Moore}. 
As a result, the sphaleron 
rate in the Higgs phase is reliably calculable at the critical temperature
for the theory with the scalar self-coupling up to $\lambda(T)/g^2(T)
<0.04$, see discussion in Section 7.1.

The evaluation of further corrections to the rate is a problem
which has not been solved, even the strategy of the necessary
computation is not known. Parametrically, they are
$O[g^2 T/M_W(T)]$.

\subsection{Real time numerical simulations}
Making use of the
conjecture on the possibility to calculate the quantum sphaleron
rate within the classical field theory, one can compute
the rate
by the numerical simulations \cite{Grigorev,grs:pl,grs:np}. The
discretization of space,
necessary for the  numerical methods,
provides a natural ultraviolet cutoff.
Probably, the most convenient discrete formulation is provided by the
lattice gauge theories.

The Monte Carlo (MC) numerical computation of the correlation
function
$\langle Q^2(t)\rangle$ consists
of
two steps. First, one should
generate a set of configurations (coordinates and momenta) in
accordance with the Boltzmann distribution $\exp(-H/T)$. Then, these
configurations are used as initial conditions for the classical
equations of motion. These equations are solved numerically, and the
topological charge is computed as a function of time. Finally, the
averaging of the quantity $Q^2(t)$ is performed. The computation
should be repeated for different volumes of the system and for
different lattice spacings; the extrapolation to the continuum limit
is to be performed at the end. Many details of the described
procedure can be found in the original
papers
\cite{grs:pl,grs:np,aaps:pl,aaps:np,Ambjorn:1994,Ambjorn:1992,Ambjorn:1995,Smit:1994,Bochkarev:1993},
here we just present some results.

Up to now, real time MC simulations of the topological number change
were performed in three different  gauge field theories. The first
one is the $1+1$ dimensional $U(1)$ Higgs model, the second is
$SU(2)$
gauge--Higgs system, and the third is a pure Yang--Mills $SU(2)$
theory.

{\bf $U(1)$ theory in 1+1 dimensions.} The Lagrangian of the $U(1)$
Higgs
model has the form:
\be
L = -\frac{1}{4}F_{\mu\nu}F_{\mu\nu} + (D_{\mu}\phi)^{\dagger}
(D_{\mu}\phi) - V(\phi),
\ee
where $F_{\mu\nu} = \partial_{\mu}A_{\nu} - \partial_{\nu}A_{\mu}$ is
the gauge field strength, $\phi$ is the charged scalar field,
$D_{\mu}=\partial_{\mu}-i e A_{\mu}$, and
$V(\phi)$ is the scalar potential,
\be
V(\phi) = \frac{\lambda}{4}(\phi^{\dagger}\phi - c^2)^2.
\ee
The topological number in this theory in $(1+1)$ dimensions is
\be
Q(t)= \frac{e}{4\pi}\int d^2x \epsilon_{\mu\nu}F_{\mu\nu}.
\ee
The static high temperature effective one-dimensional action is just
$[(D_{1}\phi)^{\dagger}(D_{1}\phi) + V(\phi)]$. It corresponds
to
finite field theory in
one
dimension.
The classical statistical
mechanics of this system is described by the Hamiltonian
\be
\fr12 E^2 + P P^* + (D_{1}\phi)^{\dagger}(D_{1}\phi) + V(\phi)
\ee
together with the Gauss constraint
\be
\partial_{1} E - ie (\phi^*P-\phi P^*)=0
\ee
imposed on the admissible states.

Let us put this system in a one-dimensional box of  length $L$,
$-L/2 < x < L/2$ and impose periodic boundary conditions on the
fields. Then in the limit $L\rightarrow \infty$ the saddle point of
this action -- sphaleron -- is a gauge transformation of the usual
kink of the scalar field theory \cite{Grigorev,grs:pl,grs:np}:
\be
\phi_{sph} = i\exp(i\pi x/L)\frac{c}{\sqrt 2}th \frac{M_H x}{2},
\ee
\be
A_1 = \frac{\pi}{e L}.
\ee
The high temperature sphaleron rate in this theory  has been
calculated
in refs. \cite{bs:two,Bochkarev:1989},
\be
\Gamma=\left[\frac{3E_{sph}}{\pi
T}\right]^{\frac{1}{2}}M_H^2
f\left(\frac{M_W}{M_H}\right)\exp\left(-\frac{E_{sph}}{T}\right),
\ee
where the sphaleron mass is given by $E_{sph} =
\frac{\sqrt{8\lambda}}{3}c^3$. The function $f(x)$ for large $x$ has
been evaluated in ref. \cite{bs:two} ($f(x)=
\frac{\sqrt{x}}{4\pi} 2^{2x-\frac{3}{4}}$) and, for arbitrary $x$, in
ref. \cite{Bochkarev:1989}.

The real time dynamics of the classical system was studied on the
lattice in refs. \cite{grs:pl,grs:np,Smit:1994,Bochkarev:1993}.
The results of the real time MC simulations show
that the rate of sphaleron transitions, indeed, does not depend on
the lattice spacing, provided it is small enough. Moreover,
quantitative
agreement with one-loop formulae is found at $T<E_{sph}$.
At very high temperatures, $T>E_{sph}$ the rate cannot be calculated
analytically, and only the numerical results exist
there \cite{deForcrand:1994}.  The
numerical simulations with Langevin-type equations replacing the real
time dynamics can be found in ref. \cite{Bochkarev}.

{\bf $SU(2)$ gauge--Higgs system.} In refs.
\cite{aaps:pl,aaps:np} the study of the sphaleron
transitions in the symmetric phase of the SU(2) gauge--Higgs theory
was performed. The sphaleron transitions were clearly observed with
different lattice spacings and volumes. However, the quality of the
lattice data did not allow the extrapolation of the results to the
continuum limit. All the data are consistent with the rate $\Gamma =
\kappa (\alpha_W T)^4$ with $\kappa >0.4$ \footnote{In ref.
\cite{aaps:np} the
corresponding constraint reads as $\kappa>0.1$; an arithmetic error
of a factor $4.4$ in this estimate was corrected in ref.
\cite{Farrar2}.}.

{\bf $SU(2)$ pure gauge theory.} Recently, the systematic study of
the
sphaleron transitions has been performed in a pure $SU(2)$ theory
\cite{Ambjorn:1995}. The use of the pure Yang--Mills theory instead
of
the full gauge--Higgs system is legitimate at the temperatures  far
above the critical one. At these temperatires
the scalar fields have  masses $\sim g T$
and decouple from the long-range fluctuations $k\sim g^2 T$  which
are believed to govern the topology change. The 3d part of the
classical model is a field theory free from ultraviolet divergences.
The authors show  nice evidence that the classical rate exists (i.e.
that it does not depend on the volume, if it is large, and lattice
spacing,
if it is small). Numerically,
\be
\Gamma=\kappa(\alpha_W T)^4, ~~~\kappa=1.09\pm 0.04.
\label{rate2}
\ee
Near the phase transition the scalar degrees of freedom have masses
$\sim g^2 T$ and do not decouple from the gauge fields. However,
since the parameter $g^2 T$ provides the only dimensionful scale of
the problem, the parametric dependence of the rate is still the same,
but the numerical coefficient may be different.

\subsection{Strong sphalerons}
We include in this section also the
discussion of other high temperature processes,
associated with anomaly, but now in QCD.
They change the chirality of quarks and, since
anomalous $B$-violation deals with left-handed
fermions, may be relevant for the discussion
of the baryon asymmetry.

It is well known that the quark axial vector current has an
anomaly and therefore is not conserved. The rate of chirality
non-conservation at high temperatures $\Gamma_{strong}$ is related to
the rate of topological transitions in QCD (``strong''
sphalerons \cite{mms:prd,Giudice:1994}),
\be
\frac{\partial Q_5}{\partial t} = - \frac{72}{T^3}\Gamma_{strong}Q_5,
\ee
where $Q_5$ is the axial charge. By analogy with
the weak sphalerons, the rate of the strong sphaleron
transitions is given by
\be
\Gamma_{strong} = \kappa_{SU(3)}(\alpha_s T)^4,
\ee
where $\kappa_{SU(3)}$ is an unknown pure number.  The
characteristic time of these transitions is therefore
\be
\tau_{strong} = \frac{1}{72 \kappa_{SU(3)}\alpha_s^4 T}.
\ee
Taking as an estimate $\kappa_{SU(3)}\sim 1$-$3$, this
time is of order $\frac{30-100}{T}$, i.e. the rate of strong
sphaleron
transitions is comparable to or even higher than the rate of
chirality-flip transitions mediated by the Yukawa
coupling of the top quark.

\subsection{Concluding remarks}
It is by now well established that there
is no suppression of the fermion number non-conservation at high
temperatures.  However, the quantitative formalism allowing for the
calculation of the rate beyond the
lowest order semiclassical approximation is still lacking.
{}From our point of view, the
most important challenge here is to establish the relation between
the quantum rate and the classical rate; the latter can then be
computed
with some kind of MC numerical analysis. Even in the framework
of the classical physics it would be important to have an analytical
understanding of the finiteness of the rate in the continuum limit.
Of course, the real time MC simulations in the broken phase of the
SU(2) gauge--Higgs system would be very important, in particular
because  the rate is now known in the Gaussian approximation.
\section{Phase transitions in gauge theories}
Potentially, phase transitions provide a source of deviations from
thermal equilibrium in the early Universe. Usually, in the theories
with scalars (such as grand unified theories or the electroweak
theory)
the symmetry is restored
at high
temperatures
and, at low temperatures,
it is broken
\cite{kir,kirlin}. If
the phase
 transition is of the first kind, it
proceeds
through the
nucleation of bubbles of a new phase \cite{VolKobOkun,CBounce}.
Depending
on the
parameters
of a model,
this process can be quite violent, the motion of the
domain walls  disturbs the plasma and may
trigger  the baryon asymmetry generation. In order to have the
detailed non-equilibrium picture of the phase transition, a number of
very difficult problems are to be solved. The questions potentially
interesting for baryogenesis include the bubble nucleation rate, the
structure of the domain walls, their velocity, the distribution
densities
of  particles near the domain walls, etc. It is
hard to
get
reliable answers to these questions, since they all deal with
complicated non-equilibrium phenomena. Moreover, even the equilibrium
treatment of the phase transitions  faces a number of
difficulties, associated with the so-called infrared problem in
thermodynamics
of the gauge fields.

There are many excellent reviews and books devoted to the study of
the phase transitions in gauge theories, see, for instance ref.
\cite{kir76,linde:rep,kapusta,Li90}. Our purpose in this section is
to report on the progress that was achieved in this area in the last
few years. Our special interest is in the study of the first order
phase transitions which are strong enough to suppress the B-violating
reactions in the Higgs phase (see Sections 3 and 7.1). The dynamics
of much weaker phase transitions may be different and is a topic of
extensive studies of refs. \cite{subcr1,subcr2,subcr3,subcr4}.

In the first subsection we shortly discuss the validity of the
equilibrium approximation to the description of the phase transition
and present some useful equations for the determination of the bubble
nucleation rate. Then we review the ``rules of the game'', which
allow
an estimate of the relevant parameters of
the phase transition with the help
of a unique function -- the perturbative effective potential for the
Higgs field. Everything contained in these subsections has been known
for
a long time
and is presented here for the sake of completeness. In
subsection 5.3 we explain why the perturbation theory  fails to
describe the high temperature phase transitions. In the following
subsections (5.4 and 5.5)
we describe the formalism which allows us to determine
reliably
the parameters of the phase transition in weakly coupled theories.
The specific results for the electroweak phase transition are
discussed in subsection 5.6. Subsection 5.7 is devoted to the
dynamics
of the phase transition.

\subsection{Equilibrium approximation}
Let us take for simplicity the Minimal Standard Model of electroweak
interactions and consider it in the cosmological context. Suppose
that the temperature of the system,  $T$,  is of the order of the
$W$-boson
mass -- the scale at which the electroweak phase transition is
expected to take place. The first question that arises
is
whether
the equilibrium description of this system is at all possible  in the
expanding Universe. In order to check this, we may compare the rates
of different particle reactions in the standard model with the rate
of the Universe expansion, $t_U^{-1}$, where the Universe age $t_U$
is
given by
\be
t_U= \frac{M_0}{T^2},
\ee
Here $M_0 = M_{Pl}/1.66 N^{\fr12}\sim 10^{18}$ GeV, $M_{Pl}$
is the Planck mass, and $N$ is the effective number of the massless
degrees of freedom. The expansion rate of the Universe is the unique
non-equilibrium parameter of the system before or some time after the
phase transition; during the phase transition another typical
non-equilibrium time scale, associated with the motion of the
bubble walls, is relevant. This time scale is smaller than the
Universe expansion rate by many orders of magnitude (see below) and
thus the deviations from the thermal equilibrium are much stronger.

Before or after
the phase transition,
the fastest perturbative reactions are those associated with
strong interactions (e.g. $q\bar{q} \rightarrow GG$); their rate is
of the order of $(\tau_{strong})^{-1} \sim \langle \sigma n v \rangle
\sim \alpha_s^2 T$. Here $\sigma$ is the cross-section of the
reaction, $n$ is the particle concentration, $v$ is the relative
velocity of the colliding particles. The typical weak reactions, say
$e\nu \rightarrow e\nu$,
occur at
 the rate $(\tau_{weak})^{-1} \sim
\alpha_W^2 T$, and the slowest reactions are those involving
chirality flips for the lightest fermions, e.g. $e_R H\rightarrow \nu
W$ with the rate $(\tau_{e})^{-1} \sim f_e^2 \alpha_W T$, where $f_e$
is the electron Yukawa coupling constant. Now, the ratio
$\frac{\tau_i}{t_U}$ varies from $10^{-14}$ for the fastest reactions
to $10^{-2}$ for the slowest ones; this means that particle
distribution functions of quarks and gluons, intermediate vector
bosons, Higgs particle and left-handed charged leptons and neutrino
are
equal to the equilibrium ones with an accuracy better than
$10^{-13}$; the largest deviation from thermal equilibrium ($\sim
10^{-2}$) is being expected for the right-handed electron.

These estimates show that the equilibrium description of the system
is a very good approximation before the phase transition, and soon
after it is completed \footnote{Clearly, this is a model-dependent
statement. For instance, if the Universe was as hot as, say,
$10^{17}$
GeV, then the equilibrium description of the (grand unified) phase
transitions would be
questionable,
since the ratio
$\frac{\tau_i}{t_U}$ would be of the order of $1$.}. Moreover, since
the
phase transition is expected to proceed
through the bubble nucleation, the
equilibrium description of the plasma is possible in the regions far
enough from the moving domain walls.

\begin{figure}
\hspace*{0cm}
\vspace{1cm}
\epsfysize=11cm
\centerline{\epsffile{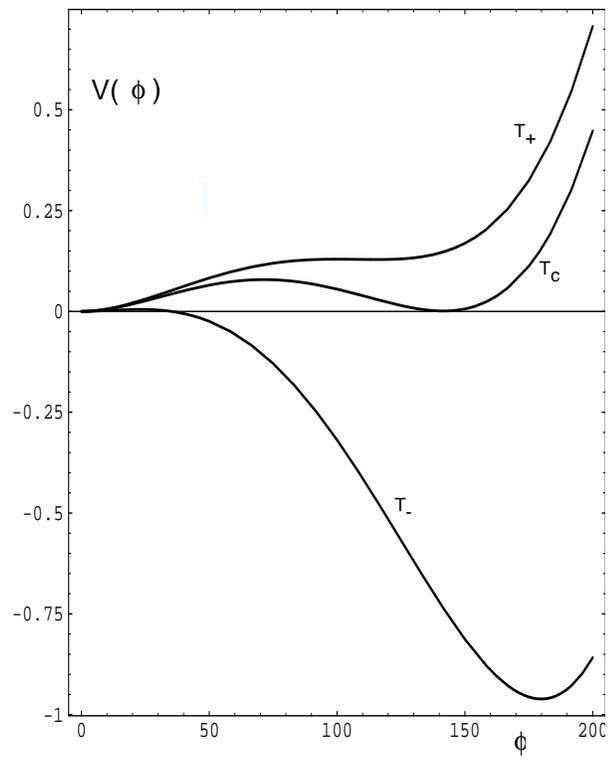}}
\vspace*{-1cm}
\caption{\protect
Effective potential evolution at first order phase transition.}
\label{t+t-tc}
\end{figure}

The above remarks suggest that the equilibrium statistical mechanics
may be applied for the evaluation of the ``static'' properties of
the phase transitions, such as  the critical temperature $T_c$,
latent heat, jump of the order parameter (expectation value of the
scalar field), interface tension (surface energy density of the plane
domain wall separating different phases). Another important
characteristics  is the  metastability range: at the upper $T_+$
(lower $T_-$ ) spinodial decomposition temperatures  broken
(symmetric)
phase ceased to exist as a metastable state (fig.
\ref{t+t-tc}).
The static
correlation lengths for various operators may help to understand
the structure of the domain walls.

The first order
phase transition in the early Universe is not an instant process, and
its total duration is of the order of the Universe age. The bubbles
of the new phase start to nucleate at temperatures somewhat lower
than $T_c$; the bubbles expand and finally fill out the Universe with
the
new phase. This happens at temperatures above $T_-$. The fraction
of the volume of the Universe occupied by a new phase, ${\cal P}
(t)$, can
be estimated
in
the following way \cite{Gu81}. Suppose that the bubble
nucleation rate per unit time
per
unit volume is ${\cal R}(T)$, and
the velocity of the bubble walls
is constant\footnote{In the hot plasma,
contrary
to the vacuum case,
there is a friction force acting on the bubble wall. This ensures the
constant velocity of the wall.} and equal to $v$. Then,
\be
{\cal P}(t)=1-\exp(-\Delta(t)),
\ee
where $V(t,t_0)=\frac{4 \pi}{3}v^3(t-t_0)^3$ is the volume that the
bubble nucleated at time $t_0$ occupies at time $t$:
\be
\Delta(t)=\int_{t_c}^t dt_0~ V(t,t_0){\cal R}(T(t_0))
\ee
and $t_{c}$ is the time corresponding to the critical temperature,
$T(t_{c}) = T_{c}$.
This equation does not take into account the red shift of the bubble
velocity,
which is in fact unimportant
at the electroweak scale.
Introducing the variable
\[
x=\frac{T_c-T}{T_c}
\]
and assuming that it is small
(this is satisfied in the electroweak case), one obtains a simplified
expression
\be
\Delta = \frac{64 \pi
v^3}{3}\left(\frac{M_0}{T_c}\right)^4\int_0^x\frac{{\cal
R}(T)}{T_c^4}
x^3~dx.
\ee
The phase transition is completed when
\be
\Delta \simeq 1.
\label{xdet}
\ee
Since
the
electroweak scale is much smaller than the Plank scale, the
probability of the bubble nucleation ${\cal R}$ at the percolation
time is
very small:
\[
x^3 \frac{{\cal R}(T)}{T_c^4} \sim \left(\frac{T_c}{M_0}\right)^4
\sim
\exp(-150).
\]
The typical bubble size is of the order of $R\sim 2 v
\frac{M_o}{T_c^2} x$, where $x$ is to be found from eq. (\ref{xdet}).

Computation of the bubble nucleation rate in the general case is a
very
complicated problem, which has not been solved. The reliable
estimates exist only in the so-called thin wall approximation, and
the
leading contribution can be read off from the Landau--Lifshitz book
on
statistical mechanics. Suppose that the temperature of the system is
$T < T_c$, and $x\ll 1$. Then the free energy of the critical bubble
can be found from the minimization condition
\be
\frac{\partial F}{\partial R}= 0,~~~
F(R)=4 \pi R^2 \sigma - \frac{4 \pi R^3}{3}\Delta p,
\ee
where $\Delta p = L x$ is the pressure
difference, $L$ is the latent heat of the transition, and
$\sigma$ is the surface tension. From
the latter relation
one immediately obtains
\be
{\cal R}=\kappa T_c^4 \exp(-S(x)),
\label{bubrate}
\ee
where the bounce action is given by
\be
S(x)= \frac{16 \pi \sigma^3}{3 T_c L^2}\frac{1}{x^2}(1 + O(x))
\label{thinact}
\ee
and corrections $O(x)$  are model dependent. Indeed, the domain wall
thickness is of the order of the typical correlation length $\xi$  in
the system, which means that the radius of the bubble is defined up
to
corrections of order $\xi$. This produces an uncertainty in the
action
$\delta S/S \sim \frac{L\xi x}{\sigma}$ and gives
an
obvious
requirement of the validity of the thin wall approximation $R \gg
\xi$.  The calculations of the bounce action in
various models can be found in ref. \cite{Li90,dlhll:pr}.

If $R\sim \xi$, then the thin wall approximation breaks down, the
nucleation rate cannot be expressed only through macroscopic
parameters of the phase
transition (latent heat and surface tension) at the critical point.
In this case the phase transition is delayed and the Universe is
supercooled in the symmetric phase. The calculation of the bubble
nucleation rate in this case cannot be done for a generic gauge
theory because of an infrared problem in the thermodynamics of the
gauge fields (see below), but it is feasible for weakly coupled pure
scalar
theory.  A detailed study of the bubble nucleation in the
scalar mean field theory defined by
potential (\ref{1loop1storder}) is contained in ref. \cite{Enq92}.
Naturally, the bubble nucleation rate receives dependence on the
scalar
correlation length $\xi$ at the phase transition, in addition to the
surface tension and the latent heat.  We refer here to
\cite{Li90,dlhll:pr,Igna} for more details.
The more complicated problem is to determine the
prefactor in the expression for the bubble nucleation rate.
Its computation in one-loop approximation in scalar models
and electroweak theory has been done in refs.
\cite{Baa93,Baa5,Baa95,klsnp}.

\subsection{Simple estimates}
To describe the high temperature phase transitions in any given
theory it is very important to have a relevant calculational
formalism.
The traditional tool is the effective potential for the scalar field
$\phi$. It is defined as
the value of the free energy of the system (pressure
with the minus sign) in  a uniform background field $\phi$. The
minima of this potential correspond to the (meta)stable states of
the system. The system undergoes a first order phase transition if
there are two degenerate minima of this potential, separated by the
energy barrier. In general, the effective potential is a gauge
dependent quantity; perturbative calculations often produce complex
terms.
However,
the values of the potential at its minima are
gauge invariant; this allows for the gauge-invariant definition of
the
critical temperature and latent heat.

The following simple strategy (the drawbacks of which we discuss
later) gives a reasonable qualitative description of the phase
transitions and often allows fairly accurate estimates
\cite{linde:np81}:

Step No. 1. Take your model and calculate the one-loop high
temperature
effective potential $V(\phi,T)$.

Step No. 2. Define from it the critical temperature, jump of the
order
parameter, latent heat and surface tension with the use of the
following equations.\\
--$T_c$ and order parameter $\phi_c$:
\be
\frac{\partial V(\phi_c,T_c)}{\partial
\phi_c}=0,~V(\phi_c,T_c)=V(0,T_c).
\ee
--Latent heat and surface tension:
\be
L=T_c\frac{\partial}{\partial T_c}(V(\phi_c,T_c)-V(0,T_c)),
\ee
\be
\sigma=\int_0^{\phi_c}\sqrt{2V(\phi,T_c)}~d\phi .
\ee

Step No. 3. Calculate the bubble nucleation rate in the thin wall
approximation and compare it with the rate of the Universe expansion.
Determine the bubble nucleation temperature and check the validity of
the thin wall approximation. If it does not work, evaluate the bubble
nucleation rate for a thick wall. To this end find O(3) symmetric
configurations extremizing the 3d action
\be
S=\frac{1}{T}\int~ d^3 x~\left[\fr12(\partial_i \phi)^2 +
V(\phi,T)\right]
\ee
with the boundary condition $\phi \rightarrow 0$ at $x \rightarrow
\infty$. The bubble nucleation rate is then ${\cal R}\sim \exp(-S)$.

An example is provided by the Minimal Standard Model. Here the
one-loop effective potential in the high temperature approximation is
(for simplicity, we take the case when the Higgs boson is
sufficiently light, and neglect the effects of the $U(1)_{Y}$
interactions):
\be
V(\phi,T)=\fr12\gamma(T^2-T^2_-)\phi^2-\fr13\alpha T\phi^3
+\fr14\lambda\phi^4.
\la{1loop1storder}
\ee
For the standard model with the top quark mass $m_\rmi{top}$
\be
\alpha=9g^3/(32\pi),
\gamma=\fr3{16}g^2+\fr12\lambda+{m_\rmi{top}^2\over 2v^2},
\label{gamma}
\ee
and the lower metastability temperature $T_-$ is related to the Higgs
mass $m_H^2=2 \lambda v^2$ through
\be
T_-={m_H\over\sqrt{2\gamma}}.
\ee
Due to the presence of the cubic term, the potential predicts
the
first
order transition with the critical temperature
\be
T_c={T_-\over\sqrt{1-\fr29 {\alpha^2\over\lambda\gamma}}}>T_{-}
\la{tc}
\ee
and the jump of the order parameter
\be
{\phi(T_c)\over T_c}=\fr23{\alpha\over\lambda}.
\la{vc}
\ee
Phase transition gets weaker when the scalar self-coupling increases.
This is  seen from the behaviour of the order parameter, latent heat,
and the surface tension, all of which decrease with the increase of
$\lambda$. At large $\lambda$ the bubble nucleation rate can be
determined
in
the
thin wall approximation, while at small $\lambda$
($\lambda\sim g^3$)
it breaks down, and the phase transition occurs with considerable
supercooling. Qualitatively, the one-loop description gives correct
results, but concrete numbers may be quite different from those
obtained by a more refined treatment. The effect of the
Debye screening on the effective potential was discussed
in refs. \cite{dlhll:pl,dlhll:pr,carr,s:msm}, and the
two-loop computation has been done in refs.
\cite{bd,ae,Heb93,Fod94,K5,La94,La95,Kri95}.
Various aspects of the phase transition were discussed
in refs. \cite{Carrington:1993,Buchmul94,Bodek4,Dos95}.

\subsection{The infrared problem and factorization}
It was  realised by Linde and Gross, Pisarski and Yaffe a
long time ago \cite{linde:pl80,gpy} that the
perturbation theory for  non-Abelian gauge
theories with small coupling constants
inevitably breaks down at high temperatures, at least in the
symmetric phase. The physical reason is that at high temperatures,
instead of the usual 4-dimensional expansion parameter $g^2$ the
relevant parameter is $\rho \sim g^2 n_B(E)$, where
$n_B(E)=[\exp(-E/T)-1]^{-1}$ is the Bose distribution function, $E$
is the typical energy of a given process in the plasma.
At $E < T$ the expansion parameter is greater
than that at zero temperatures, namely, $\rho \sim \frac{g^2 T}{E}$,
accounting for  typical Bose amplification of the amplitudes. In the
symmetric phase, gauge bosons are massless in perturbation theory,
there is no infrared cutoff, and the expansion parameter can be
arbitrarily large. In the broken phase the infrared cutoff is
provided by the vector boson mass, and perturbation theory converges
provided $\frac{g^2 T}{m_W} \ll 1$. Below we will give a more formal
description of the infrared catastrophe.

The fact that the perturbation theory breaks down at $T \neq 0$ poses
non-trivial difficulties for the description of the phase transition.
Indeed, the phase transition occurs when the free energy of the
broken phase is equal to that of the symmetric phase; but the latter
cannot be calculated perturbatively. The latent heat of the
transition
receives contributions from both the symmetric and broken phases, and
the same is true for the surface tension. In
the Universe the phase transitions occur when it is cooling, so that
the
initial phase is the one
in which perturbation theory breaks down. For strongly first order
phase transitions, the leading contribution to the above
parameters comes from the broken phase, where perturbation theory is
applicable; in that case the perturbative description may be
reliable. However, the infrared problem is fatal for an attempt of
the
perturbative quantitative study of the weakly first order phase
transitions. Unfortunately, direct Monte Carlo lattice
simulations of high temperature gauge theories are not possible at
present for realistic theories, containing chiral fermions, due to
well known difficulties in discretisation of the chiral fermion
determinant. The purely bosonic sector of the models can be put on 4d
lattice, and extensive 4d numerical simulations have been
carried out in refs. \cite{Bunk92,Bunk93,Fo94,Fo95,4d1},
for a summary of results see a nice review by Jansen \cite{Jans95}.

Recently, an approach was suggested, which allows for a solution
of the equilibrium problem of phase transitions in weakly coupled (at
zero temperatures) gauge theories \cite{K5,K3,K2,K1}. It combines
both
perturbative analysis and numerical Monte Carlo methods. The main
idea of the method is the factorization of the different scales
appearing in the description of the high temperature plasma. At the
first stage, a much simpler effective theory, incorporating all
essential non-perturbative dynamics of the phase transition, is
constructed by perturbative methods. The idea of this construction,
known as dimensional reduction, goes back to the papers by Ginsparg
\cite{Gins80}, and by Appelquist and Pisarski \cite{App81}.
It was developed
in refs. \cite{K5,K3,K2,K1} in application to the phase transitions
and applied later to hot QCD in ref. \cite{Bra1,Bra2,Bra3,Bra4,Bra5}.
Different aspects of dimensional reduction were studied
in refs. \cite{Jack81,nadkarni,Lands89,Jako94,Jak94,Jak95}.
At the second stage the effective 3-dimensional theory is analysed by
non-perturbative methods (MC lattice simulations)
\footnote{Whenever the comparison between 3d and 4d simulations
possible, they are in agreement, indicating the correctness of the
dimensional reduction beyond perturbation theory.
Generically the errors in 4d simulations are considerably
larger than those in 3d \cite{Jans95}, because of rather stringent
requirements on the lattice size in 4d, see ref. \cite{K3}.}.
In the discussion
below we follow ref. \cite{K2}.

The idea of dimensional reduction comes from an
observation
that {\em
equilibrium} finite temperature field theory is equivalent to
Euclidean zero temperature field theory defined on a finite ``time''
interval $\beta=1/T$ supplied with periodic boundary conditions for
bosons and antiperiodic ones for fermions.

Periodic and antiperiodic boundary conditions enable one to decompose
Bose ($\phi$) and Fermi ($\psi$) fields in Fourier series with
respect to the finite time interval,
\be
\phi(x,\tau)=\sum_{n=-\infty}^{\infty}\phi_n(x)\exp(i
\omega^b_n \tau),
\ee
\be
\psi(x,\tau)=\sum_{n=-\infty}^{\infty}\psi_n(x)\exp(i
\omega^f_n \tau),
\ee
where $\omega^b_n=2 n \pi T,~ \omega^f_n=(2n+1) \pi T $. Therefore,
4d finite temperature field theory is equivalent to the 3d theory
with an infinite number of fields, and 3d boson and fermion masses
are just frequencies $\omega^b$ and $\omega^f$. One can easily
recognize here a perfect analogy to Kaluza--Klein theories with
compact higher-dimensional space coordinates. The equilibrium
dynamics of the theory is completely characterized by the set of
Matsubara (imaginary time, or Euclidean) Green's functions,
$G_n(\omega_i, \vec{k_i})$, where $\omega_i$ are discrete
frequencies,
$n$ is the number of legs. The static bosonic Green's functions
(fermionic Green's functions are never static, since the fermion
frequences are odd) play an important role for the phase transition.
For example, the expectation value of the scalar field is just
$G_1(0,0)$; the static correlation lengths can be extracted from
two-point Green's functions, etc.

Assume now that the theory is weakly coupled, and that the
expectation value of the
Higgs field in the broken phase is small enough, so that the vector
boson masses are much smaller than the temperature. Then the
description of the phase transition (i.e. static bosonic
Green's
functions) can be derived
within
a simpler 3d theory, which contains
only bosonic fields corresponding to the $n=0$ sector of Fourier
decomposition. In loose terms,``superheavy'' (we keep the word
``heavy''
for other fields defined below) fields are integrated out. This
theory is valid up to momenta $k \ll T$. To specify the dynamics of
the effective theory one writes down the most general 3d
super-renormalizable Lagrangian, containing zero modes only, and
determines its parameters by the matching condition. This condition
requires that the
2-, 3- and 4-point
one-particle irreducible Green's
functions, evaluated in the effective theory and in the full 4d
theory
are the same up to some power of the coupling constant. The effective
theory approximately describes then {\em all} static Green's fuctions
of the high temperature 4d theory. As discussed in ref. \cite{K2},
the
maximum accuracy that can be reached with a {\em
super-renormalizable} 3d theory is
\be
\frac{\Delta G}{G} \sim O(g^4).
\label{accurh}
\ee

To be more explicit, take as an example the Minimal Standard
Model
Lagrangian. Then the 3d effective theory
is
an $SU(2)\times U(1)$
bosonic theory, which contains the Higgs doublet, the scalar triplet
(zero
component of the $SU(2)$ gauge field), and
the
scalar singlet
(zero component of the $U(1)$ field) with
the action
\ba
S & = & \int\! d^3x \biggl\{
\frac{1}{4}G^a_{ij}G^a_{ij}+ \frac{1}{4}F_{ij}F_{ij}+
(D_i\Phi)^{\dagger}(D_i\Phi)+
m_3^2\Phi^{\dagger}\Phi+\lambda_3
(\Phi^{\dagger}\Phi)^2 +\lambda_B B_0^4\nonumber \\
& &
 +\frac{1}{2}
(D_iA_0^a)^2+\frac{1}{2}m_D^2A_0^aA_0^a+
\frac{1}{4}\lambda_A(A_0^aA_0^a)^2
+\frac{1}{2} (\partial_iB_0)^2+\frac{1}{2}m_D'^2B_0B_0 \nonumber\\
& &
+ h_3\Phi^{\dagger}\Phi A_0^aA_0^a
+ h_3'\Phi^{\dagger}\Phi B_0B_0
-\frac{1}{2}\lambda_{AB}B_0 \Phi^{\dagger}A_0^a\tau^a\Phi
+\tilde{\lambda}_{AB}B_0^2A_0^aA_0^a
\,\,\biggr\} ,
\label{action}
\ea
where
$G^a_{ij}=\partial_iA_j^a-\partial_jA_i^a+g_3\epsilon^{abc}A^b_iA^c_j$,
$F_{ij}=\partial_iB_j-\partial_jB_i$,
$D_i\Phi=(\partial_i-ig_3\tau^aA^a_i/2+ig_3'B_i/2)\Phi$,
$D_iA_0^a=\partial_iA_0^a+g_3\epsilon^{abc}A_i^bA_0^c$, and
$\Phi=(\phi_3+i\phi_4,\phi_1+i\phi_2)^T/\sqrt{2}$.
The $\tau^a$'s are the Pauli matrices.
The factor $1/T$ multiplying the action has been scaled into
the fields and the coupling constants, so that the fields have
the dimension GeV$^{1/2}$ and the couplings $g_3^2$, $\lambda_3$,
$\lambda_A$, $\lambda_{AB}$  have the dimension GeV.

The complete one-loop calculation of the 3d coupling constants and
two-loop calculation of the 3d masses of the effective theory can be
found in ref. \cite{K5,K2}; we present here only
simple tree relations for the
coupling constants
\be
g_3^2=g^2 T,~g_3'^2=g'^2 T,~ \lambda_3=\lambda T,~ h_3=\fr14 g^2 T~,
\ee
\[
h_3'=\fr14 g'^2 T,~
\lambda_A=\lambda_B=\tilde{\lambda}_{AB}=0,~\lambda_{AB}=g g'T
\]
and one-loop relations for the doublet scalar mass
\be
m_3^2(\mu) =  -\fr12 m_H^2
+T\biggl(\frac{1}{2}\lambda_3+\frac{3}{16}g_3^2+\frac{1}{16}g_3'^2+
\frac{1}{4}g_Y^2\biggr)
\label{m3}
\ee
and triplet and singlet masses
\ba
m_D'^2 & = & \biggl(\frac{1}{6}+\frac{5n_F}{9}\biggr) g'^2 T^2, \\
m_D^2 & = & \biggl(\frac{2}{3}+\frac{1}{6}+\frac{n_F}{3}\biggr) g^2
T^2.
\label{mD}
\ea
Here $m_H$ is the zero temperature Higgs mass, $g_Y$ is the Yukawa
coupling
constant
corresponding to the $t$-quark, $n_F=3$ is the number of
fermion generations.

Clearly, is is much easier to study this 3d model than the initial 4d
theory, since it contains much less degrees of freedom. Another
important fact is that the only remnant of fermions is their
contribution to the 3d masses and coupling constants.

An inspection of relations (\ref{m3}) and (\ref{mD}) shows that
further simplification of the effective theory is possible. Indeed,
at the tree level at $m_3^2<0$ the symmetry is broken, and at
$m_3^2>0$ it is restored.
 In other words, the phase transition takes place
near $m_3^2=0$. At this point the hierarchy
$m_3^2 \ll m_D^2$ and $m_3^2 \ll m_D'^2$ holds, allowing for the
construction of the effective theory containing the Higgs doublet and
$SU(2)\times U(1)$ gauge fields only. The ``heavy'' scale $\sim g T$
is
integrated out, and the super-renormalizable Lagrangian of the
effective theory is merely
\be
L  =
\frac{1}{4}G^a_{ij}G^a_{ij}+ \frac{1}{4}F_{ij}F_{ij}+
(D_i\Phi)^{\dagger}(D_i\Phi)+
\bar{m}_3^2\Phi^{\dagger}\Phi+\bar{\lambda}_3
(\Phi^{\dagger}\Phi)^2 .
\label{univers}
\ee
In the one-loop approximation
\be
\bar{m}_3^2(\mu)  = m_3^2-\frac{1}{4\pi}\biggl(3 h_3 m_D +
h_3' m_D'\biggr)
\ee
and $\bar{\lambda}_3=\lambda_3$ at the tree level. Higher order
corrections to these relations can be found in ref. \cite{K5,K2}.
In fact, the
possibility of
integrating out
the heavy scale has a general
character and can be applied to any gauge theory.
The effective description in terms of the
{\em super-renormalizable} Lagrangian for the ``light'' modes only
provides an accuracy
\be
\frac{\Delta G}{G} \sim O(g^3).
\label{accurh1}
\ee
for Green's functions for ``light'' (static gauge bosons and
scalars) fields. This theory is valid for $k \ll g T$, but $k$ may be
as large as $g^2 T$. The effective potential for the scalar
field in 3d coincides with the hard loop resummed potential
in 4d at high temperatures \cite{ae,Fod94,K5}.

\begin{figure}
\vspace*{0cm}
\hspace{1cm}
\epsfysize=11cm
\centerline{\epsffile{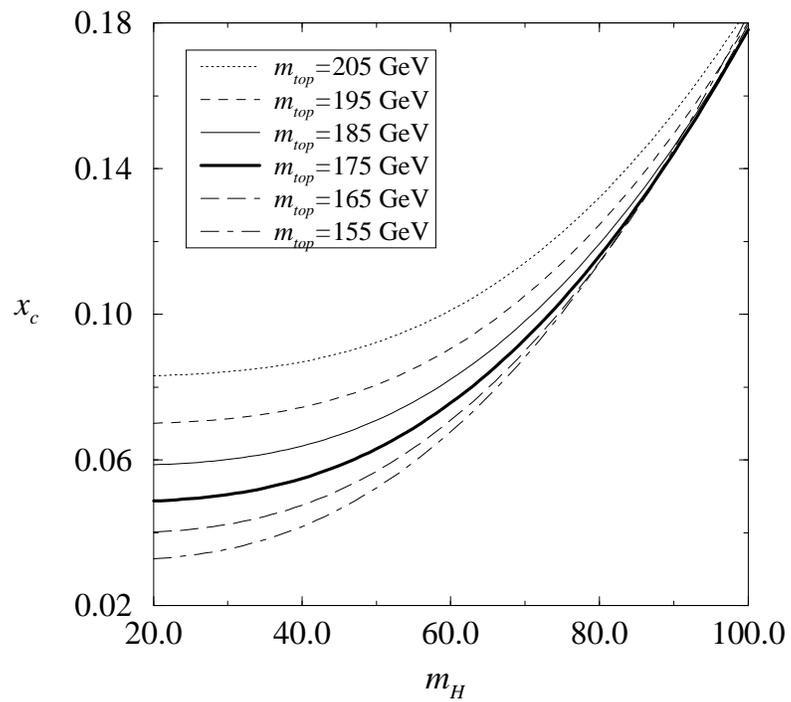}}
\vspace*{-3cm}
\caption[a]{\protect
The critical value $x_c=\lambda_3/g_3^2$ as a function of the
physical Higgs
mass $m_H$ and the top quark mass $m_{\rm top}$. In general,
$x$ depends on the Higgs mass, the top mass and logarithmically
on the temperature. From ref. \cite{K1}}
\label{xmhdep}
\end{figure}

Since all the four parameters of the 3d theory are dimensionful,
the theory is uniquely fixed by three dimensionless
ratios:
\be
\qquad x\equiv {\bar{\lambda}_3\over \bar{g}_3^2},\qquad
y\equiv \frac{\bar{m}_3^2}{\bar{g}_3^4},\qquad
z\equiv \left({\bar{g}'_3\over \bar{g}_3}\right)^2
\la{3dvariables}
\ee
and
the
overall scale $g_3^2$, on which the physics does not depend. The
temperature variation changes the parameter $y$, while $x$ and $z$
depend
on $T$ only through logarithmic corrections. So, we arrive
at the important conclusion that the dynamics of the electroweak
phase
transition depends on one dimensionless number only ($x$) since $z$
is related to the known
weak mixing angle, $z \simeq 0.3$, and $y$ is
fixed by the requirement that the temperature is equal to the
critical one. The dependence of the parameter $x$ on the
physical Higgs mass in MSM was found in ref.
\cite{K2}, see fig. \ref{xmhdep}.

In fact, the 3d theory defined by (\ref{univers}) plays the role of
the
universal theory describing the phase transition not only in the
Minimal Standard Model, but in many of its extensions,  including
sypersymmetry, two Higgs doublets, etc. \cite{K2}. Let us take as an
example the two Higgs doublet model. The integration over the
superheavy modes gives a 3d $SU(2)\times U(1)$ theory with an extra
Higgs
doublet in addition to the theory considered above. Construct now the
one-loop scalar mass matrix for the doublets and find the
temperatures
at which one of its eigenvalues is
zero. Take the higher temperature; this
is the temperature near which the phase transition takes place.
Determine the mass of the other scalar at this temperature.
Generally, it is of the order of $g T$, and therefore, is heavy.
Integrate
this heavy scalar
out together with the heavy triplet and singlet -- the
result is the simple $SU(2)\times U(1)$ model. In the case when both
scalars are light near the critical temperature, a more complicated
model, containing two scalar doublets, should be studied. It is
clear, however, that this case requires  fine tuning. The
consideration of the phase transitions in the two Higgs doublet model
at the one-loop level can be found in
refs.
\cite{bks:higgs,tz:higgs,Land92,Jain93,Lee94,Dav94,Cline95,Zar95}.

The same strategy is applicable to the Minimal Supersymmetric
Standard Model (MSSM). If there is no breaking of colour and charge
at high
temperature (breaking is possible, in principle, since the theory
contains squarks), then all degrees of freedom, excluding those
belonging to the two Higgs doublet model, can be integrated out. We
then return back to the case considered previously. The conclusion in
this case is similar to the previous one, namely that the phase
transition in the MSSM can be described by a 3d $SU(2)\times U(1)$
gauge--Higgs model, at least in a considerable
part of the parameter space. A
one-loop analysis of this theory was carried out in refs.
\cite{Giu92,My92,Espi93,Brig94}.

Of course, effective super-renormalizable 3d theories do not
provide an exact description of the phase transitions.
Parametrically, $\frac{\Delta G}{G} \sim O(g^4)$ or $\sim O(g^3)$
depending on the level of dimensional reduction. The numerical
estimate of uncertainty is model dependent. For the standard
electroweak theory, corresponding estimates were given
in  ref. \cite{K2}, with
the result that the effective description provides an accuracy $\sim
1\%$ for Higgs masses from $30$ GeV up to few hundred GeV. For
smaller Higgs masses the phase transition is too strong and the
vector boson
masses are of the order of temperature, i.e. the assumption of
the scale hierarchy does not hold; if the Higgs mass is close to the
unitarity bound, the perturbation theory, used in the construction of
the effective Lagrangian, breaks down.

To summarize this discussion, in order to study phase transitions in
weakly coupled gauge theories, one may construct simpler 3d theories,
parameters of which can be perturbatively calculated. This
calculation is free from any infrared divergences, and
its ultraviolet divergences are removed by the usual counterterms of
the zero-temperature perturbation theory.
A unique 3d theory
plays a role of universal theory for
the
description of the phase
transition in many 4d models.
In particular,
the
strength
of the
electroweak phase transition depends on the unique number,
$x=\bar{\lambda}_3/\bar{g}_3^2$.

\subsection{Phase structure of 3d theory}
The construction of the effective 3d theory is free from infrared
divergences. However, the perturbative calculations in 3d are
infrared divergent in the symmetric phase. This is easy to see by
simple
power counting. For example, the $N$-th loop contribution to
the vacuum energy in 3d (with dimension GeV$^3$) must contain the
factor
$(g_3^2)^{N-1}$. Therefore, starting from $N=5$ some dimensionful
parameter (say, the typical momentum scale $k$) should appear in the
denominator of each term of the expansion, i.e. the expansion
parameter is $g_3^2/k$. In the symmetric phase the infrared
cutoff
is
absent, and perturbation theory breaks down. The first
(logarithmically) divergent contribution in the vacuum energy appears
in four loops and is of the order of $(g_3^2)^3$, giving a
contribution of the order of $g^6 T^4$ to the free energy.

This means that the 3d theory should be treated in a non-perturbative
way. At present, non-perturbative studies of the complete 3d
$SU(2)\times U(1)$ model are absent, and most results were derived
for
an $SU(2)$ gauge--Higgs system. Let us consider first the general
structure
of the phase diagram of this theory. In the $SU(2)$ gauge--Higgs
model there
are no known local gauge invariant order parameters, which would
acquire a non-zero value only in one of the two phases of the model.
In other words, there is no symmetry breaking or restoration in this
model, and the gauge symmetry is always
intact (see, for instance refs. \cite{Fradk79,Banks79}).
An ordinary jargon, saying that the vector bosons are massless
in the symmetric phase and massive in the broken one is incorrect,
since vector degrees of freedom are massive at any choice of
parameters.
A good statistical analogue of this system is an ordinary liquid,
which
may undergo the
first order liquid--vapour phase transition without
restoration or breaking of any symmetry. The phase diagram of the
latter system usually has a critical point, where the line of the
first order phase transition  ends. At this point the phase
transition is of the second order, and there is a massless scalar
excitation in the physical spectrum.

\begin{figure}
\vspace*{0cm}
\hspace{1cm}
\epsfysize=17cm
\centerline{\epsffile{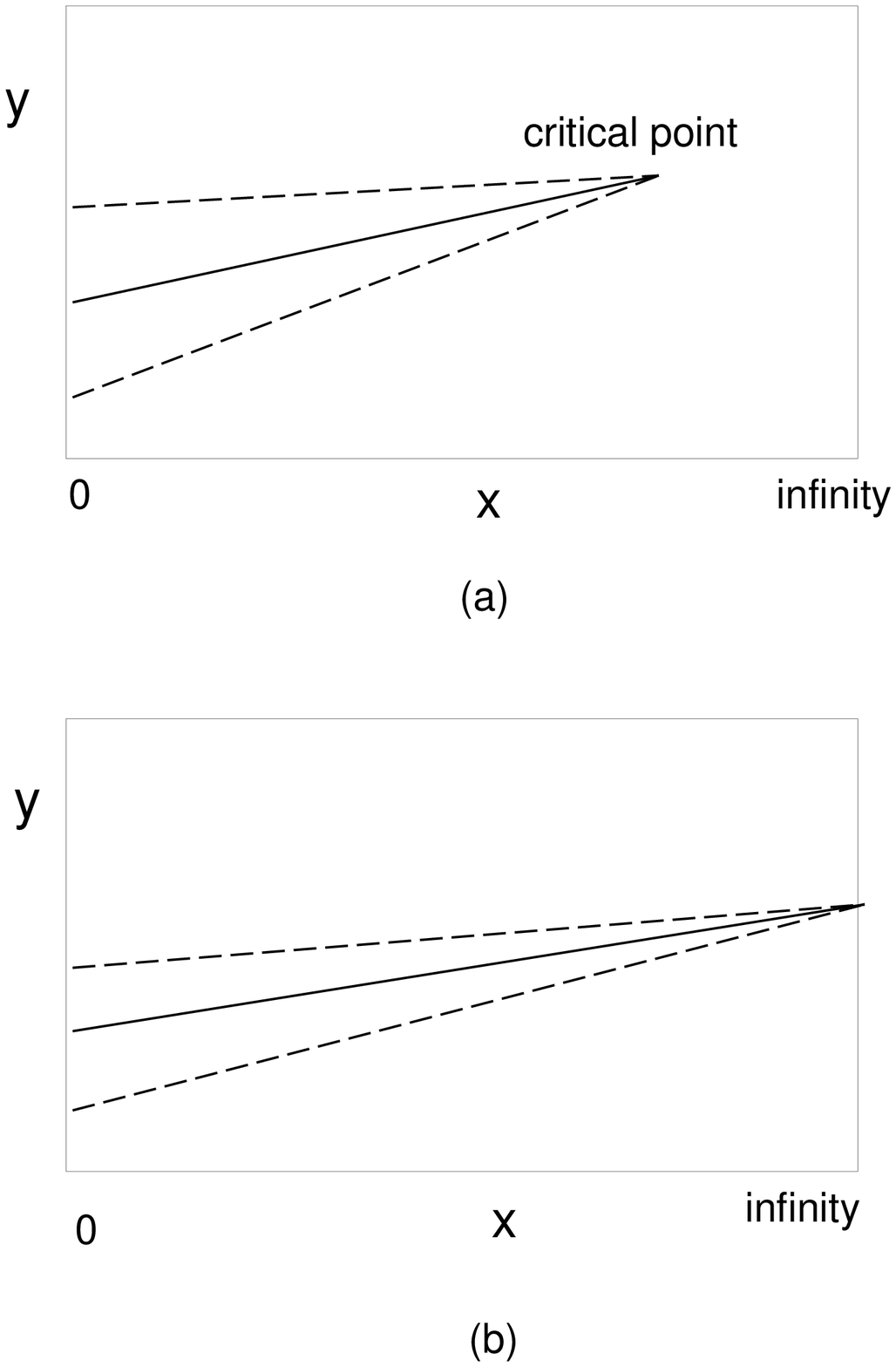}}
\vspace{-2cm}
\caption[a]{\protect
The schematical phase diagrams for gauge-Higgs SU(2) system.}
\label{phdiagr}
\end{figure}

The absence of the true order parameter in the gauge--Higgs system
suggests  two possible phase diagrams in the $(x,y)$ plane, shown
schematically in fig. \ref{phdiagr}. In the first case the line
of the first order
phase transitions has an end point at finite $x$, while in the second
case the end point is at $x=\infty$. By dotted lines we show
the upper and lower spinodial decomposition temperatures. In the
first case there exists a critical scalar self-coupling above which
there is no phase transition at all, while in the second case a first
order phase transition occurs for all physical parameters. It is
important for cosmological applications to clarify the type of the
phase diagram. For example, if there exists an end point of the line
of
the first order phase transition, and the physical value of $x$ is to
the right of this end point, then there is a smooth cross-over
transition without any strong deviations from thermal equilibrium. In
the latter case there are no observational consequences from the
electroweak epoch.

The following approaches were applied to $SU(2)$ gauge--Higgs system:
$\epsilon$ expansion \cite{Arn94}, exact renormalization group
approach
\cite{Reut93,Buc94,Be95,Ber95},
Schwinger--Dyson equation \cite{Buch5},
perturbation theory \cite{K5},  and lattice Monte Carlo
simulations \cite{K1,Ilg95,Karsch95}. The first two approaches
favour the second type
of the phase diagram, while the third one argues in favour of the end
point of the critical line. In general, perturbation theory
describes well different characteristics of the phase transition
at small scalar self-coupling constants, while $\epsilon$-expansion
is more suitable in the regime
when $\lambda_3/g_3^2\gg 1$ \cite{Ya5,Arno4,Ya4}.
The lattice MC simulations cannot resolve
the order of the phase transition at large scalar self-coupling
constant.

We shall review here the results from the lattice 3d MC
simulations only \cite{K6,K4,K3,K1,Ilg95,Gur95}.
The comprehensive presentation of all existing
approaches would make the discussion too lengthy;
our emphasis on MC simulations is mostly because
this
is the only approach which is based on the first principles and does
not require any extra assumptions. For example, $\epsilon$ expansion
relies on the hope that $\epsilon =1$ may appear to be a small
expansion parameter. Exact renormalization group approach requires
some truncation of the equations, and Schwinger--Dyson equations are
constructed with the use of the perturbation theory which may appear
to be not applicable. The lattice MC simulations produce a set of
``true'' characteristics of the phase transition (provided the
quality of the data is such that extrapolation to the continuum limit
is possible), which then can be used in cosmological applications. Of
course, MC simulations do not provide an analytical understanding of
the dynamics of the transition, but they may be considered as an
``experimental'' basis for construction or testing of different
theories of the phase transitions. The lattice MC simulations fail to
describe very weakly first order phase transitions. However, only
sufficiently strong first order electroweak phase transition is
of interest for cosmology, i.e. the lattice simulations can cover the
entire interesting range of parameters.

\subsection{The lattice formulation}
The 3d gauge theory defined by the action
\be
S  =  \int\! d^3x \biggl\{
\frac{1}{4}G^a_{ij}G^a_{ij}+
(D_i\Phi)^{\dagger}(D_i\Phi)+
m_3^2\Phi^{\dagger}\Phi+\lambda_3
(\Phi^{\dagger}\Phi)^2 \biggr\}
\label{lagr}
\ee
has a number of remarkable properties. It contains only dimensionful
coupling constants, and, therefore, it is super-renormalizable. The
only
ultraviolet infinities are those in the scalar mass renormalization,
while
the $\beta$ functions for $g_3^2$ and $\lambda_3$ are equal to zero.
The exact $\beta$ function for the mass parameter is known \cite{K5},
\be
\mu \frac{\partial m_3^2(\mu)}{\partial \mu}= -\frac{1}{16
\pi^2}f_{2m},
\ee
where
\be
f_{2m} ={51\over16}g_3^4+9\lambda_3 g_3^2-12\lambda_3^2.
\la{f2m}
\ee

The lattice version of the continuum theory \nr{lagr} is
defined by the following action:
\begin{eqnarray}
S&=& \beta_G \sum_x \sum_{i<j}\left(1-\fr12 \tr P_{ij}\right) +
\nonumber \\
 &-& \beta_H \sum_x \sum_i
\fr12\tr\Phi^\dagger(\bfx)U_i(\bfx)\Phi(\bfx+i) +
\la{lagrangian} \\
 &+& \sum_x
\fr12\tr\Phi^\dagger(\bfx)\Phi(\bfx) + \beta_R\sum_x
 \bigl[ \fr12\tr\Phi^\dagger(\bfx)\Phi(\bfx)-1 \bigr]^2.
\nonumber
\end{eqnarray}
Here $U_i(x)$ and $P_{ij}$ are the link and plaquette variables, and
$\Phi$ is the scalar field. The action depends on the three
dimensionless parameters $\beta_G, \beta_H,\beta_R$ (recall that
the continuum dynamics is completely described by two numbers, $x$
and $y$). The fact that the theory under consideration is
super-renormalizable enables one  to find an exact (in the continuum
limit)
matching between the lattice variables and the physical parameters.
The
corresponding calculation has been carried out in
refs. \cite{K3,Laine95}; it
provides the relation between the lattice renormalization scheme and
the
$\msbar$ scheme in continuum. The continuum limit is $\beta_G
\rightarrow \infty,~ \beta_H\rightarrow \fr13,~\beta_R\rightarrow 0$
in such a way that the physical parameters $x$ and $y$, defined as
\begin{eqnarray}
x & = & {1\over4}\lambda_3 a \beta_G =
{\beta_R\beta_G\over\beta_H^2}, \la{betar}\\
y & = &
{\beta_G^2\over8}\biggl({1\over\beta_H}-3-
{2x\beta_H\over\beta_G}\biggr)+{3\Sigma\beta_G\over32\pi}
(1+4x)+
\nonumber\\
&&+{1\over16\pi^2}\biggl[\biggl({51\over16}+9x-12x^2\biggr)
\biggl(\ln{3\beta_G\over2}+\zeta\biggr)+\eta+ \bar{\eta}x\biggr]
\la{y}
\end{eqnarray}
stay constant. Equation~(\ref{y}) depends on several constants
arising
from lattice
perturbation theory; $\Sigma=3.17591$, $\zeta=0.09$, 
$\eta=5.0$ and $\bar{\eta}=5.2$  were computed in  refs.
\cite{K3,Laine95}.

If we denote by $a$ the lattice spacing which has dimension of
length, then the overall continuum scale $g_3^2$ is related to it as
\be
g_3^2a  = {4\over \beta_G},
\ee
i.e. the continuum limit corresponds to $a \rightarrow 0$. The
relation
of the lattice variables to the continuum fields in this limit is
\be
\Phi=VR_L,\quad R_L^2= {2a\over \beta_H}\phi^{\dagger}\phi=
\fr12\tr\Phi^\dagger\Phi,
\quad U_i=\exp\left(\fr12 i a g_3 \tau_a A_i^a\right),
\ee
where $V$ is a unitary SU(2) matrix, $R_L$ is the radial mode of the
Higgs field, $\tau_a$ are the Pauli matrices.

Equations (\ref{betar}) and (\ref{y}) serve as the basis for relating
the
results of the lattice simulations to the physical reality. We
are not going to discuss here  technical details of the lattice
simulations \cite{K1}. Instead, we review the general
strategy for the lattice study of the phase transition. The reader
not
familiar with MC computations may consider the computer as a device
for calculating the Euclidean functional integral
\be
\langle O(A,\phi)\rangle =\frac{\int~ DA D\phi~ O(A,\phi)\exp(-S)}
{\int DA D\phi \exp(-S)},
\ee
where $O(A,\phi)$ is some gauge-invariant functional of the gauge and
scalar fields, and $S$ is the lattice action. The integral should be
computed for different volumes of the system and different lattice
spacings, and extrapolation of the results to the infinite volume and
zero $a$ must be taken at the end. By choosing the
 specific $\delta$-function form
of this functional,
\be
O=\delta(c-G(A,\phi))
\ee
one can construct probability distributions for various order
parameters $G(A,\phi)$, essential for the study of the phase
transition.

\begin{figure}
\hspace*{0cm}
\vspace{1cm}
\epsfysize=20cm
\centerline{\epsffile{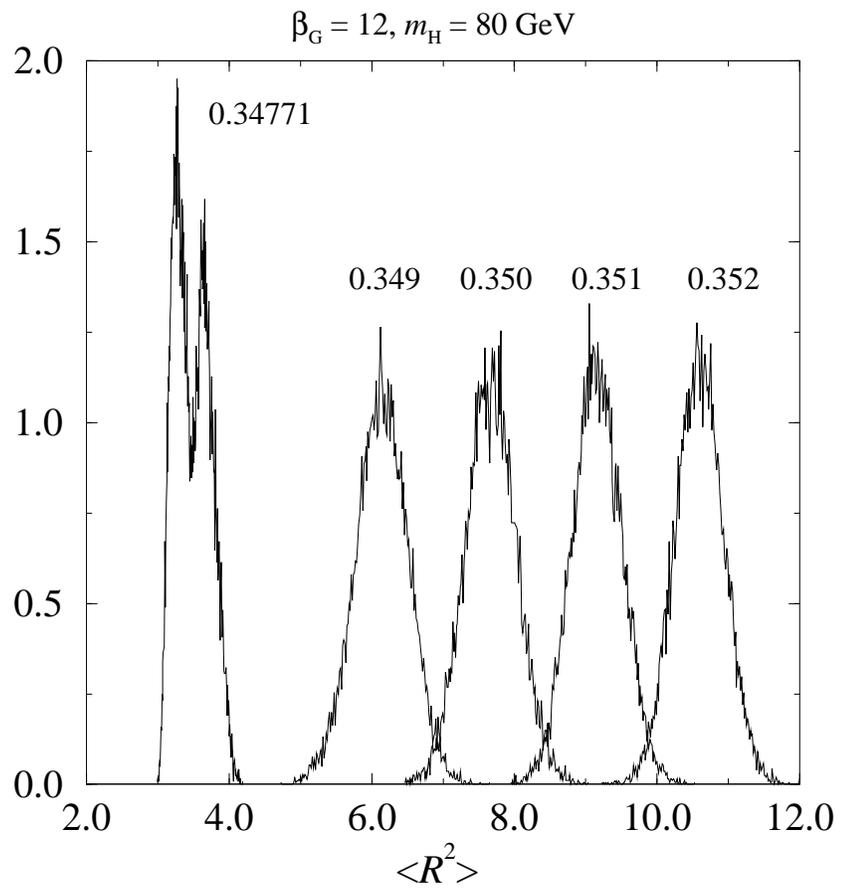}}
\vspace{-6cm}
\caption[a]{\protect
The evolution of the distribution of $\langle R_L^2\rangle$ with
$\beta_H$ (temperature). From ref. \cite{K3}.}
\label{evol}
\end{figure}

{\bf The critical temperature.} Let us fix $\beta_G$ (i.e. the
ultraviolet cutoff) and vary parameters $\beta_H$ and $\beta_R$ in
such a
way that the physical variable $x$  stays constant. In 4d
language this corresponds to
varying the temperature.
The existence of
the first order phase transition means that at some critical value of
$\beta_H$ there are two degenerate  ground states with
different properties. In particular, expectation values of the gauge
invariant observables are different in distinct states. For example,
the average value of $R^2$ is expected to be smaller in the symmetric
phase than in the Higgs phase. This means that the probability
distribution at the critical point has non-trivial form and contains
two peaks rather than one. The typical picture of the distribution
evolution is shown in fig. \ref{evol}. At large $\beta_H$ (low
temperatures) there is a unique value of the order parameter $R^2$,
corresponding to the Higgs phase, while at some $\beta_H$ a double
peak stucture develops, signalling about the first order phase
transition. The critical value of $\beta_H$ can be found from the
requirement that the areas under the two peaks are
equal to each other.
Of course, it must be checked that the double peak structure is not a
lattice artefact and passes different tests singling out the
first order behaviour \cite{K1}.
After the critical value of $\beta_H$
is found, it may be converted to
the
critical temperature
of
the
underlying 4d theory.

{\bf The latent heat.}
The latent heat $L$ --- the energy released in the transition --- can
be calculated from
\be
  \fr{L}{T} = \fr{d\Delta p}{dT} =
  \fr{T}{V}\fr{d}{dT}\Delta\log Z = \fr{T}{V} \fr{d}{dT} \Delta P\,,
  \la{latent}
\ee
where the derivatives are evaluated at the critical temperature,
$\Delta p$ is the difference of the pressures in the symmetric and
broken phases, and $\Delta P$ is the difference in the probabilities
of the phases in volume $V$. In \eq\nr{latent}, $T$ is the physical
(4d) temperature. The quantity $\Delta P$ is directly proportional to
the difference of the areas
under
the two peaks in the order parameter
distributions near $T_c$; it may be computed by performing
simulations
at $\beta_H$ close to the critical one.

{\bf The jump of the order parameter.} The order parameter usually
discussed in the study of the effective potential is the vacuum
expectation value $v$ of the Higgs field. This quantity, however, is
not gauge invariant. The gauge-invariant substitute of it is the
scalar condensate $\pdp$. This is a composite operator,
whose
expectation
value  contains linear divergence at the one-loop level and
logarithmic
divergence at the two-loop level. These divergences can be removed,
e.g. by the $\msbar$ prescription; the resulting condensate is then
dependent on the scale parameter $\mu$. Because of the fact that the
3d
theory is super-renormalizable, an {\em exact} relation between the
lattice quantity $\langle R^2\rangle$ and $\pdp$ can be found
\cite{K3},
\be
{\pp{\mu}\over g_3^2}=\fr18\beta_G\beta_H\biggl(\langle R^2\rangle-
{\Sigma\over\pi\beta_H}\biggr)-
{3\over(4\pi)^2}\biggl(\log{3\beta_G g_3^2\over2\mu}+\zeta
+\fr14\Sigma^2-\delta\biggr)+O\left(\frac{1}{\beta_G}\right).\la{rl2}
\ee
Numerically $\zeta+\fr14\Sigma^2-\delta=0.67$. Thus, the
extrapolation of lattice measurements of the quantity $\langle
R^2\rangle$ to the limit $\beta_G\rightarrow \infty$ allows one to
determine an ``exact'' value of the scalar condensate. The comparison
of the lattice results with known two-loop perturbative expansion
allows the extraction of the magnitude of the higher order
perturbative
terms. In this way a three-loop correction to the effective potential
has been numerically determined in ref. \cite{K1}.
For estimates, the relation $\fr12
\frac{v^2}{T^2}=\pp{T}/T$
can be used. The effective potential for the gauge-invariant
condensate was constructed in ref. \cite{Buch94}.

The jump of the quantity $\pdp$ at the phase transition (difference
between condensates in the broken and symmetric phases  at the
critical temperature) is a finite and scale independent quantity,
which can be directly measured on the lattice as the distance between
the positions of
the two
 peaks in the $R^2$ distribution,
\be
\Delta (\langle \phi^{\dagger}\phi \rangle)
= \fr18 g_3^2\beta_G \beta_H \Delta (\langle R^2\rangle).
\ee
It can be shown that $\Delta (\pdp)$ is directly related to the
latent
heat of the transition \cite{K3,K1,Buch95}.

\begin{figure}[tb]
\vspace*{-1cm}
\centerline{\hspace{-3.3mm}
\epsfxsize=15cm\epsfbox{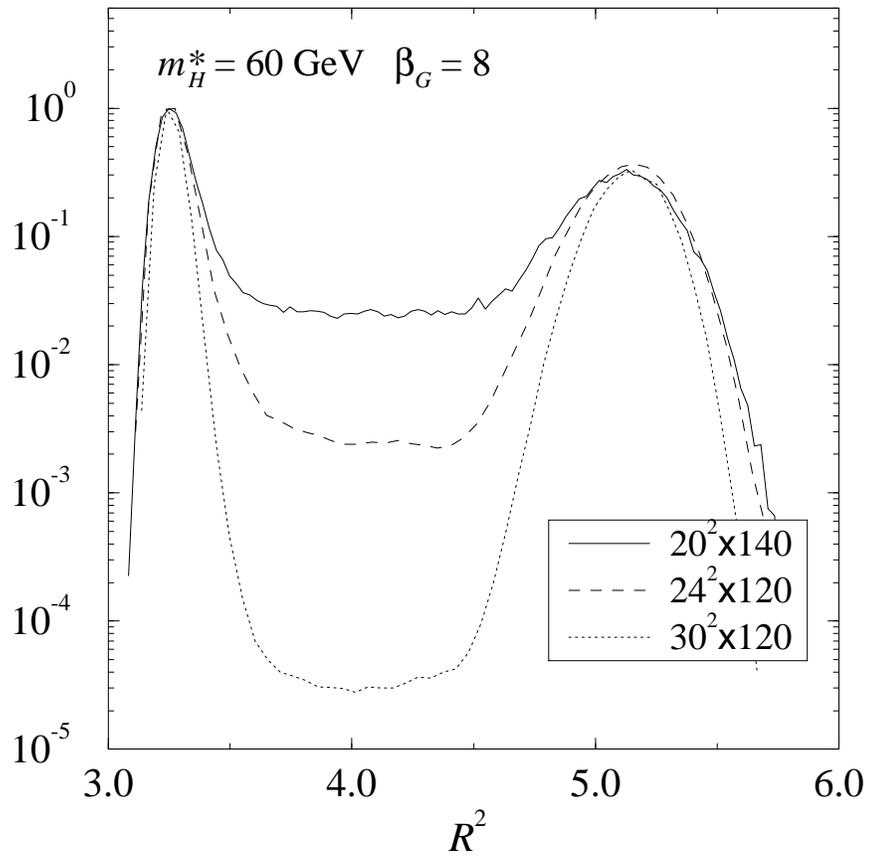}}
\vspace*{-5cm}
\caption[a]{\protect
The probability distribution of the average Higgs
length squared $R^2$ used for a surface tension determination. From
ref. \cite{K1}.}
\label{m60}
\end{figure}

{\bf The interface tension.} The interface tension is one of the most
important quantities which
characterize the strength of the phase transition. It can be measured
by constructing probability distributions of some local order
parameter.
At the critical temperature, a system in  finite
volume predominantly resides in either the Higgs or the symmetric
phase,
but it can also exist in a mixed state consisting of domains of the
two states. The probability of the mixed state is suppressed by the
extra free energy associated with the interfaces between the phases.
This causes the typical
two-peak
structure of the probability
distribution of the order parameter at the critical temperature (see
fig.~\ref{m60}): the midpoint between the peaks corresponds to a
state which consists of equal volumes of the symmetric and broken
phases. Because of the associated extra free energy, the area of the
interfaces tends to minimize. Assuming  lattice with periodic
boundary conditions and geometry $L_x^2\times L_z$, where $L_x \le
L_z$, the minimum area is $2\times A = 2(L_x a)^2$ --- the
factor 2
appearing because there are two separate interfaces. The interface
tension $\sigma$ can be extracted from the limit
\be
\fr{\sigma}{T} =
\lim_{V\rightarrow\infty} \fr{1}{2A} \log \fr {P_{\rm max}}{P_{\rm
   min}}\,
\la{sigma}
\ee
where $P_{\rm max}$ is the maximum of the probability distribution
in the peak and $P_{\rm min}$ is the
minimum of distribution
between the peaks in fig.~\ref{m60}.
At sufficiently large size of the system in the $z$ direction, the
probability distribution has a characteristic plateau,
corresponding to the translational zero mode of the domain wall.
Other methods for determination of the surface
tension are discussed in \cite{Fo95,4d1}, the estimate
of the higher order perturbative corrections is
contained in ref. \cite{Krip5}.

\subsection{Some lattice results}
The 3d lattice simulations have been made for four values of the
continuum parameter $x$, namely, 
$x=0.01830,~0.06444,~0.08970$ and $x=0.1188$ 
\cite{K1,Gur95}. These values correspond to
different 4d physical parameters
in different models. For definiteness, let us take the $SU(2)$ sector
of the standard model with top quark mass $m_t=175$ GeV. Then the
first
value of $x$ cannot be realized with any value of the Higgs mass (see
fig. \ref{xmhdep}), while the others correspond
to $m_H\approx 51.2,~68.0$ GeV and $\approx 81$ GeV, respectively.
The latter numbers come from the one-loop relations between
the $\msbar$ and physical parameters of the standard model and the
one-loop dimensional reduction \cite{K2}. Because of the large value
of the top
Yukawa coupling constant the one-loop corrections are quite
substantial for
small Higgs masses, for example $\delta m_H/m_H \sim 15\%$ for
$m_H\approx 50$ GeV.  A naive estimate of the two-loop corrections
(taken as square of the one-loop contribution) indicates that the
accuracy of the one-loop computation of the physical masses at fixed
3d parameter $x$ is $\sim 2\%$ for $x=0.06444$ and better for larger
$x$.

For the ``large'' value of the Higgs
mass ($81$ GeV) it was not possible to resolve the order of the
phase transition on the lattices up to $48^3$, and the data are
compatible with the smooth cross-over, second order phase transition
or very weakly first order phase transition. For smaller Higgs masses
the transition is of the first order.

The most complete study has been done for $x=0.06444$. This value may
be quite realistic for extended versions of the electroweak theory,
but is excluded experimentally in the MSM (it corresponds to
$m_H\approx 51.2$ GeV). Below we
present the specific numbers for MSM with $m_t=175$ GeV, derived
with the use of one-loop relations.  The critical
temperature of the phase transition is $T_c=89.79$ GeV, and the
vev-to-temperature ratio at $T_c$ is $v/T_c=0.64$. The domain wall
separating the broken and symmetric phases has the surface tension
$\sigma\simeq 0.002 T_c^{3}$, and the latent heat is $L/T_c^4=0.12$.
At $T>T_{+}= 89.93$ GeV only the symmetric phase is stable, and
at $T<T_{-}=89.36$ only the broken phase is stable, while
at temperatures between $T_{-}$ and $T_{+}$
both phases can exist simultaneously. The scalar correlation lengths
in the symmetric and Higgs phases are $\sim 6/T_c$ and $\sim
8/T_c$, respectively.
The statistical errors of the lattice numbers are
$0.015\%$ for the critical temperature, $1\%$ for the
expectation value of the Higgs
field and the latent heat, and $\sim 20\%$ for the interface tension.
The two-loop corrections may introduce extra uncertainties in these
numbers of the order of $2\%$.

The bubble nucleation temperature $T_{\rm bubble}$ lies somewhere
between $T_c$ and $T_{-}$ and may be estimated with the use of the
surface tension and latent heat found on the lattice. Inserting the
lattice numbers to
the relations (\ref{xdet}), (\ref{thinact}), (\ref{bubrate})
gives an estimate $\frac{\Delta T}{T_c}\simeq 0.0004$,
i.e. the bubble nucleation temperature is
very close to the critical one. The smallness of ${\Delta T}/{T_c}$
is due to fact that the ratio $\sigma^3/L^2T_c\sim 2 \cdot 10^{-6}$
is so small~\cite{ikkl}. Since $\Delta T/(T_c-T_{-})\simeq 0.1$ is
also small, one is in the thin wall regime; indeed, the size of the
bubbles when they nucleate is at least $R_c \simeq 110/T_c$, which is
much larger than the scalar correlation length in either the broken
or
the symmetric phase at $T_{\rm bubble}$. Because $T_{\rm bubble}$ is
very close to the critical temperature, the expectation value
of the Higgs field at
$T_{\rm bubble}$ is almost the same as at $T_c$. In other words, the
transition is very weakly first order.

The lattice results can be compared with the perturbative
calculations of the two-loop effective potential. The predictions of
the
critical temperature, latent heat, and the jump of the order
parameter appear to be quite reasonable and are within a few per cent
for these quantities. However, perturbation theory fails to describe
the surface tension (and, therefore, the bubble nucleation rate) at
least for $x=0.06444$: the perturbative value is about 3 times larger
than the lattice one.
\subsection{Dynamics of the phase transition}
In the case of the ``vacuum phase transitions'' -- false vacuum decay
at zero temperatures -- the energy stored in the
metastable vacuum transforms
into kinetic energy of the domain walls \cite{VolKobOkun,CBounce}.
As a result, the velocity of a domain wall increases and approaches
the speed of light. Then in the collisions of domain walls, their
kinetic energy is released and transforms into heat, and the system
is reheated up to a certain temperature.

\begin{figure}[tb]
\vspace*{-1cm}
\centerline{\hspace{-3.3mm}
\epsfxsize=15cm\epsfbox{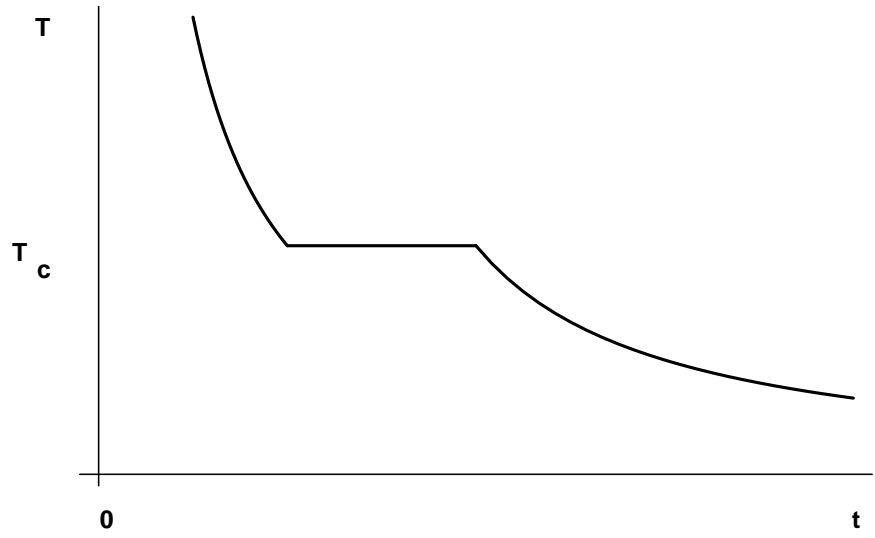}}
\vspace*{-10cm}
\caption[a]{\protect
The temperature evolution at the electroweak phase transition in the
adiabatic case.}
\label{ideal}
\end{figure}

At non-zero temperatures nucleated bubbles expand in the medium, and
the latent heat of the transition may be released by many different
mechanisms. In the idealistic case of the very slow expansion of the
Universe, the temperature stays constant during the phase transition,
fig. \ref{ideal} (see, e.g. ref. \cite{Enq92}).
Every point on the plateau
corresponds to a mixed state containing domains of broken and
symmetric phase; the left point of the plateau corresponds to the
pure symmetric phase, and the right point to the pure broken phase.
The
release of the latent heat of the transition is accommodated by the
Universe expansion.

In fact, the Universe expands not so slowly at the electroweak scale
and is  supercooled in the symmetric phase. One can distinguish
several different epochs in the phase transition. The first one is
the
bubble nucleation, the second one is the bubble growth, the third one
is the bubble percolation -- the period when  different bubbles
collide. At this stage the Universe may be reheated up to the
critical temperature (fig. \ref{ideal}). If it happens, then the
later evolution may be
close to an ideal case described above.

We have already discussed  the bubble nucleation rate and
determination of the bubble nucleation temperature $T_b$. Let us now
consider the bubble expansion from macroscopic
point of view in more detail
\cite{Ste82,Gyu83,Enq92,Kaj92,Carrington:1993,L994,ikkl,Kurki95,Kurk95}.

Initially, the bubble of a new phase is a microscopic object with
a size of several correlation lengths. At the bubble nucleation
temperature the energy density in the symmetric phase is larger than
that in the broken phase,
\be
\epsilon_{sym}= \frac{\pi^2}{30}N_{eff}T^4,~~
\epsilon_{Higgs}= \frac{\pi^2}{30}N_{eff}T^4-L.
\ee
Here $N_{eff}=N_b+\fr78 n_f$ is the effective number of the massless
degrees of freedom and $L$ is the latent heat. For the Minimal
Standard Model $N_{eff}=106\fr34$. The general hydrodynamical
consideration of the bubble evolution leads to  two possible types
of bubbles, known as deflagration and detonation
bubbles \cite{Ste82,Gyu83}.

Consider an {\em isolated} macroscopic bubble. For a large enough
bubble, its curvature may be neglected and the interface may be taken
as a planar domain wall \footnote{As was shown in \cite{Hue993},
the domain walls are stable against small perturbations.}.
Let us proceed in the rest frame of the domain
wall and denote by $v_1$ the velocity of the medium falling on the
domain wall (symmetric phase) and by $v_2$ the velocity of the medium
inside the bubble. Then, if $v_1 < c_s$, where $c_s\simeq
\frac{1}{\sqrt{3}}$ is the velocity of sound in the medium, then the
phase transition proceeds through deflagration. In this case the
medium is accelerated when it passes through the domain wall,
$v_2>v_1$. If, on the contrary, $v_1 > c_s$, then $ v_1>v_2$, and we
have detonation bubbles. The realization of
one of the two mechanisms of
the bubble walls propagation depends on the relationship between the
latent heat of the transition, surface energy density, and the rate
of entropy generation on the phase boundary \cite{Gyu83}.
The latter is to be found from the microscopic analysis
of the interaction of particles with the bubble
walls \cite{Tu92,dlhll:pr,khl:wall,Liu92,Ar93,M95,Mo95}.

If we now choose the reference frame to be the rest frame of the
plasma
before the bubble has nucleated, then the medium should be at rest in
the centre of the bubble and far from it. Then, for the deflagration
bubbles the medium in front of the bubble wall is
accelerated by the motion of the wall, and there is a shock wave in
the symmetric phase moving with the velocity $v_{shock}>v_b$, and
$v_b=v_2$ is the bubble wall velocity. The
velocity of the symmetric phase plasma between the
shock front and the bubble wall is given by
$v_{sym}=(v_2-v_1)/(1-v_1v_2)$. The temperature of the medium
between the shock front and the bubble wall $T_{shock}$ and the
temperature inside the bubble $T_r$ are  different from the bubble
nucleation temperature (temperature outside the shock front) $T_b$.
They
depend in general on the distance from the bubble centre. Usually the
inequality $T_b < T_r < T_{shock}$ holds true \cite{Gyu83},
but $T_r < T_b$ at high rates of entropy generation at the domain
walls \cite{ikkl}. The latent heat of the transition
transforms
into the kinetic energy of the plasma in the symmetric phase, and
to  heating of the plasma inside the shock wave front.

The detonation bubbles have a different structure. The velocity of
the domain wall is larger than the speed of sound, the symmetric
phase plasma is at rest right in front of the domain wall and has
temperature $T_b$. The plasma just behind the wall (in the Higgs
phase) is accelerated by it and has velocity
$v_{br}=(v_1-v_2)/(1-v_1v_2)$. Finally, the plasma is stopped at some
surface inside the bubble by a rarefaction wave. As in the previous
case, the temperatures inside the front of the rarefaction wave $T_r$
and behind the bubble wall $T_{br}$ are different from $T_b$.

The recent analysis carried out in ref. \cite{Mo95,M95}
suggests that for a sufficiently
wide range of the parameter space of the standard model, $m_H< 90$
GeV,
an isolated bubble expands as a weak deflagration (``weak'' means
that
the velocity of the bubble wall is subsonic). The velocity $v_b$ was
found to be in the range $0.38<v_b<0.45$, while the velocity of the
shock front is close to the speed of sound in the medium,
$v_{shock}\simeq c_s\simeq 0.58$. When the shocks coming from
different bubbles begin to collide, the single bubble
approximation breaks down. At this time, roughly, a $(v_b/c_s)^3
\simeq 0.3$ part of the Universe volume is in the broken phase. The
subsequent evolution of bubbles depends on the temperature in the
symmetric phase, which will emerge as a result of shock waves
interaction. This temperature can be estimated
as follows \cite{ikkl,Heck95}. Suppose
that all latent heat of the transition is immediately released in
form of heat in the broken phase. Then the reheating temperature can
be found from energy conservation condition,
\be
\frac{\pi^2}{30}N_{eff}T_b^4= \frac{\pi^2}{30}N_{eff}T_r^4-L,
\ee
so that
\be
\frac{T_r-T_b}{T_c}= \frac{15}{2\pi^2N_{eff}}\frac{L}{T^4}.
\ee
Now, if $\frac{T_c-T_b}{T_c}\gg\frac{T_r-T_b}{T_c}$, then the
reheating process may be neglected. However, in the opposite case the
heat release is important and the expansion of the bubbles of a new
phase should slow down.

Simple estimates can be done in a thin wall approximation. The bubble
nucleation temperature is given by
\be
\frac{T_c-T_b}{T_c}= \sqrt{\frac{16\pi\sigma^3}{3L^2T_c S_0}},
\ee
where $S_0\simeq 160$ is the action for the critical bubble and
$\sigma$ is the surface tension.
For example, for the Minimal Standard Model with
$m_H=51.2$ GeV \footnote{Of course,  no such Higgs boson
exist in minimal standard model because of the experimental
constraint.
This value is
taken because this is the highest MSM Higgs mass for which
the magnitude of the surface tension is known reliably
from the lattice simulations, see section 5.5. Being
unrealistic for MSM, this example is phenomenologically
acceptable for the extensions of the standard model.}
and $m_t = 175$ GeV we get $T_c=88.93$ GeV, $L/T^4=0.124$,
$\sigma/T^3=0.0023$ \cite{K1}, and
\be
\frac{T_r-T_b}{T_c}=9\cdot 10^{-4}>\frac{T_c-T_b}{T_c}=3\cdot10^{-4}.
\ee
So, for this choice of parameters the Universe should reheat to the
critical temperature, the speed of the domain walls is greatly
reduced,
the broken phase bubbles expand slowly due to the Universe expansion.
At the final stage of the phase transition the remnants of the
symmetric phase shrink, again due to the Universe expansion.
This picture is also true for higher values of the
MSM Higgs masses, provided the phase transition is still
of the first order.

An estimate of the bubble wall velocity at a late stage of the phase
transition (under the
assumption that the Universe is reheated up to the critical
temperature)
can be found from a simple thermodynamical
consideration \cite{KK86} (see also ref. \cite{Heck95}).
Suppose that the average bubble size is $R_b$, and that the part of
the volume of the space occupied by the broken phase is ${\cal P}$.
Then the requirement that the Universe expands adiabatically is given
by
\be
\frac{{\dot{R_b}}}{R_b} {\cal P} L= s H,
\ee
where $s$ is the entropy density in the symmetric phase,
$s=\frac{2\pi^2}{45}N_{eff}T^4$,
and $H$ is the Hubble constant. The average bubble size can be found
from the following consideration. The (non-normalized) distribution
of the bubbles in sizes at moment $t$ is given by
\be
P(R,t)dR \sim {\cal R}(t_1),
\ee
where  ${\cal R}(t_1)$ is the bubble nucleation probability at time
$t_1$, and $t_1$ is related to $R$ and $t$ by the  obvious
condition that $t_1=t-\frac{R}{v_b}$. Then
\be
\langle R \rangle(t) = \frac{\int~ R~ P(R,t)~dR}{\int ~P(R,t)~dR}.
\ee
In the thin wall approximation
\be
{\cal R}(t) \sim \exp \left(-\frac{AT_c^2}{4(T_c-T)^2}\right)
  \sim \exp \left(-\frac{At_c^2}{(t-t_c)^2}\right)
\ee
with $A= \frac{64 \pi \sigma^3} {3 L^2 T_c}$, and $t_c$ is the time
corresponding to the temperature $T_c$. The average bubble
radius at time $t$ is
\be
\langle R \rangle (t) \simeq \fr12 v \frac{(t-t_c)^3}{A t_c^2}
\ee
and at the percolation temperature it is about $ \langle R \rangle
(t)\simeq \fr12 v_b t_c \frac{\sqrt{A}}{\sqrt{S_0^3}}$.  For example,
for the numerical values of the parameters given above one finds
$A\simeq 5.3\cdot 10^{-5},~HR_b\simeq 10^{-6}$ and
\be
{\dot{R}}=v_b\simeq \frac{s}{L} \frac{H R_b}{{\cal P}} \sim 10^{-3}
\ee
for ${\cal P}\sim 0.3$. This value gets larger if the scalar
self-coupling is decreased. For
example, for  stronger phase transition the reheating up to the
critical temperature may occur, and the velocity of the domain wall
from the estimate of ref. \cite{Heck95} may be higher
by a factor $\sim 10$.

Of course, the estimates given above are rather rough since they are
based on the assumption of the instantaneous latent heat release.
Nevertheless, they show that it is quite plausible that the slow
stage of the phase transition takes place for the interesting range
of
the parameters of the underlying theory \cite{Heck95}.
\section{Survival of primordial baryon asymmetry}
The anomalous electroweak processes are rapid at sufficiently high
temperatures. Their rate $\Gamma_{sph}$ exceeds the rate of the
Universe expansion $\frac{T^2}{M_0}$ in the standard Big Bang
scenario in the following interval of temperatures:
\be
100 \mbox{GeV} \sim T^*< T < T^{**}\simeq
\alpha_W^4 M_{Pl}\simeq 10^{12} \mbox{GeV},
\ee
where the lower temperature $T^*$ is to be found from the condition
of
decoupling of the sphaleron processes in the broken phase of the EW
theory \cite{s:m^14,s:sm87},
\be
\frac{M_{sph}(T^*)}{T^*} \simeq 45.
\label{strength}
\ee
Clearly, the equilibrium character of $B$-violating reactions has an
important impact on the survival of the primordial baryon
asymmetry.  Several different cases can be distinguished, depending
on
initial conditions and on the rate of $B$ and $L$ non-conservation
due to processes other than those associated with sphalerons.\\

(i) Suppose that the Universe is asymmetric with respect to the
anomaly free fermionic charges $\Delta_i=L_i-\frac{1}{n_f} B$ of the
standard model at $T>T^{**}$, and assume that at $T<T^{**}$
there is no $B$ or $L$
violating interactions besides the electroweak
anomalous processes. The origin of the primordial asymmetry
is not essential here. Then anomalous reactions convert the initial
asymmetry to the baryonic one at $T=T^*$. For the minimal standard
model the relationship is given by
\cite{krs87,ks} (see also \cite{Harvey,Dreiner}),
\be
\Delta_0 = \frac{8n_f+4}{22n_f+13} \Delta _{B-L}
- K \frac{4}{13 \pi^2} \sum_{i=1}^{n_f} \frac{m_i^2(T^*)}{(T^*)^2}
\Delta_i .
\label{bfin}
\ee
where $m_i^2$ is the lepton mass of a given generation, $K \approx
1$.
The first term in the right hand side of this equation tells that
$(B-L)$ asymmetry is reprocessed into the baryon asymmetry, while
$(B+L)$ tends to be washed out; the second term is the correction
coming from slightly different behavior of quarks with different
masses in the plasma.
If the initial value of $(B-L)$ is non-zero (coming, say, from the
GUT
physics) then the baryon asymmetry, up to a possible contribution
from the EW phase transition (see below), has a primordial character.
If, on the contrary, the initial $B-L$ asymmetry is absent, we can
rely
only on the second term in (\ref{bfin}).
For three lepton generations one gets a suppression
$\Delta_0 \simeq 3 \times 10^{-6} \Delta_3$.
So, to have a non-negligible effect, the initial asymmetry $\Delta_3$
must be very large, or the standard theory should be extended by
adding heavy leptons.

(ii) Suppose now that there are some reactions, which do not
conserve all $\Delta_i$, and which are in thermal equilibrium for
some period between $T^*$ and $T^{**}$. At this intermediate epoch
$B$ and $L$ are
non-conserved separately, and according to the third Sakharov
condition all baryonic and leptonic asymmetries are washed out.
Hence, the existence of these reactions is fatal for the primordial
baryon asymmetry. If the baryon asymmetry
is not produced at a later time, the
requirement of the absence of these reactions may appear to be a
powerful tool for constraining the properties of new particle
interactions \cite{Fuku90,Harvey,Ne90,Fis91}. However, some time
ago  it was realized that
most of these constraints are drastically weakened due to the
smallness of some Yukawa coupling constants in the standard model or
its supersymmetric extensions \cite{Iban92,Ca92,Cam92,Cl2,Cl1,Cl3}.

Let us discuss the main idea of these estimates on the example of
lepton number violating interactions, leading to the Majorana
neutrino masses $m_{ij}$ \cite{Cl1}. We take for simplicity the
Minimal Standard
Model and add to it lepton number violating interactions. The
$SU(2)\times U(1)$ symmetric low energy
Lagrangian with $\Delta L =2$ has the
form
\be
\frac{1}{v^2}m_{ij}(\bar{L_i}\phi)(\tilde{\phi}^{\dagger}L_j^c),
\ee
where $L_i$ and $L_j^c$ are  lepton doublet and its charge
conjugate, respectively, $\phi$ is the scalar doublet, $v$ is the
vacuum expectation value of the Higgs field $v=246$ GeV. The rate of
$L$ non-conserving reactions $L \phi \rightarrow L^c \phi^*$ at high
temperatures has been found in  ref. \cite{Cl1}
\be
\Gamma \simeq \frac{9}{\pi^5}\frac{T^3}{v^4}\bar{m}_{\nu}^2,
\ee
where $\bar{m}_{\nu}$ is an average Majorana neutrino mass,
\be
\bar{m}_{\nu}^2=\fr53|m_{ee}|^2 + |m_{e\mu}|^2+|m_{e\tau}|^2.
\ee
These reactions were initially required \cite{Harvey} to be out of
thermal equilibrium  at $T<T^{**}$; this leads to a very stringent
constraint
\be
\bar{m}_{\nu} <\frac{v^2}{\sqrt{M_0 T^{**}}}\simeq 10^{-2} \mbox{eV}.
\label{neutr}
\ee
An implicit assumption in the derivation above is that the set of
conserved numbers $\Delta_i$ is a complete one below $T=T^{**}$. In
fact, this is not true due to the smallness of the right-handed
electron
Yukawa coupling constant. In the limit when this constant is zero,
the right-handed electron number is conserved, and the asymmetry in
it
propagates to the asymmetry in baryon number. The rate of reactions
not conserving the right-handed electron number (say,
$e_L H \rightarrow e_R W$) is of the order of
\be
\Gamma_R \sim \alpha_W f_e^2 T
\ee
where $f_e$ is the electron Yukawa coupling constant.
These reactions are out
of equilibrium at $T>T_R\simeq 3$ TeV \cite{Cl1}. It is this
temperature
which should be used in eq.(\ref{neutr}) instead of $T^{**}$. Thus,
we
arrive at  much weaker constraint \cite{Cl2,Cl1,kimmo}
\be
m_{\nu}< 8 \mbox{KeV}.
\ee
which must be satisfied in any case because of the known laboratory
limits and other cosmological considerations.

The same type of considerations apply to other possible interactions
breaking lepton and baryon numbers. The general conclusion is
that the initial charge asymmetry can survive during the epoch at
which anomalous reactions are at thermal equilibrium. Moreover,
initial asymmetries in fermionic quantum numbers, different from the
baryon number, are usually transferred to baryon asymmetry towards
the end of the equilibrium sphaleron period.

We barely know the history of the Universe at very high temperatures
(say, at $T\gg 1$ TeV).
It may well be  that the Universe was symmetric with respect to all
fermion  charges at $T>10^{12}$ GeV. This assumption, being
a bit arbitrary, may be in fact a natural consequence of inflation,
which exponentially dilutes the densities of all global quantum
numbers (e.g. baryonic or leptonic). If true, baryon asymmetry
should be produced at relatively late stages of the Universe
expansion.
As pointed out in section 3, this may happen either at intermediate
temperatures (1 TeV $< T < 10^{12}$ GeV) or at the electroweak
temperature
($T \sim$ (a~few)$\times 100$ GeV).

Our main topic is the discussion of baryogenesis in the case where
the
only relevant source of $B$ and $L$ non-conservation is the
electroweak
anomaly. In a sense, this is the most conservative possibility, since
it relies only on  physics we trust experimentally. We further
constrain ourselves and consider the Minimal Standard Model or its
natural extensions, such as the two Higgs doublet model or
supersymmetry.
In these models
the only known possibility to generate the observed baryon asymmetry
is that associated with
the electroweak
phase transition. Further extensions of the standard theory,
containing topological defects (such as strings) can also lead to
baryogenesis via anomalous reactions. The discussion of this
interesting possibility can be found in refs.
\cite{Bran89,Bran91,Bran93,Bran94,Trod95}.
\section{Electroweak baryogenesis}
\subsection{Strength of the phase transition}
There is a general condition that should be satisfied in any
particle physics model used for the generation of the baryon
asymmetry
at the electroweak scale
\cite{s:m^14,s:sm87,bs:higgs}. Namely, the baryon asymmetry
created by some mechanism
must not be erased by the anomalous reactions. In other words, the
sphaleron processes should be out of thermal equilibrium
immediately after the
electroweak phase transition (in the broken phase), i.e.
inequality (\ref{strength}) must be satisfied.
The temperature $T^*$ can be as large as the critical temperature
$T_c$, if the Universe is reheated up to it, or as small as the
bubble nucleation temperature $T_b$. The requirement (\ref{strength})
places a strong
constraint on the strength of the phase transition, and, therefore,
on the parameters of the electroweak theory. The recent
discussion of this bound, incorporating the results of numerical
simulations of the electroweak phase transition is contained in ref.
\cite{K1}. We sketch here the main points of the analysis.

As  discussed in Section 5, the 3d $SU(2)\times U(1)$ gauge--Higgs
theory
plays a role of universal theory of the electroweak phase transition
in the
Minimal Standard Model
and a number of its extensions. The 3d effective theory is
characterized by a unique parameter,  $x=\lambda_3/g_3^2$, completely
defining its dynamics; in particular, the effective sphaleron mass is
a function of it. In the one-loop approximation
\cite{Baacke1,Baacke2},
\be
\frac{E_{sph}(T)}{T}=
B\left(\frac{\lambda_3}{g_3^2}\right)\frac{2 \pi T^{1/2}}{g_3}
\frac{\phi}{T}
\label{sph1loop}
\ee
where $\phi$ is the scalar field expectation value  determined
from the one-loop effective potential. The two-loop corrections to
the
sphaleron mass are unknown; parametrically
\be
\frac{\delta E_{sph}(T)}{E_{sph}}=
     A \left(\frac{g_3^2}{\pi m_T}\right)^2.
\ee
where $m_T$ is defined by eq. (\ref{masses}).
The perturbative and numerical analysis of various quantities
in the broken phase (such as free energy, correlation lengths, vacuum
expectation value of the Higgs field) suggests that the ``true''
expansion parameter is $\kappa g_3^2/(\pi m_T)$ with $\kappa \sim 1$.
So, it is natural to assume that $|A| \sim 1$. Then, from
eqs. (\ref{strength}) and (\ref{sph1loop}) one obtains
$v/T>1.2$ or  $v/T > 1.5$, depending on the sign of $A$. To establish
a
conservative upper bound, a bubble nucleation temperature $T_b$,
which is  somewhat smaller than the critical temperature, should be
taken. If the perturbative description of the bubble nucleation based
on
two-loop effective potential is valid,
then at $T_b$ the ratio $v/T$  is about
$20\%$ larger than at the critical temperature and we may require
$v(T_c)/T_c > 1$.  Now, the ratio $v/T$ at the critical temperature
is a function of $x$.
The use of the lattice MC results, together with perturbation theory,
allows us to determine this function quite reliably\footnote{The
complete
$SU(2)\times U(1)$ gauge-Higgs model in 3d has never been simulated
on the
lattice, so that the treatment of the $U(1)$ factor is
perturbative.},
and the lower limit on the vacuum expectation value is converted to
the upper limit on the  ratio of constants,
$x<0.043$. For an extreme opposite case, when $A>0$ and the Universe
is reheated up to the critical temperature, the constraint is
somewhat stronger, $x<0.026$. To summarize, electroweak baryogenesis
requires that the parameter $\lambda_3/g_3^2$ in the 3d
$SU(2)\times U(1)$ gauge--Higgs effective theory is bounded from
above,
\be
\lambda_3/g_3^2< 0.026-0.043.
\label{bound}
\ee
In order to obtain the constraints, following from this requirement,
on
the particle spectrum of the underlying 4d theory, one has to express
this ratio through the physical parameters of the 4d theory at the
critical temperature. This computation may be quite
involved~\cite{K2}, but it is very clean from the physics point of
view and does not contain any infrared
divergences. An essential point is that only {\em one-loop graphs
have to
be computed} in weakly coupled gauge theories, such as the MSM or
MSSM.

The application of the constraint of eq.~\nr{bound} to the case of
the MSM follows from fig. \ref{xmhdep}. For experimental numbers
$m_t=175\pm 20$
GeV, and $m_H>65$ GeV one finds $x>0.07$, which is inconsistent with
eq. (\ref{bound}). Moreover, for $m_t=175$ GeV {\em no Higgs mass}
can
ensure the necessary requirement of eq.~(\ref{bound})\footnote{The
upper bound on the Higgs mass was evolving in time quite a bit. The
first estimate of the critical Higgs mass, based on the one-loop
effective potential for the scalar field with small mass of the top
quark (which was unknown at the time), gave a value of $m_H< 45$ GeV
\cite{s:sm87,bs:higgs}. The accounting for the
large top quark mass and Debye
screening effects in the one-loop effective potential reduced this
number to $m_H< 35$ GeV \cite{dlhll:pr}. The two-loop effects
\cite{bd,ae} somewhat relaxed this condition, while the assumption
about
the large non-perturbative effects in the symmetric
phase \cite{s:condensate} allowed a sufficiently strong first order
phase transition with
experimentally allowed Higgs boson. The lattice
simulations \cite{K1,Fo95} reduced
all uncertainties substantially.}. So, in the MSM the baryon number
non-conservation is in
thermal equilibrium after the phase transition.
This points to new physics at the
electroweak scale, which may strengthen the first order nature of the
electroweak phase transition.

The two Higgs doublet model has more freedom, and the results of
refs. \cite{bks:higgs,tz:higgs} show that the constraint
(\ref{bound}) can be satisfied there. 
The reason is that the effective 3d scalar
self-coupling constant is a complicated combination of the different
scalar and pseudoscalar masses and mixing angles.
The
extensions of the standard model (supersymmetric or not), including
scalar singlets, can also help \cite{And92,Choi93,Es993}.

According to refs. \cite{Giu92,Espi93,Brig94} the phase transition in
the MSSM in the most part of the parameter space occurs in the same
way as it does in the MSM. Here the MSSM also fails in preserving the
baryon asymmetry after the phase transition. However, in a 
recent paper \cite{cqw} a specific portion of the parameter space of
the MSSM, where electroweak baryogenesis is possible, has been
found\footnote{We thank M. Carena, M. Quiros and C.E.M. Wagner for
describing their results prior to publication.}. What is most
interesting is that quite strong constraints on the masses of the
Higgs boson and squarks were derived.

In order to explain the idea of ref. \cite{cqw} in a simplest way 
let us add to the minimal standard model an $SU(2)$ singlet
but colour triplet scalar field (scalar quark) $\chi$ with the potential
\be
U(\chi,\Phi)= -\fr12 m_H^2 \Phi^{\dagger}\Phi +
\lambda(\Phi^{\dagger}\Phi)^2
+ m^2\chi^*\chi + 2 h\chi^*\chi \Phi^{\dagger}\Phi+\lambda_s
(\chi^*\chi)^2.
\ee
Assume now that the expectation value of the field $\chi$ is zero at
all temperatures (this is possible at some particular choice of
parameters). Then the contribution of this field to the
effective high temperature Higgs potential is
\be
-\frac{2\cdot 3}{12\pi}(m^2(T) + h \phi^2)^{\fr32}.
\ee
Now, if the effective high temperature mass $m(T)$ is small near
the electroweak phase transition, $m^2(T_c) \simeq 0$, then this term
increases the magnitude of the cubic coupling $\alpha$ in the
effective potential (\ref{1loop1storder}), $\alpha \rightarrow
9g^3/(32\pi)+ 3h^{\fr32}/(2\pi)$. This, in turn, makes the phase
transition stronger first order, and the value $\phi(T_c)/T_c$ (see
eq. (\ref{vc})), crucial for the electroweak baryogenesis,
increases\footnote{A similar idea of introducing $SU(3)\times
SU(2)\times U(1)$ scalar singlets in order to enhance the strength of
the electroweak phase transition has been suggested in ref.
\cite{And92}. A new element of \cite{cqw} is accounting for the high
temperature contributions to the SU(2) singlet scalar mass.}.

In the case of MSSM the role of SU(2) singlet is played by the right
handed light stop \cite{cqw}. Its high temperature effective mass
$m^2(T)$ contains two essential contributions. The first one is the
soft supersymmetry breaking mass, and the second is a positive
temperature contribution $\sim g_s^2 T^2$, where $g_s$ is the strong
gauge coupling constant. To make the idea work, the soft SUSY
breaking mass must be negative and approximately equal to the high
temperature contribution at the critical temperature. Previously, the
negative values of that mass have not been considered because of the
danger of colour breaking; the authors of \cite{cqw} have shown that
it is possible to satisfy simultaneously the requirements of the
absence of colour symmetry breaking, strongly enough first order
phase transition together with experimental bounds on SUSY particles.
The region of parameters allowing for electroweak baryogenesis requires
that the Higgs mass is smaller than the $Z$ mass, lightest stop mass
is smaller than the top mass, and $\tan \beta <3$. This range of masses
is accessible
for experimental search  at LEP2 and Tevatron.

Yet another possibility to have strongly enough first order phase
transition may be realized in the models of dynamical electroweak
symmetry breaking \cite{dynamic1,dynamic2}. Here the phase transition
occurs in a strong coupling regime both in the symmetric and Higgs
phases and high temperature 3d description does not work. These type
of models predict new (strongly interacting) physics at the TeV
scale.

\subsection{Sources of CP-violation in the EW theory and its
extensions}
To produce the baryon asymmetry, the particle interactions must
break C and CP symmetry.  C symmetry is broken due to the chiral
character of electroweak interactions. In the Minimal Standard
Model, the conventional
source of  CP violation is that associated with
Kobayashi--Maskawa (KM) mixing of quarks.
The Yukawa interaction of quarks with the Higgs boson in the MSM has
the
following form,
\be
{\cal L}_Y = \frac{g_W}{\sqrt{2}M_W} \{\bar{Q}_L K M_d
D_R\phi +
\bar{Q}_L M_u U_R\tilde{\phi} + \mbox{h.c.}\},
\label{Yukawa}
\ee
where $M_u$ and $M_d$ are diagonal mass matrices of up and down
quarks, $K$ is the KM mixing matrix, containing one CP violating
phase $\delta_{CP}$. The MSM contains yet another source of
CP-violation, associated with the QCD vacuum angle $\theta$. It is
constrained experimentally, $\theta <10^{-9}$.

A popular extension of the MSM is a model with two Higgs doublets,
$\varphi_1$ and $\varphi_2$. In order to suppress flavour changing
neutral currents, the interaction of Higgs bosons with fermions is
chosen in such a way that $\varphi_1$ couples only to right-handed
up quarks
while $\varphi_2$ couples only to down quarks. The other possibility
is that $\varphi_2$ decouples from fermions completely and
$\varphi_1$ gives masses to all the fermions. In addition to the KM
mixing, this model contains CP violation in the Higgs sector. The
scalar potential has the form \cite{Gun86}:
\[
V=\lambda_{1}(\varphi_{1}^{\dagger}\varphi_{1} - v_1^2)^2 +
\lambda_{2}(\varphi_{2}^{\dagger}\varphi_{2} - v_2^2)^2 +
\]
\[
\lambda_{3}[(\varphi_{1}^{\dagger}\varphi_{1} - v_1^2) +
            (\varphi_{2}^{\dagger}\varphi_{2} - v_2^2)]^{2}+
\]
\be
\lambda_4[(\varphi_{1}^{\dagger}\varphi_{1})(\varphi_{2}^{\dagger
}
\varphi_{2}) -
(\varphi_{1}^{\dagger}\varphi_{2})(\varphi_{2}^{\dagger}\varphi_{
1})]+
\label{pot}
\ee
\[
\lambda_5[\mbox{Re}(\varphi_{1}^{\dagger}\varphi_{2}) -
v_1v_2\cos\xi]^2
+
\lambda_6[\mbox{Im}(\varphi_{1}^{\dagger}\varphi_{2}) -
v_1v_2\sin\xi]^2,
\]
$\xi$ being a CP-violating phase.

In the supersymmetric extensions of the standard model the Higgs
potential is CP invariant and  CP is violated by the soft
supersymmetry breaking terms.  In the simplest version of MSSM there
are
two extra CP phases and the relevant interaction has
the form \cite{Ell82}, (for a review see ref. \cite{Nill84})
\be
[\mu\hat{H}\hat{H}']_F + m_g [A (\hat{\bar U} \xi_U
\hat{Q}\hat{H}
+\hat{\bar D} \xi_D \hat{Q}\hat{H}' + \hat{\bar E} \xi_E
\hat{L}\hat{H}') + \mu_B \hat{H}\hat{H}']_A + \mbox{h.c.}
\ee
where $\hat{U},~\hat{D},~\hat{Q},~\hat{L},~\hat{E},\hat{H}$ and
$\hat{H}'$ are the quark, lepton and Higgs superfields
respectively, parameters $\mu$ and $A$ are complex and flavour
matrices $\xi$ are assumed to be real, $m_g$ is the gravitino mass.
In
this model extra CP-violating phases appear in the vertices
containing superpartners of ordinary particles.

\subsection{EW baryogenesis: how to state the problem}
Switching off  the sphaleron transitions in the broken phase
is, clearly, not enough for the asymmetry production. According to
the third Sakharov condition, baryogenesis requires deviations from
thermal equilibrium in reactions that break CP. This is provided by
the first order nature of the phase transition, which proceeds
through
the bubble nucleation. Before the bubbles percolate, the largest
deviations from thermal equilibrium (e.g. in the particle number
densities) are at the fronts of the shock waves of the deflagration
bubbles and near bubble walls. At the time of percolation,
deviations from
thermal equilibrium arise because of collisions of the domain walls
and shock fronts. The latter effect cannot give  substantial
contribution to the baryon asymmetry since it is proportional to the
fraction of the Universe volume occupied by domain walls,
\be
\frac{\xi}{R_b}\sim 10^{-10},
\ee
where $\xi$ is the thickness of a domain wall (scalar correlation
length), $R_b$ is the typical bubble size; for numerical estimates
of $\xi$ and $R_b$ see section 5.
The shock fronts propagate in the symmetric phase, where the rate of
fermion number non-conservation is higher than the rate of the
Universe expansion, or in the broken phase, where B-nonconservation
is
switched off. So, baryogenesis cannot happen near shock
fronts, and the
only possibility is to associate it with domain walls.

For the most part of their life, bubbles of the broken phase are
macroscopic (their size is much larger than the typical correlation
length); the domain walls move with  constant velocity before
percolation; after it the Universe may or may not be reheated up to
the critical temperature, depending on parameters of the model. If it
does, the velocity of the bubble walls drops considerably down to
values $v\sim 10^{-2}$-$10^{-3}$, and then slowly varies depending
on the bubble size. Therefore, the picture of a planar domain wall,
``eating up'' the symmetric phase with some velocity $v$, is a good
approximation to the problem.

How should the solution of the baryogenesis problem look like? In
very general terms, the answer is: Write down the kinetic equations
accounting for all relevant processes, supply them with appropriate
boundary conditions (equilibrium in the symmetric phase far from the
domain wall) and then determine the baryon number deeply in the
broken
phase by solving  these equations. Besides the two obvious types of
processes (1. $B$ non-conservation, which
is rapid in the symmetric phase and slow in the broken one;  2.
CP-violating interactions of various particles with the domain wall
and
with each other), a number of other reactions should be taken into
account.

The first group of phenomena deals with $B$ and $L$ conserving
processes. 1. Ordinary strong and weak interactions tend to make
momentum dependence of distribution functions for quarks, leptons,
gauge bosons and Higgses to be an equilibrium one. These processes
govern the diffusion of the CP asymmetries in fermion number, created
in the vicinity of the domain walls. 2. Chirality flip interactions
of quarks and leptons, coming from interactions with Higgses and from
strong sphalerons. These reactions tend to make the concentrations of
left-handed and right-handed particles  equal to each other.
Since anomalous $B$ and $L$
non-conservation deals with left-handed
fermions, the left-right transitions
influence the $B$-violating reactions. 3. Debye screening of the long
range gauge forces, which tend to damp any non-trivial distribution
of the dynamical charges, such as hypercharge \cite{khl:Y} (see
also \cite{ckn:1992}).

The second group of phenomena deals with the description of the
relevant degrees of freedom at high temperatures: 1. High temperature
physical excitations are different from those at zero temperatures.
Therefore, the corresponding kinetic equation should deal with
quasi-particles rather than particles \cite{s:msm,Farrar1,Farrar2}.
2. Simultaneous
interaction of quasiparticles with the heat bath and varying scalar
field results in the mixing, analogous to that of the neutrino in
matter
\cite{Wolf78,Mikh85}. So, quasi-particles
are to be described by the density
matrices rather than the particle number
distributions \cite{Dolg81,Stod87,Barb91,Sigl93}. 3. Quasi-particles
in the plasma have  finite lifetime, i.e. they should be
characterized by an energy and momentum simultaneously. In general,
the kinetic equation should be able to account for this.

To our knowledge, the complete programme outlined above has never
been carried out. The main difficulty is the construction of the
kinetic equation incorporating all necessary features. Some of the
effects mentioned above were taken into account, but the complete
picture is still missing. So, we consider, at the qualitative level,
various ideas and estimates of the baryon asymmetry produced at
the EW phase transition.

\subsection{Uniform scalar fields}
A good theoretical laboratory, allowing an understanding of physical
processes giving rise to the charge asymmetry, is the consideration
of the uniform but time dependent scalar fields. Probably, this
situation is never realized, but this case is much simpler than that
of the bubble wall propagation.

Suppose that we have a kind of spinodial decomposition phase
transition, in which case the scalar field is initially near $\phi=0$
and
the system is in the symmetric phase. Sphaleron processes are in
thermal equilibrium. Then the scalar field uniformly rolls down to
the
true vacuum, where the $SU(2)\times U(1)$ symmetry is broken and
sphaleron
processes are suppressed. The first rough estimates of the baryon
asymmetry in this case were given in ref. \cite{s:sm88}, and a
lot of work on
this subject has been done in refs.
\cite{McL89,Tur90,Tur91,Mc91,dhss,ckn:1991} and many others,
for relatively recent reviews see refs. \cite{turrev,ckn:1993} and
references therein.

We will consider the main idea on the example of the two Higgs
doublet
model.
Our scalar fields $\varphi_1$ and $\varphi_2$ are uniform in space
but
change from $\varphi=0$ to $\varphi=\varphi_c$ during
time $\Delta t$ of the
spinodial decomposition phase transition. Suppose that this time is
small enough, $\Delta t/\tau_{top}\ll 1$ where $\tau_{top}$ is a
typical
time of top quark chirality flip (the top quark is most important
since it has the largest Yukawa coupling constant). Then the top
quark distribution has no time to adjust itself to the changing
scalar
field. So, it may be integrated out  with the use of the equilibrium
Matsubara technique. This was carried out in ref. \cite{Mc91} with
the
result that the effective action has the following form:
\be
S_P = \mu N_{CS},
\label{pbreak}
\ee
where
\be
 \mu= - i \frac{7}{4} \zeta (3) \left(\frac{m_t}{\pi T}\right)^2
\frac{2}{v_1^2}{\cal O}(\varphi_1),~~
{\cal O}(\varphi_1)=(\varphi_1^{\dagger}{\cal D}_0 \varphi_1 - ({\cal
D}_0
\varphi_1)^
{\dagger}\varphi_1),
\ee
and $m_t$ is the mass of the top quark, $\zeta$ is the
Rieman $\zeta$-function.

The effective bosonic action now breaks P and CP simultaneously, with
CP violation in the scalar field potential, and P violation in the
term (\ref{pbreak}). This allows us to generate the non-zero value of
the
topological charge $Q$, which is P- and CP-odd \footnote{The presence
of fermions is essential here. The purely bosonic tree action
conserves P, and the net topological charge cannot be produced.}.
Note also that in more complicated models the operator $\mu$ may appear
to be P- and CP-even (e.g., 
$\mu \sim \partial_0 (\varphi^{\dagger}\varphi)$). In the latter case
the effective action (\ref{pbreak}) itself breaks P and CP
simultaneously. Estimates of the baryon asymmetry produced in this case
were made in refs. \cite{s:sm88,dhss}.

The zero-temperature bosonic effective action of this model also
contains parity odd term $~\theta(x)q(x)$, where $\theta(x)$ is the
relative phase of the scalar fields. This important fact was
discovered by Turok and Zadrozny and applied to baryogenesis in ref.
\cite{Tur90}.

If $\Delta t$ is not too small, $\Delta t\cdot m(T)\gg 1$, where
$m(T)$ is
the typical mass scale at high temperatures, then the term
(\ref{pbreak}) may be considered as the chemical potential for the
Chern--Simons number and the number density of fermions created
during
the transition is
\be
n_B = n_f \int_0^{\infty} dt \Gamma_{sph}(t) \mu(t),
\label{bau}
\ee
where $\Gamma_{sph}$ is the time-dependent rate of the sphaleron
transitions, and $n_f = 3$ is the number of fermion generations.
The sphaleron rate $\Gamma_{sph}$ rapidly decreases when the mass of
the vector boson increases.
A natural way to estimate it in the entire range of $W$-boson
masses is to
use eq. (\ref{rate}) for  some $M_W > M_{crit}$ and (\ref{rate2})
for the opposite case where $M_{crit}\simeq 7 \alpha_W T$
is found\footnote{A
factor $2$-$3$ instead of $7$ was found in ref. \cite{scott94} from
other
considerations. Clearly, these estimates are qualitative rather than
quantitative.}  from  the relation $\Gamma_{br} = \Gamma_{sym}$.
In this approximation,
\be
n_B \simeq n_f (\alpha_W T)^4 \mu(t^*),
\ee
where $\mu(t^*)$ is the chemical potential at the ``freezing'' time
$m_W(t^*) = M_{crit}$. The asymmetry was estimated
in ref. \cite{Mc91} and lately
corrected in the detailed analysis of ref. \cite{Kaz92} for the
spinodial
decomposition phase transition. The asymmetry reads
\be
\Delta \sim \frac{45}{2\pi^2 N_{eff}}\kappa n_f \alpha_W^6 \sin^3
2\alpha \lambda_{CP} \frac{m_t^2T_c^2}{v_1^3v_2},
\label{2doub}
\ee
with $N_{eff}$ being the number of effectively massless degrees of
freedom, $\lambda_{CP} = (\lambda_5 - \lambda_6) \sin 2\xi_0$, and
\be
\tan \alpha =
\frac{m_1^2(T_c)}{m_2^2(T_c)},
\label{theta}
\ee
with $m_i(T_c)$ being the temperature-dependent scalar masses
at the moment of the phase transition (see, e.g.
ref. \cite{bks:higgs}). An analogous dependence on the coupling
constants was
found in ref. \cite{scott94}. In spite of the rather high power
of the coupling
constants, this estimate can give an asymmetry consistent with
observations \footnote{One can obtain
a similar estimate for the asymmetry
from  different consideration \cite{s:sm88,Tur90,Tur91}
dealing with non-perturbative
fluctuations of the Chern-Simons number in the symmetric phase
\cite{Ambjo9}.}.

This consideration can be easily generalized to  more complicated
models. First, one calculates an effective bosonic action, which
breaks
P and CP and defines an effective potential for the CS number. Then,
an estimate of the net  production of fermions is given by
(\ref{bau}).

An essential assumption in the estimates presented above is that the
time of the phase transition is shorter than that of kinetic
reactions.
An ``exact'' solution to the problem in the opposite case for quite a
specific situation has been suggested in ref. \cite{ckn:1991}.
Since this example is  instructive,  we reproduce here
the main idea of this paper, using correct coefficients
from  ref. \cite{Giudice:1994}.

Let us take again the two Higgs doublet model. For simplicity, we set
all Yukawa couplings, except for that of the $t$-quark, to zero, i.e.
the Yukawa interaction is assumed to be
\be
L_Y = f_t \bar{Q_3}U_3 \varphi_1.
\label{yukawa}
\ee
Here $Q_i$ are the left-handed fermion doublets, $U_i,~D_i$
are right-handed
quark fields, and $i$ is the generation index.
We also neglect the reactions with quark chirality flip associated
with strong sphalerons.

We put $\lambda_5=\lambda_6=0$ in eq. (\ref{pot}); with these
couplings
the potential has an extra global U(1) symmetry, which is
spontaneously
broken. Let us consider this model in an extermal Higgs background of
a
special form, namely $\theta = \dot{\theta} t$  at $t>0$ and $\theta
=0$ at $t<0$, where $\theta$ is the Goldstone mode,
\be
\tan \theta = \frac{\mbox{Im}(\varphi_{1}^{\dagger}\varphi_{2})}
{\mbox{Re}(\varphi_{1}^{\dagger}\varphi_{2})}.
\ee
 Suppose that at $t<0$ the system was in thermal equilibrium and was
charge
symmetric. We want to determine the baryon number of the system at $t
\rightarrow \infty$. The density matrix $\rho(t)$ of the system obeys
the Liouville equation
\be
i\frac{\partial\rho(t)}{dt}=[H(t),\rho(t)],
\ee
where $H(t)$ is the time-dependent Hamiltonian of the system in the
background field. Now, one can make an anomaly free
hypercharge rotation of the fermion fields in such a way that the
time dependence disappears from the Yukawa coupling (\ref{yukawa}).
Because of the global U(1) symmetry this converts the time dependent
Hamiltonian to a time independent
one, $H(t) \rightarrow H_{eff}=H - \dot{\theta}Y_F$ where $Y_F$ is
the fermionic hypercharge operator,
\be
Y_F = \sum_{i=1}^3\left[ \frac{1}{3}\bar{Q_i}\gamma_{0}Q_i
+ \frac{4}{3}\bar{U_i}\gamma_{0}U_i
- \frac{2}{3}\bar{D_i}\gamma_{0}D_i
- \bar{L_i}\gamma_{0}L_i
- 2 \bar{E_i}\gamma_{0}E_i\right] .
\ee
At $t\rightarrow \infty$ the system must be in thermal equilibrium,
$\frac{\partial\rho(t)}{dt}=0$. Since in the new representation the
Hamiltonian is time independent, the density matrix is
\be
\rho(\infty)  = \frac{1}{Z}\exp{\left[ - \frac{1}{T}(H_{eff} -
\mu_i X_i)\right] },
\label{density}
\ee
where $X_i$ is a  {\em complete} set of conserved charges (operators
commuting with the Hamiltonian). Their average must be equal to zero.
This requirement fixes the chemical potentials $\mu_i$ and allows the
unambiguous determination of the baryon number of the system. A
complete
list of the conserved charges can be found in ref.
\cite{Giudice:1994},
and we quote
here the final result for the baryon number only,
\be
\langle B \rangle
=\frac{n_s}{6+11n_s}T^2\dot{\theta}(1+O(m_t^2/t^2)),
\label{bspont}
\ee
where $n_s=2$ is the number of the scalar doublets.

The result (\ref{bspont}) is quite amazing. It does not contain
the Yukawa coupling constant, the scalar vacuum expectation value, or
the
rate of sphaleron transitions. One might even think that it is wrong,
since if $f_t$ or $v$ or $\Gamma_{sph}$ is zero, then, obviously, one
must have $\langle B \rangle=0$. Nevertheless, it is correct. The key
point is that the time at which the asymptotic value of the baryon
number is reached tends to infinity when the above mentioned
quantities tend
to zero. Many conclusions based on the straightforward analysis of
the
perturbation theory break down at large times, when the application
of
the kinetic theory is essential, and this is one of the examples. For
 typical values of the parameters, the top quark chirality
equilibration
time is $\tau_t\sim 30/T$, and the $B$ non-conservation time is
$\tau_{sph} \sim 10^5/T$; the result (\ref{bspont}) is valid only for
$t>\tau_{sph}$. The discussion of intermediate time $\tau_t \ll t \ll
\tau_{sph}$ is contained in ref. \cite{ckn:1991,Giudice:1994}.

It is worth noting that high temperature
sphalerons, and other chirality flip reactions, may change the
estimate (\ref{bspont}). In particular, strong sphalerons, discussed
in
Section 4 have the physical effect of maintaining the same chemical
potential for left- and right-handed baryonic numbers and diminishing
the set of conserved quantum numbers in the system. This leads to the
suppression of the baryon number by a factor $\sim (\frac{m_t}{\pi
T})^2$ \cite{Giudice:1994}. Other aspects of influence of strong
sphalerons on baryon asymmetry were discussed in ref.
\cite{Mohapatra}.

We conclude this discussion by remarking that the use of U(1) global
symmetry was essential in the derivation of eq. (\ref{bspont}). Without
it the hypercharge rotation would not remove the time dependence from the
Hamiltonian, and the solution of the Liouville equation could
not be found so
straightforwardly. 
This is discussed in more detail in refs. \cite{scott94,Jo94}.

\subsection{Asymmetry from fermion--domain wall interactions}
In reality, though, the phase transition goes through the bubble
nucleation rather than as spinodial decomposition. This is an
additional challenge, since the baryon number (or, in general,
asymmetries in particle number densities) can now be distributed in a
non-uniform way and depend on the distance from the domain wall.
Correspondingly, the analysis of the kinetic equations is much more
complicated.

Two different cases are usually considered, depending on the
relation between the mean free paths of particles and domain wall
thickness. The physics of  thick wall baryogenesis
was originally considered in refs. \cite{Mc91,dhss} 
and has much in common with the quasi-adiabatic case of uniform fields 
discussed in the previous subsection. P or CP non-invariant 
interaction of fermions with the moving domain wall together with
CP-breaking scalar dynamics induces P- and CP-odd terms in the bosonic 
effective action, which bias the sphaleron transitions inside the domain
wall. The excess of quarks generated in this way is absorbed then by
the expanding bubble. (Another, equivalent way to say this 
\cite{ckn:1991} is that
the particle densities of fermions gradually adapt to scalar 
background changing in space and time in such a way that an excess of right 
quarks and left antiquarks is created. Left antiquarks are
destroyed by the sphaleron reactions while the right fermion number is
intact and is converted into baryon number at the end.)

A nice physical picture of thin domain wall baryogenesis was suggested
by Cohen, Kaplan and Nelson \cite{ckn:1990,ckn:1991a,ckn:1992a}. Since
the masses of fermions are different in the symmetric and broken
phases, they  scatter on domain walls (are reflected  or
transmitted). CP violation manifests itself in different 
reflection coefficients for
particles and antiparticles. So, the moving domain wall acts 
like a separator for
different types of fermion numbers, filling the bubble with fermions
and outer space with antifermions (or vice versa, depending on the
sign of CP violation). Of course, the interactions of fermions with
domain wall conserve the fermion number, i.e. the number of fermions
flying
into the broken phase is equal to the number of anti-fermions moving
into
the symmetric phase. Antifermions, injected into the symmetric phase,
participate in the anomalous reactions that change fermion number,
while fermions injected into the broken phase do not. As a result,
non-zero baryon and lepton asymmetries are established 
in the broken phase.

Clearly,  reliable calculations of the effect in realistic
theories are quite complicated because of the large number of
different particle species participating in interactions. Moreover, a
number of  effects, discussed above, should be taken into
account. A number of papers is devoted to the study of the
origin of the CP-violating fermion currents and their propagation in
front of domain walls \cite{Aya94,Farrar95,Funak94,Khleb1995};
the most recent (and probably most elaborate) treatment can be
found in  \cite{ckn:1994,Joyclass,Jo94,Joy94,Cline95}.

Below we discuss the qualitative features of the domain wall
baryogenesis. Our consideration is by no means complete, and the
reader may consult the original papers for  details.

At sufficiently small velocities of the domain walls we can divide
the problem in two parts \footnote{If the diffusion tails (see below)
are comparable with the thickness of the domain walls, this is not
possible.}. The first one is the microscopic calculation of
various fermion currents {\em at the} domain wall. The second one
is the consideration of the diffusion of the particle number
densities in
front of the wall and their dissipation in different processes.

We begin with the first part \cite{ckn:1990,ckn:1991a,ckn:1992a}.
Let us ignore for a moment any
high temperature effects. The simplest case is that of the thin
domain wall moving with some constant velocity $v$. Let us choose the
rest frame of the wall and consider scattering of fermions on it. For
example, left-handed fermions incident in the symmetric phase may be
reflected back to the symmetric phase as right-handed
fermions (because of
the spin conservation) or can be transmitted through.
The transmission and reflection coefficients $r_{ij}$ ($i$ is the
label
of an incident fermion and $j$ is that of the final state)
can be found from the Dirac equation:
\be
\left( \begin{array}{cc}
\omega+ i\frac{\partial}{\partial x}
&M\\
M^{\dagger}&\omega -
i\frac{\partial}{\partial x} \end{array} \right) \cdot
\left( \begin{array}{c}L\\ R \end{array}\right) =0
\label{dirac}
\ee
with appropriate boundary conditions\footnote{A method for 
high precision numerical evaluation of reflection coefficients
was considered in ref. \cite{Farrar2}.}. Here $L$ and $R$ correspond to
up and down components of two-dimensional Weyl spinors. The $x$
dependent matrix $M$ is complex, giving rise to CP-violation. In
general, $r_{ij}$ for particles are different from $\bar{r}_{ij}$ for
anti-particles. This leads to non-zero fermionic currents:
\be
\langle J_i \rangle =\int \frac{d\omega k_{||} dk_{||}}{(2\pi)^2}
(n_F(\omega_+)-n_F(\omega_-))\left[r^{\dagger}r-\bar{r}^{\dagger}
\bar{r}\right]_i ,
\label{ascurppar}
\ee
where  $n_F^i$ is the Fermi distribution for the incident
particles, $\omega_{\pm}=\omega \pm v p_t$, $p_t$ and $p_{||}$ are
the momenta of fermions tangential and parallel,
respectively, to the domain
wall. This expression vanishes if the domain wall is at rest ($v=0$)
or if there is no CP violation. The total baryonic current $J_{CP} =
J_L + J_R$ which results from the solution of the Dirac equation
(\ref{dirac}) vanishes, but the currents of left-handed ($J_L$) and
right-handed
($J_R$) fermions are non-zero.

The construction of the Dirac equation for quasiparticles, accounting
for leading high temperature effects, was done
in ref. \cite{Farrar1,Farrar2}, where
generalized expressions for the particle currents can be found. The
major qualitative effect  of high temperature corrections is that the
currents of left-handed and right-handed
fermions do not compensate each other and
the total baryon current is produced. Physically, this happens
because left-handed
particles participate in the weak interactions but right-handed
particles do not, and $J_{CP} \sim \alpha_W J_L$.

The thin wall description of the fermion scattering is applicable
only if the mean free path of fermions at high temperatures is much
larger than the domain wall thickness. This allows us to use
distribution functions of fermions
undisturbed
by the domain wall
and impose
ordinary boundary conditions for the scattering problem  at spatial
infinity. The thick wall case (the mean free path is small compared
with
the thickness) is much more complicated. Clearly, the scattering
description is not adequate in that case.
The physical phenomenon to be taken into
account is the modification of the particle distributions across the
domain wall.  A number of interesting effects arising in
the latter situation are discussed in ref.
\cite{Joyclass,Jo94,ckn:1994}.

The second problem is the particle transport.
Several approaches were applied to the consideration of it.
The first one is that of Monte Carlo simulations of the injected
flux of particles \cite{ckn:1992a}, the second  one is diffusion
equations
\cite{Farrar2,Jo94,Joy94,ckn:1994,Cline95,Joy95}.
Limitations of the diffusion approximation were considered
in ref. \cite{Clin94}. In the discussion below we closely
follow ref. \cite{Farrar2} where the analytical approximation to the
problem was constructed for a simple case.

Given the flavour and chirality structure of the fermionic currents,
the diffusion equations should be written for all particle species.
Fermions participate in many processes on both sides of the wall with
different time scales. In order to understand what the relevant time
scales are, let us consider the fate of a particle after it has been
reflected from the domain wall towards the unbroken phase. Roughly,
its
typical distance from the bubble wall is given by $\sqrt{D t} - vt$,
$D$ is the diffusion coefficient. The first term describes the random
walk of the particle in the rest frame of the plasma and the second
term describes the motion of the bubble wall. This particle will be
trapped by the bubble after the time interval $t_D \sim D/v^2$, so
that all processes with characteristic time $\tau < t_D$ must be
taken into account. The examples of the relevant processes include
$B$-violation, the elastic scattering of quarks and gluons, chirality
flipping transitions of heavy quarks, strong sphalerons.

In order to get a better feeling of the physics involved, consider
the
simplest case when the total baryonic current originated from
CP  non-invariant interactions is not zero and neglect all processes
besides the elastic scattering of fermions and anomalous $B$ and $L$
non-conservation.
Let us take a planar domain wall which moves through the plasma with
sufficiently small velocity $v$ (we shall see below how small it
should be in order
that this consideration  works). We take a reference frame
associated with the domain wall; let the
broken phase be to the right and the
symmetric one be to the left, and $x$ be the distance from the domain
wall. Assume that the thickness of the domain wall is small enough
(again, we shall see below what this means). We denote by $n_B (x,t)$
and $n_L(x,t)$ the densities of baryon and
lepton numbers in the rest frame of the wall. The diffusion
equations for the broken phase, where sphalerons do not operate, are
\be
\frac{\partial}{\partial t}\left( \begin{array}{c}n_B\\ n_L
\end{array}\right)=
\left( \begin{array}{cc}
 D_B \frac{\partial^2}{\partial x^2}- v \frac{\partial}{\partial x} &
0\\ 0&D_L \frac{\partial^2 }{\partial x^2}- v \frac{\partial
}{\partial x} \end{array}\right)
\left( \begin{array}{c}n_B\\ n_L  \end{array}\right).
\label{diffun}
\ee
For $x<0$ we have
\be
\frac{\partial}{\partial t}\left( \begin{array}{c}n_B\\ n_L
\end{array}\right)=
\left( \begin{array}{cc}
 D_B \frac{\partial^2}{\partial x^2}- v \frac{\partial}{\partial
x}-\frac{3}{2}\Gamma & -\Gamma\\ -\frac{3}{2}\Gamma&D_L
\frac{\partial^2
}{\partial x^2}-
v \frac{\partial }{\partial x}-\Gamma \end{array}\right)
\left( \begin{array}{c}n_B\\ n_L  \end{array}\right) ,
\label{diffbr}
\ee
where $\Gamma = 9 \Gamma_{sph}/T^3$, $D_B$ and $D_L$ are diffusion
constants for quarks and leptons respectively. An estimate of these
gives \cite{Jo94,Joy94} $D_B \sim \frac{6}{T}$, $D_L\sim
\frac{100}{T}$.

We are looking for a steady state (time independent) solution to
these equations. In the broken phase the only solution consistent
with the boundary conditions is constant density,
\be
n_B=n_L = \mbox{const}=B_+.
\ee
In the symmetric phase, the solution is a combination of dying
exponentials. We present it in two limiting cases. The first one
deals with ``large'' velocities,
\[
\rho = 3D_B \Gamma /v^2 \ll 1 .
\]
Physically, this corresponds to the situation when an extra
antibaryon,
injected into the symmetric phase, is exposed to the $B$-violating
reactions for  short times $t\sim D_B/v^2$, before it is trapped by
the
moving domain wall. One finds \cite{Farrar2}
\be
n_B=C_1 \exp \left(\frac{v x}{D_B}\right),
        ~~~ n_L=C_2 \exp \left(\frac{v x}{D_L}\right),
\ee
corresponding to the diffusion of quarks and leptons to the distances
$
D_B/v$ and $ D_B/v$ respectively. In the opposite case of low
velocity,
the
transitions from quarks to leptons due to sphaleron processes are
essential, and the solution reads
\be
n_B=C_3 \exp \left(\frac{3 v x}{5 D_L}\right)
     + C_4 \exp \left(\sqrt{\frac{5 \Gamma}{2
          D_B}}x\right),
\ee
\[
n_L=-\fr32 C_3 \exp \left(\frac{v x}{D_L}\right)
          +C_4 \frac{D_B}{D_L}
     \exp \left(\sqrt{\frac{5 \Gamma}{2 D_B}}x\right).
\]
One of the requirements for the validity of the diffusion
approximation
is that the diffusion tail (the shortest one is for the quarks) is
much longer than the domain wall thickness $l$, namely $l \ll D_B/v$.

The constants $C_1$--$C_4$ can be determined from the matching
conditions at the domain wall. If we denote by $J_{CP}$ the
total CP-odd baryonic current originating from interactions
of quarks with the
domain wall and assume that CP asymmetry in the leptonic current is
zero,
then
\be
B_+= \frac{12}{5}J_{CP}f_{sph}(\rho),
\ee
where  $f_{sph}(\rho) = 1$ for $\rho \gg 1$ and $f_{sph}(\rho) =
\frac{5}{6}\rho$ for $\rho \ll 1$.

The asymmetry inside the bubble has an interesting velocity
dependence. If $J_{CP}\sim v$, as in the quantum-mechanical
consideration of the thin wall case, then the maximum asymmetry is
produced at $\rho \sim 1$, i.e. $v \sim \sqrt{3 \Gamma D_B}\simeq
0.01$. It is worth noting
that these small velocities are quite possible at
the final stage of the phase transition if the Universe is reheated
up to the critical temperature. The analysis of the consequences of
this
scenario can be found in ref. \cite{Heck95}. For a more
realistic thick wall
case one effectively has  $J_{CP}\sim v^2$
\cite{Joyclass,Jo94,Joy94} \footnote{We thank
Michael Joyce for clarification of this point.}, and the asymmetry is
velocity independent. The same conclusion has been reached also in
ref. \cite{ckn:1994}.

In the realistic case of many particle species this consideration
must
be generalized. Instead of the CP-violating flavour independent
baryonic current considered above, many CP-odd currents appear,
resembling the flavour dependent interaction of fermions with the
domain
wall. The left- and right-handed currents must be distinguished,
since
particles of different chiralities have different interactions with
the
heat bath and sphalerons.
Quantitatively, the results are model dependent. For example, in some
schemes the lepton interactions with domain walls produce more
asymmetry than quark interactions \cite{Joy4}. Serious investigations
of
realistic models have been carried out in very interesting
papers \cite{ckn:1994,Joy94,Jo94,Cline95},
and we refer to them for more detail.

\subsection{Strength of CP violation and baryon asymmetry}
Extensions of the standard model, having strong enough first order
phase transition, may not contain any new source of CP violation.
The question arises whether the KM source of
CP violation (or QCD vacuum angle)
alone can be responsible for the baryon asymmetry. Let us begin with
the
KM mechanism of
CP violation\footnote{ Of course, this possibility is rather
unnatural: the only known way to strengthen the phase transition is
to
add extra scalar particles. This means that new scalar interactions
appear. In general, they contain complex phases and lead to CP
violation.}.

An important property of the interaction (\ref{Yukawa}) is that the
CP-violating phase $\delta_{CP}$ can be rotated away by phase
transformation of the fermion fields if there is a degeneracy in
the up or down quark sectors, or if some mixing angle between
different generations is zero. In other words, in the standard model,
CP violation vanishes together with the Jarlskog determinant
\cite{jarlskog},
\be
d_{CP} =
\sin(\theta_{12})\sin(\theta_{23})\sin(\theta_{13})\sin\delta_{CP}
\times
\ee
\[
(m_t^2 - m_c^2)(m_t^2 - m_u^2)(m_c^2 - m_u^2)
\cdot (m_b^2 -m_s^2)(m_b^2 - m_d^2)(m_s^2 - m_d^2),
\]
where $\theta_{ij}$ are the mixing angles and $m_i$ are the quark
masses.

The structure of the KM mechanism of
CP violation makes baryogenesis a very
non-trivial problem. Indeed, the electroweak phase transition, where
strong deviations from thermal equilibrium are expected, occurs at
temperatures of order $100$ GeV. It seems, therefore, that quark
masses
(maybe with the exception of the $t$-quark) can be treated as
perturbations, so that the dimensionless measure of CP violation is
just
$\delta_{CP} \sim d_{CP}/T^{12} \sim 10^{-20}$ \cite{s:m^14,s:sm87}.
Clearly, this number is too small to account for the observed
asymmetry. However, there may be loopholes in this argument and a
number of dynamical mechanisms, in which the KM source of
CP-violation may be enhanced, have been suggested
\cite{s:sm87,Farrar1,Farrar2,Nas94}.

If there is a dynamical spontaneous CP violation in the
electroweak theory
before \cite{s:sm87} or during \cite{Nas94,Khleb1995} the electroweak
phase transition, then
the Universe contains domains with different CP-parity at some stage
in its evolution. The small explicit CP violation breaks the
degeneracy between the different CP states, so that more
energetically favorable domains "eat" those with the opposite
CP parity. The fact that the age of the Universe at $T\sim 100$ GeV
is
macroscopic, gives rise to an enhancement factor $\sim M_{Pl}/T_*\sim
10^{16}$ \cite{s:sm87}. In this mechanism, the baryon asymmetry
does not depend on the
magnitude of CP violation but does depend on its sign
\cite{s:sm87,Nas94}, and a power-counting estimate of the effect
gives
\be
\Delta \sim \frac{1}N_{eff}\alpha_W^3.
\ee
We should stress, however, that the possibility of spontaneous
CP breaking at high temperatures in electroweak
theory is very speculative.

Another possible caveat in the no-go argument presented above has
been discussed in ref.  \cite{Farrar1,Farrar2}. The  ``no-go''
theorem relies on the
applicability of the perturbation theory in quark masses and
makes use of the
assumption that the typical energy scale relevant to the estimate
of the asymmetry is the temperature of the phase transition. In fact,
these assumptions break down when the interaction of fermions with
domain walls is considered. Namely, if the energy of the quark in the
unbroken phase is smaller than its mass in the broken phase, it will
be reflected from the domain wall with unit probability independently
of the value of its mass. The perturbation theory does not work only
in a small fraction of the phase space determined by the quark
masses, but the loss in the phase space factor may be smaller
than the gain in CP-violating amplitude. The maximum asymmetry comes
from the strange quark complete reflection: the GIM cancellation
does not occur for it,
and what is left over from $d_{CP}$ is just the product of
mixing angles and CP-violating phase\footnote{In refs.
\cite{Gav1,Gav2,Gav3,Huet95} an opposite conclusion was
reached. The calculational procedure of these works was criticized in
ref. \cite{Farr94,Far94}, where it was argued that the claims of
ref. \cite{Gav1,Gav2,Gav3,Huet95} are
not justified.}
of order $10^{-5}$. The estimates of
the asymmetry presented in ref. \cite{Farrar2} are rather
uncertain, $\Delta \sim
10^{-10}$--$10^{-18}$, but they indicate that the KM mechanism
of CP violation
cannot be discounted as a source of the baryon asymmetry of the
Universe.

The possible impact of the strong CP violation (associated with QCD
vacuum angle $\theta$) on the electroweak
baryogenesis has been analysed in
ref. \cite{kst}. It was concluded that it does not play any
significant
role
because strong CP effects at the
electroweak phase transition temperature are
suppressed by at least semi-classical exponent
$\exp(-\frac{2\pi}{\alpha_s})\sim 10^{-25}$, yet other suppression
factors come from Yukawa couplings. The same statement holds true in
the models with axions, where effective angle $\theta$ may be of the
order of $1$ at the electroweak scale.

While convincing arguments showing that the standard model CP
violation is sufficient to generate the baryon asymmetry
are absent, the extensions of
the electroweak theory naturally provide  new sources of CP
violation.
In spite of the fact that the existing estimates of the baryon
asymmetry are valid, probably, within an order of magnitude, it is
clear that extended versions of the electroweak theory can
accommodate the observed baryon asymmetry of the Universe.
The specific estimates for the two Higgs doublet model
can be found in ref. \cite{ckn:1994,Jo94,Joy94,Cline95}
and for the sypersymmetric theories in
ref. \cite{Cohen92,comel,Ab93,Huet:1995}.
\section{Instanton-like processes in high energy collisions.}

As discussed in section 2, instanton-like transitions may occur at
unsuppressed rates at sufficiently high energies. This possibility is
definitely realized at high temperatures. It is natural to ask
whether
collisions of highly energetic particles may lead to baryon and
lepton
number non-conservation with exponentially unsuppressed cross
sections.
The relevant energy scale, $E_{sph}\sim m_{W}/\alpha_{W}$ (10 TeV in
the
electroweak theory), is not too far from collider energies, so this
problem
is not  of academic interest only.

In this section we outline the current status of this very
complicated
problem. We summarize the results of perturbative calculations about
the
instanton, which show that the cross section indeed increases
exponentially
with energy at $E\ll E_{sph}$. However, the perturbation theory about
the
instanton is unreliable in the most interesting energy region
$E\sim E_{sph}$, so the perturbative calculations cannot tell whether
the
instanton factor (\ref{I6*}) is overcome. We present below a rather
general
argument based on unitarity and conventional perturbation theory,
which
shows that the instanton-like transition rates are most likely
exponentially small at all energies. This argument, however, does not
exclude the possibility that the suppression disappears
asymptotically at
$E\to\infty$, and that the actual values of the suppression factor
are not
too small at realistic values of coupling constants and energies. So,
the
computation of the instanton-like transition probabilities remains an
interesting problem whose solution requires non-perturbative
approaches.
We outline in this section
one of these approaches and first non-perturbative results;
these results indicate that the instanton-like cross sections are
indeed
unobservably small at all energies.

In spite of the considerable progress in understanding the
instanton-like
processes in high energy collisions, this problem is still not solved
completely.
The non-perturbative techniques adequate to this problem are still
being
developed. It is worth pointing out that electroweak $B$ and $L$
non-conservation in high energy collisions belongs to a wider class
of
processes, which includes false vacuum decay induced by
particle collisions,
induced decays of metastable solitons \cite{FarhiSolitons}, and,
notably,
production of multiparticle final states in the trivial vacuum sector
(without instantons)
\cite{Cornwall1,Goldberg,VoloshinPHI4,VolPhi4,Brown}. We
shall not discuss the latter problem, which is also of potential
phenomenological interest, and refer to a review
\cite{VoloshinReview}. However, we stress that its solution also
requires
novel non-perturbative techniques which should have much in common
with the
approaches relevant to instanton-like processes.

\subsection{Summary of perturbative analysis about the instanton}

The perturbative analysis about the instanton was started in refs.
\cite{Ringwald:1990,Espinosa:1989} and
has lead to the picture of exponentially
increasing total instanton-like cross sections at relatively low
center-of-mass energies,
$E\ll E_{sph}$. It also suggested the functional form
of the
cross section which indicated that the cross section may be
calculable in a
semiclassical way \cite{KRTFormalism,YaffeHoly,ArnoldMattisHoly}. We
do
not consider the technical details of the perturbative calculations
which
are reviewed in refs.
\cite{MattisReview,TinyakovReview,KonishiReview}
and
present only basic ideas and main results.

\begin{figure}[p]
  \begin{center}
    \mbox{\epsffile{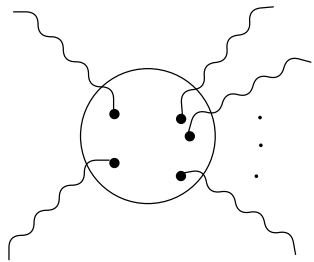}}
  \end{center}
  \caption{Leading order contribution to the instanton-induced
amplitude.}
\label{fig?}
\end{figure}

Consider a process where two $W$-bosons scatter into $n$ $W$-bosons,
and
the system simultaneously makes the transition from one vacuum of
fig.1 to
a neighbouring vacuum. The topological number $N[A]$ of the relevant
field
configurations, given by eq. (\ref{I2*}),
should be equal to $1$ (in fact, in
the case of a finite number of incoming and outgoing particles it is
more
appropriate to measure the change in the vacuum number $n[\omega]$ by
the
winding number of the Higgs field \cite{FarhiTransitions}). Let us
disregard fermions --- they play a minor role as
far as the cross sections are
concerned \cite{EspinosaFermions} --- and consider the bosonic sector
of
$SU(2)$ theory with one Higgs doublet, i.e. the bosonic sector of the
simplified model introduced in section 2.  To evaluate the amplitude
of this
process we begin with the $(2+n)$-point Euclidean Green's function
\begin{equation}
       G_{n+2}(x_{1},x_{2},y_{1},\dots,y_{n}) =
       \int~DA~ \e^{-S[A]} A(x_{1})A(x_{2})A(y_{1})\dots A(y_{n})
\label{E5+}
\end{equation}
where  spatial and group
indices are omitted. Since the instanton is a minimum of the
Euclidean action (in fact, we have to deal with constrained
instantons
considered in section 2), we make use of the semiclassical
approximation and
obtain in the leading order
\begin{equation}
       G^{leading}_{n+2}(x_{1},\dots,y_{n}) =
       \int~d^{4}x_{0}~\frac{d\rho}{\rho^{5}}~\mu(\rho)
       \e^{-\frac{8\pi^{2}}{g^{2}} - \pi^{2}\rho^{2}v^{2}}
       A^{inst}(x_{1}-x_{0};\rho)\dots A^{inst}(y_{n}-x_{0};\rho).
\label{E5*}
\end{equation}
Here $x_{0}$ and $\rho$ are the usual collective coordinates of the
instanton and the measure is the same as in eq. (\ref{I7*}).
The constrained
instanton configuration is described by eqs. (\ref{I7a*}) and
(\ref{I7**}).
One should also integrate over the instanton orientations; this
integral is
not explicitly written in eq. (\ref{E5*}). The Green function
(\ref{E5*})
may be picturized as shown in fig.\ref{fig?}.
It is worth pointing out that the dependence on coordinates in
eq. (\ref{E5*}) factorizes, up to the integration over the instanton
position ensuring the overall momentum conservation. This means that,
in
the leading semiclassical order, the Green function has a point-like
structure. So, in this order, the cross section of the process
$2W \to nW$ will exhibit a power law increase with energy.

To obtain the cross section, one performs the analytical continuation
into
Minkowski space-time and then makes use of the LSZ procedure.  This
can be
done easily, as the only dependence on coordinates in eq.(\ref{E5*})
is
through $A^{inst}$. Since the constrained instanton field decays in
Euclidean space-time according to eq.(\ref{I7**}), its Fourier
transform
indeed has a pole at $p^{2} = - m_{W}^{2}$. The residue at this pole
can in
fact be obtained directly from eq. (\ref{I7a*}): the Higgs mechanism
shifts
the pole of the instanton field from $p^{2}= 0$ to $p^{2} = -
m_{W}^{2}$
but does not change the residue, up to small corrections. In this way
one
obtains the residue
\begin{equation}
       R({\bf p}; \rho) =
		\frac{1}{g}\rho^{2} |{\bf p}|
\label{E6*}
\end{equation}
where we omitted the tensor structure depending on the instanton
orientation. Thus, the amplitude has the following form,
\[
 A^{leading}_{2 \to n}
 ({\bf k}_{1}, {\bf k}_{2}, {\bf p}_{1},\dots, {\bf p}_{n})
 \sim
\]
\begin{equation}
        \int~\frac{d\rho}{\rho^{5}}~\mu(\rho)
       \e^{-\frac{8\pi^{2}}{g^{2}} - \pi^{2}\rho^{2}v^{2}}
       R({\bf k}_{1};\rho)\dots R({\bf p}_{n};\rho)
       \delta(k_{1} + k_{2} - p_{1} - \dots - p_{n})
\label{E7a*}
\end{equation}
where $k_{1}$, $k_{2}$
and
$p_{1},\dots, p_{n}$
are the momenta of incoming and outgoing particles, respectively.
The integration over $\rho$ is straightforward, while the integration
over
orientations, implicit in eq. (\ref{E7a*}),
is quite complicated. Ignoring the
latter complication we obtain the following estimate for the
amplitude,
\begin{equation}
 A^{leading}_{2 \to n} \sim
       \e^{-\frac{8\pi^{2}}{g^{2}}}
       (n+2)! \left(\frac{1}{gv^{2}}\right)^{n+2}
       |{\bf k}_{1}|
       |{\bf k}_{2}|
       |{\bf p}_{1}| \dots
       |{\bf p}_{n}|
       \delta(K - p_{1} - \dots - p_{n})
\label{E7*}
\end{equation}
where $K=(E,0)$ is the total centre-of-mass  four-momentum. Equation
(\ref{E7*}) leads to the following estimate for the $2W \to nW$
instanton-like cross section at $n\gg 1$,
\[
 \sigma^{leading}_{2\to n}(E) =
	\frac{1}{(k_{1}\cdot k_{2})} \frac{1}{n!}
	\int~\prod_{i=1}^{n}
	\frac{d^{3}p_{i}}{2\omega_{p_{i}}(2\pi)^{3}}
	|A^{leading}_{2 \to n}|^{2}
\]
\[
    \sim \frac{1}{n!} \left(\frac{\const \cdot
E^{2}}{gv^{2}n}\right)^{2n}
     \e^{-\frac{16\pi^{2}}{g^{2}}}
\]
where we assumed for simplicity that the outgoing particles are
relativistic, i.e. $E/n \gg m_{W}$. As expected, the cross section
exhibits the power law growth with
energy \cite{Ringwald:1990,Espinosa:1989}.
The {\em total} cross section, in the leading semiclassical order,
grows
exponentially,
\[
 \sigma^{leading}_{2\to any}(E) =
   \sum_{n}
 \sigma^{leading}_{2\to n}(E)
\]
\begin{equation}
     \propto
     \exp\left[-\frac{16\pi^{2}}{g^{2}} +
     \const
\left(\frac{E^{4}}{g^{2}v^{4}}\right)^{\frac{1}{3}}\right]
\label{E8*}
\end{equation}
where the number of particles at which this sum is saturated is of
order
\begin{equation}
	n \sim
      \left(\frac{E^{4}}{g^{2}v^{4}}\right)^{\frac{1}{3}}
\label{E8**}
\end{equation}
 The exponential growth of the cross section was found first
\cite{McLerranVV} in the context of multi--Higgs final states, which
have turned out to  be subdominant at relatively low energies, where
the
leading order calculations are reliable. This type of behaviour is
inherent
in all models with instantons \cite{TinyakovReview} and also for the
processes of multiparticle production without instantons
\cite{Libanov1,Libanov2,VoloshinReview}.

In the case of $2W \to nW$ processes, there exist at least two ways
of
actually calculating the cross section including
the constant in eq. (\ref{E8*}).
One is to make use of
instanton--anti-instanton configurations \cite{ZakharovLO,Porratti}
and another is based on the coherent state formalism
\cite{KRTFormalism}.
Both techniques lead to the same result:
\begin{equation}
    \sigma^{leading}_{2\to any}(E) \propto
    \exp \left[ \frac{4\pi}{\alpha_{W}}\left( -1 +
    \frac{9}{8}
\left(\frac{E}{E_{0}}\right)^{\frac{4}{3}}\right)\right],
\label{E9*}
\end{equation}
where $E_{0} = \sqrt{6}\pi m_{W}/\alpha_{W} \sim 15$ TeV is of order
of the
sphaleron energy in the electroweak theory. In writing eq.
(\ref{E9*}) we
made use of the  fact that
\[
      \frac{1}{\alpha_{W} E_{0}^{\frac{4}{3}}}
      \sim
      \left(\frac{1}{g^{2}v^{4}}\right)^{\frac{1}{3}}.
\]
For the same reason, the number of $W$-bosons produced can be written
as
follows,
\begin{equation}
	   n \sim
\frac{1}{\alpha_{W}}\left(\frac{E}{E_{0}}\right)^{\frac{4}{3}}.
\label{E9+}
\end{equation}
Note that the average energy per outgoing particle is of order
\begin{equation}
  |{\bf p}| \sim \frac{E}{n} \sim
	    m_{W}\left(\frac{E}{E_{0}}\right)^{-\frac{1}{3}}.
\label{E9a*}
\end{equation}
Therefore, at $E\ll E_{0}$ the final particles are relativistic,
while
at $E\sim E_{0}$ they are soft, $E/n \sim m_{W}$.
Note also that the typical instanton size is of order
\begin{equation}
 \rho \sim \frac{\sqrt{n}}{E} \sim
	    \frac{1}{m_{W}}\left(\frac{E}{E_{0}}\right)^{\frac{2}{3}}.
\label{E9a+}
\end{equation}
This estimate follows from eqs. (\ref{E5*}), (\ref{E6*}) and
(\ref{E9+}).

We conclude that the {\em leading order} total cross section becomes
unsuppressed exponentially at $E\sim E_{0}$, and at these energies
the
number of final particles becomes of order $1/\alpha_{W}$.

\begin{figure}[p]
  \begin{center}
    \mbox{\epsffile{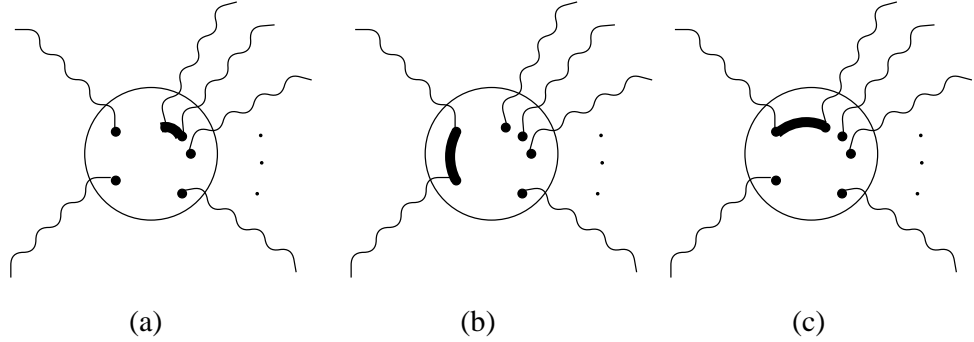}}
  \end{center}
  \caption{First corrections to the amplitude
    $2W \to nW$. Dots represent the residues of the instanton field,
solid
    lines correspond to residues of the propagator, in the instanton
    background, on the mass shell for both momenta.
    \label{fig??}}
\end{figure}

Clearly, the actual cross section of the instanton induced process
$2W \to nW$ should not be described
at all energies by the leading order result
(\ref{E9*}): at $E \gsim E_{0}$ this expression contradicts
unitarity.
So, corrections to the leading order formula must be large at least
at
$E \gsim E_{0}$. These corrections appear when the gauge field in
eq. (\ref{E5+}) is written as
\[
     A = A^{inst} + \delta A
\]
and $\delta A$ is treated as quantum field. In the next-to-leading
order,
the action in eq. (\ref{E5+}) is quadratic in $\delta A$, $n$ fields
in the
integrand remain $A^{inst}$ and  two fields are $\delta A$. Upon
integration over $\delta A$ one obtains the first correction to the
Green
function, which is similar to eq. (\ref{E5*}) but with two of the
instanton
fields substituted by the propagator in the instanton background,
$D_{inst}(z-x_{0},z'-x_{0})$ where $z$, $z'$ are any two of the
coordinates
$x_{1},x_{2},y_{1},\dots,y_{n}$. The first correction to the
amplitude is
then determined by the residue of the Fourier transform
$D_{inst}(q,q')$ at
the double pole $q^{2}=-m_{W}^{2}$, $q'^{2}=-m_{W}^{2}$. The
diagrammatic
representation of this correction is shown in fig. \ref{fig??}.

There are basically  three types of corrections, which we discuss in
turn.

i) {\em Soft--soft corrections.} These appear, in the
next-to-leading order,
when two instanton fields in eq. (\ref{E5*}) corresponding to {\em
final}
particles are substituted by the propagator, as shown in fig.
\ref{fig??}a. As
compared with the leading term (\ref{E7a*}) this contribution is
suppressed
by $g^{2}$ because $A^{inst}$ is proportional to $1/g$ while
$D_{inst}$ is
$O(g^{0})$, but is enhanced by the combinatorial factor $n^{2}/2$,
the
number of legs in fig.\ref{fig??}a that can be joined.
The residue of the propagator
at the double pole at low momenta is of order
\cite{ArnoldMattisCorr,DiakonovPetrovCorr,MuellerCorr}
\[
   {\mbox{Res}} D_{inst}(p_{i},p_{j})
	      \sim \rho^{2}
	      \sim
	      \frac{g^{2}}{\rho^{2} {\bf p}^{2}}
	      R({\bf p}_{i};\rho)
	      R({\bf p}_{j};\rho)
\]
where $R$ is the residue of the instanton field, eq.(\ref{E6*}).
Combining
all factors and recalling eqs. (\ref{E9+}), (\ref{E9a*}) and
(\ref{E9a+})
one finds that the soft--soft correction to the amplitude
at relevant $n$ is
of order
\begin{equation}
	A^{soft-soft} \sim
	      A^{leading}\cdot \frac{g^{2}n^{2}}{\rho^{2}{\bf p}^{2}}
	      \sim
	      A^{leading}\cdot \frac{1}{\alpha_{W}}
	      \left(\frac{E}{E_{0}}\right)^{2}.
\label{E13+}
\end{equation}
We see that this correction exceeds the leading order amplitude even
at
$E\ll E_{0}$. However, it has been shown
\cite{KRTFormalism,YaffeHoly,ArnoldMattisHoly},
that the soft--soft
corrections to the total cross section exponentiate, so that the
total
cross section with these corrections included has the form
\begin{equation}
    \sigma^{leading~+~soft-soft}_{2\to any} \propto
              \exp\left[\frac{4\pi}{\alpha_{W}}\left( -1 +
        \frac{9}{8}\left(\frac{E}{E_{0}}\right)^{\frac{4}{3}}
       - \frac{9}{16}\left(\frac{E}{E_{0}}\right)^{2}\right)\right]
\label{E13*}
\end{equation}
where we inserted numerical a coefficient 9/16 calculated in
refs.\cite{KhoseRCorr,ArnoldMattisCorr,DiakonovPetrovCorr,MuellerCorr}

\begin{figure}[p]
  \begin{center}
  \mbox{\epsffile{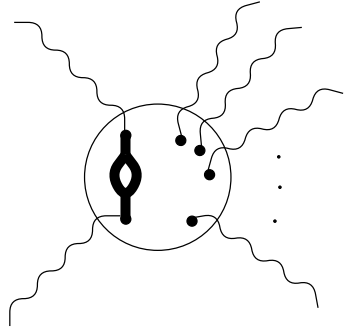}}
  \end{center}
  \caption{Higher order hard-hard correction.
   \label{fig3?}}
\end{figure}

Higher order soft--soft corrections have also been shown to
exponentiate
so the total cross section with all soft--soft
corrections included has the  following functional form
\cite{KRTFormalism,YaffeHoly,ArnoldMattisHoly},
\begin{equation}
    \sigma^{leading~+~soft-soft}_{2 \to any} \propto
      \exp\left[\frac{4\pi}{\alpha_{W}}
      F\left(\frac{E}{E_0}\right)\right]
\label{E14*}
\end{equation}
where $F(E/E_{0})$ is an unknown function which, at small $E/E_{0}$,
is
represented by a series whose first terms are given by eq.
(\ref{E13*}).
It has been found \cite{KRTFormalism} that the soft--soft
contributions into
the exponent $F(E/E_{0})$ come from {\em tree} diagrams about the
instanton, while loops contribute to the pre-exponential factor only.

ii) {\em Hard--hard corrections} \cite{Mueller2,Mueller3}. They are
due to
diagrams of fig. \ref{fig??}b.
There is no combinatorial enhancement of these
diagrams, but they  produce
large contributions nevertheless. The reason is
that the residue of the propagator is large
\footnote{This property holds for a wide class of models
\cite{VoloshinPropagator}.} at high momenta $|{\bf k}| \sim E$,
\[
{\mbox{Res}} D_{inst} \sim
         g^{2}\rho^{2} (k_{1}\cdot k_{2})
	 \ln(k_{1}\cdot k_{2}) R({\bf k}_{1};\rho) R({\bf k}_{2};\rho).
\]
Therefore, the first hard--hard correction into the amplitude at
relevant
$n$ is of order
\begin{equation}
   A^{hard-hard} \sim
   A^{leading} \cdot \frac{1}{m_{W}^{2}}
   \left(\frac{E}{E_{0}}\right)^{\frac{4}{3}} E^{2} \ln E^{2}
   \sim
   A^{leading} \cdot \frac{1}{\alpha_{W}}
   \left(\frac{E}{E_{0}}\right)^{\frac{10}{3}} \ln E^{2},
\label{E15*}
\end{equation}
which is again large even at $E\ll E_{0}$. Higher order hard--hard
corrections come from {\em loop} diagrams like those shown in fig.
\ref{fig3?}.
There exist strong arguments showing that hard--hard and hard--soft
corrections exponentiate \cite{Mueller3,KhlebT}, i.e., the total
cross
section has the form (\ref{E14*}). Unfortunately, the complete proof
of the
exponentiation is still lacking.

iii) {\em Hard--soft corrections}. These come from diagrams of
fig. \ref{fig??}c. They
contain both the combinatorial and energy factors. The estimate
analogous
to eqs. (\ref{E13+}) and (\ref{E15*}) is, up to logarithms,
\[
  A^{hard-soft} \sim
  A^{leading}\cdot \frac{1}{\alpha_{W}}
  \left(\frac{E}{E_{0}}\right)^{\frac{8}{3}}.
\]
As we already pointed out, these corrections are also likely to
exponentiate.

To summarize, the perturbation theory about the instanton strongly
suggests that the total cross section has the exponential form
\begin{equation}
    \sigma^{inst}_{2 \to any} \propto
      \exp\left[\frac{4\pi}{\alpha_{W}}
      F\left(\frac{E}{E_0}\right)\right].
\label{E16+}
\end{equation}
The exponent is perturbatively calculable at $E/E_{0}\ll 1$, where it
is
represented by a series in $(E/E_{0})^{2/3}$ (up to logarithms),
\begin{equation}
 F\left(\frac{E}{E_{0}}\right) = -1 +
        \frac{9}{8}\left(\frac{E}{E_{0}}\right)^{\frac{4}{3}}
       - \frac{9}{16}\left(\frac{E}{E_{0}}\right)^{\frac{6}{3}}
       + \dots
\label{E16*}
\end{equation}
Hard--soft corrections contribute at the order $(E/E_{0})^{8/3}$;
this order
has been studied in refs. \cite{KhlebT,Bal,Diak}. Hard--hard corrections
begin at
the order $(E/E_{0})^{10/3}$. While soft--soft corrections to the
exponent
$F$ come from tree diagrams about the instanton, hard--hard and
hard--soft
contributions include all loops.

Clearly, the series (\ref{E16*}) blows up at $E/E_{0} \sim 1$.
Therefore,
the perturbative calculations about the instanton cannot tell whether
the
exponential suppression disappears at $E \sim E_{0}$
or if it persists at all
energies. The analysis of this most interesting problem requires
entirely
non-perturbative techniques.

\subsection{Unitarity}

Before presenting non-perturbative approaches to the evaluation of
the
cross sections of instanton-like processes, let us give a general
argument
in favour of the exponential suppression of these cross sections at
all
energies except, maybe, exponentially high ones. This argument is in
the
spirit of ref. \cite{ZakharovUnitarity} and
is based on unitarity
and conventional perturbation theory at low momenta (see also
refs. \cite{Maggiore,Veneziano,KonishiReview}).

\begin{figure}[p]
  \begin{center}
  \mbox{\epsffile{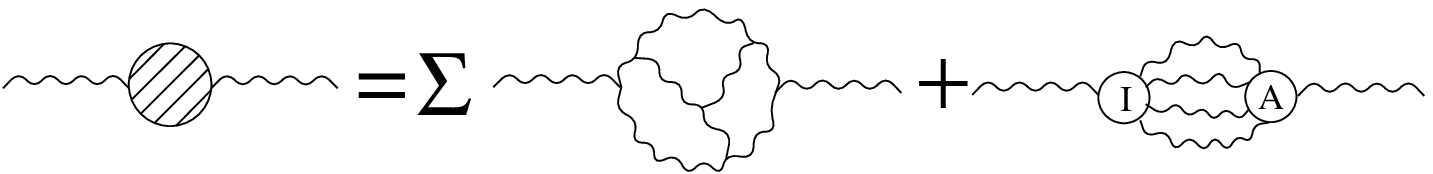}}
  \end{center}
  \caption{$W$-boson propagator at low $Q^2$ with perturbative and
   non-perturbative (instanton--anti-instanton) contributions.
   \label{fig4?}}
\end{figure}

\begin{figure}[p]
  \begin{center}
  \mbox{\epsffile{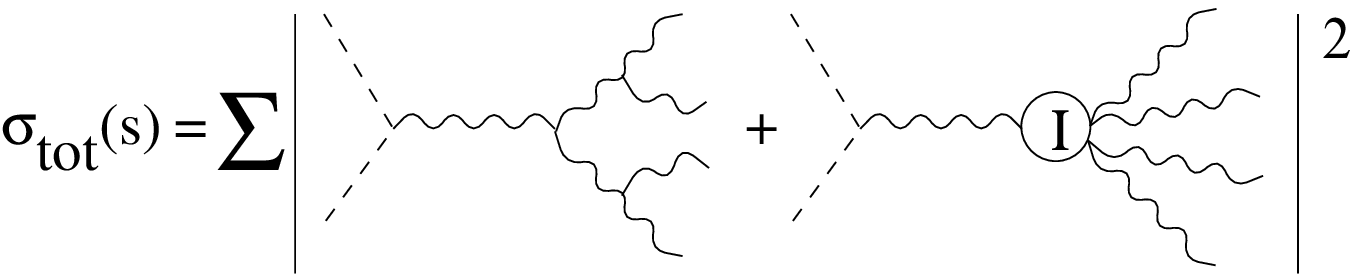}}
  \end{center}
  \caption{Total cross section with perturbative and instanton
           contributions.
   \label{fig5?}}
\end{figure}

Let us consider the full propagator of the
$W$-boson as shown in fig. \ref{fig4?},
and take the momentum of virtual $W$ to be Euclidean and small, say
$Q^{2}  = m_{W}^{2}$. Then the dispersion relation relates this
propagator
to the total cross section of ``$\nu_{e}e$-annihilation'' at
c.m.energy
$\sqrt{s}$ into an arbitrary number of $W$-bosons,
\begin{equation}
 G(Q^{2} \sim m_{W}^{2}) \propto
	     \int~ds~ \frac{\sigma_{tot}(s)}{s + Q^{2}}
\label{E19*}
\end{equation}
Here ``$\nu_{e}e$--annihilation'' means just the production of a
virtual
$W$-boson by an external probe, as shown in fig. \ref{fig5?}.

The left-hand side of eq. (\ref{E19*})  is believed to be a nice
asymptotic
series  in $\alpha_{W}$ whose finite number of terms, say,
$k\ll 1/\alpha_{W}$, are given by ordinary perturbation theory
(perturbative diagrams in fig. \ref{fig4?}).
Indeed, at low $Q^{2}\sim m_{W}^{2}$
there is no reason to suspect that non-perturbative contributions
like
instanton--anti-instanton shown in fig. \ref{fig4?}
are not exponentially suppressed.
Thus,
\[
       G(Q^{2}\sim m_{W}^{2}) =
	 \sum_{i=1}^{k} C_{k} \alpha_{W}^{k} + O(\alpha_{W}^{k+1})
\]
where $C_{k}$ are determined by conventional perturbation theory and
$k$ is
fixed in the limit $\alpha_{W} \to 0$. These $k$ terms are precisely
matched
by perturbative contributions to the total cross section on the
right-hand
side of eq. (\ref{E19*}) (perturbative graphs
in fig. \ref{fig5?}) which include the
production of $k$ final particles or less. So, the instanton
contribution
into the right-hand side of eq. (\ref{E19*}) is small,
\begin{equation}
	 \int~ds~ \frac{\sigma_{tot}^{inst}(s)}{s + Q^{2}} <
	 \const \cdot \alpha_{W}^{k}~~~~~{\mbox{at}}~~ Q^{2}\sim
m_{W}^{2},
	 ~~k\ll \frac{1}{\alpha_{W}}.
\label{E21*}
\end{equation}
This certainly excludes the possibility that the instanton-like cross
sections are large (of order $\alpha_{W}^{n}$ with finite $n$) at
energies
of order $E_{0} \sim m_{W}/\alpha_{W}$. It is straightforward to
generalize
this argument to collisions of two real vector bosons and other
processes
with a small number of incoming particles. In all cases the relation
like
eq.(\ref{E21*}) must hold.

This argument is consistent with the expected functional form of the
total
instanton-like cross section, eq. (\ref{E16+}). It shows that the
exponent
$F$ is negative at all energies, i.e. the instanton-like processes
are
always exponentially suppressed. This general argument does not,
however,
exclude a still very interesting possibility that $F$ tends to zero
as
$E \to \infty$, in which case the cross section may not be
numerically
small at still reasonable energies. In any case, a
theoretical understanding
of an actual exponential behaviour is definitely of interest.

Let us note in passing that the same argument implies the exponential
suppression of the production, in a trivial vacuum, of a large number
of
final particles, $n\sim 1/g^{2}$. These processes have been actively
studied in recent years; for a review see ref. \cite{VoloshinReview}.

Except unitarity, the above argument assumes the validity of the
ordinary
perturbation theory as an asymptotic expansion in $\alpha_{W}$ for
few-point exact Green's functions at low $Q^{2}$. So, there remains a
logical possibility that the perturbation theory is badly wrong
starting at
some finite order in $\alpha_{W}$. We shall see below that this
logical
possibility, which would be revolutionary for the entire quantum
field
theory, is not supported by existing (albeit limited) calculations.

\subsection{From $many \to many$ to $few \to many$}

The exponential form of the instanton-like cross section looks
semiclassical, the failure of the naive semiclassical procedure being
reflected by the fact that the inverse coupling constant enters not
only
the overall factor $4\pi/\alpha_{W}$ but also the characteristic
energy
scale $E_{0} \sim m_{W}/\alpha_{W}$. So, it is natural to expect that
there
exists an appropriate modification of the naive semiclassical
procedure,
which would enable one to evaluate the exponent of the cross section.
Since $F(E/E_{0})$ is determined both by tree graphs about the
instanton
(soft--soft contributions) and by loops (hard--hard and hard--soft
contributions), the correct {\em semiclassical} procedure should
incorporate the relevant part of {\em loops}. Clearly, the very
existence
of  a semiclassical scheme, which would incorporate loops is far from
obvious, and it is a challenging problem to invent such a scheme.

The latter difficulty may be rephrased in the following way. At
energies
of interest, the final particles are soft --- their average energy
per
particle does not contain the large parameter $1/\alpha_{W}$ --- and
numerous, $n \sim 1/\alpha_{W}$. So, it is natural that these
particles may
be described in classical terms:  roughly speaking, these are
classical
waves.  On the other hand, the initial particles are few in number,
and
carry a momentum proportional to the large parameter $1/\alpha_{W}$.
The
problem is that these energetic initial particles are hard to
describe in
(semi)classical terms.

On the basis of the above observations, the following approach to
$few \to many$ instanton-like transitions was suggested
\cite{RTMany,TinyakovMany}. As an intermediate step, consider
$many \to many$ transitions where the number of incoming particles is
\[
     n_{i} = \frac{\nu}{\alpha_{W}}
\]
where $\nu$ is a  variable parameter. In the  limit
\begin{equation}
	\alpha_{W} \to 0,~~~ \frac{E}{E_{0}} = {\mbox{fixed}},~~~
	\nu = {\mbox{fixed}},
\label{E25*}
\end{equation}
the number of incoming particles is large, their energy per particle
is of
order $m_{W}/\nu = $ independent of $\alpha_{W}$, so these particles,
as
well as outgoing ones, may be described in (semi)classical terms. We
shall
see that the total probability for the optimium choice of the initial
state
at given $n_{i}$ has the exponential form
\begin{equation}
    \sigma^{inst}(E, n_{i}) \propto
      \exp\left[\frac{4\pi}{\alpha_{W}}
      F\left(\frac{E}{E_0}, \nu \right)\right]
\label{E25**}
\end{equation}
and the exponent
$F(E/E_{0},\nu)$
is semiclassically calculable.

One argues that
$F(E/E_{0},\nu)$ decreases (becomes more negative) when $\nu$ (i.e.
the
number of incoming particles) decreases. Indeed, processes with fewer
incoming particles may be viewed as a subset of processes with larger
$n_{i}$ with some incoming particles  not participating in the
scattering.
So, $many \to many$ transitions provide an upper bound on $few \to
many$,
\begin{equation}
  \sigma_{2 \to any}^{inst}(E) <
  \sigma^{inst}\left( E,n_{i}=\frac{\nu}{\alpha_{W}}\right)~~
  {\mbox{for~~any}}~~\nu
\label{E26+}
\end{equation}
in the regime (\ref{E25*}). Furthermore, one argues that the exponent
$F(E/E_{0})$ for $few \to many$ probability is obtained from
$F(E/E_{0},\nu)$ in the limit $\nu \to 0$, i.e. with exponential
accuracy one has
\begin{equation}
    \sigma^{inst}_{2 \to any}(E) \propto
      \exp\left[\frac{4\pi}{\alpha_{W}}
      F\left(\frac{E}{E_0}, \nu \to 0\right)\right].
\label{E26*}
\end{equation}
This expectation has been confirmed by explicit perturbative
calculations
about the instanton \cite{MuellerMany}. The reason for eq.
(\ref{E26*}) is
that in the limit $\nu \to 0$ the overlap between the two-particle
initial
state and initial states with $n_{i} = \nu/\alpha_{W}$ particles is
not
exponentially small \cite{TinyakovReview}.

Thus, the idea of
refs. \cite{RTMany,TinyakovMany}
is to evaluate
$F(E/E_{0},\nu)$ semiclassically and then study the limit $\nu \to 0$
to
obtain the exponent for $2 \to any$ cross section.

To implement this idea, one considers the quantity
\cite{RTMany,TinyakovMany}
\begin{equation}
   \sigma^{inst}(E,n_{i}) = \sum_{i,f} |\bra{f} \hat{S} \ket{i}|^{2}.
\label{E27*}
\end{equation}
The sum runs over all initial states obeying the constraints
\[
   \hat{N}\ket{i} = n_{i}\ket{i}
\]
\[
   \hat{H}\ket{i} = E\ket{i}
\]
\[
   \hat{{\bf P}}\ket{i} = 0
\]
where $\hat{N}$, $\hat{H}$ and $\hat{{\bf P}}$ are the particle
number
operator, Hamiltonian and operator of total spatial momentum in the
Fock
space of initial states. The final states in the sum (\ref{E27*}) are
arbitrary, and $\hat{S}$ is the $S$-matrix in the instanton sector.
The quantity $\sigma(E,n_{i})$ may be viewed as the
``microcanonical''
probability for instanton transitions from states with  given c.m.
energy
and number of incoming particles. It is this quantity that enters
eqs.(\ref{E25**}) and (\ref{E26+}).

A convenient basis in the space of initial states is provided by
coherent
states $\ket{a}$. Let us recall that the $S$-matrix
is given, in the coherent state representation,
by the following functional integral
\cite{Berezin,Faddeev}
\begin{equation}
 \bra{b} \hat{S} \ket{a} = S(b^{*},a) =
   \int~D\phi_{i} D\phi_{f} D\phi~
   \exp \left[ B_{i}(\phi_{i},a) + B_{f}(\phi_{f},b^{*}) +
   iS_{T_{i},T_{f}}\right]
\label{E29*}
\end{equation}
where the integration is over initial ($t=T_{i}$), final ($t=T_{f}$)
and
intermediate values of the field(s), the boundary terms are
\[
    B_{i}= -\frac{1}{2} \int~d^{3}k~ a_{k}a_{-k}
\e^{-2i\omega_{k}T_{i}}
	   -\frac{1}{2} \int~d^{3}k~ \omega_{k} \phi_{i}({\bf k})
	   \phi_{i}(-{\bf k})
	   + \int~d^{3}k~ \sqrt{2\omega_{k}} a_{k} \phi_{i}(-{\bf k})
	   \e^{-i\omega_{k}T_{i}}
\]
\[
    B_{f}= -\frac{1}{2} \int~d^{3}k~ b^{*}_{k}b^{*}_{-k}
              \e^{2i\omega_{k}T_{f}}
	   -\frac{1}{2} \int~d^{3}k~ \omega_{k} \phi_{f}({\bf k})
	   \phi_{f}(-{\bf k})
	   + \int~d^{3}k~ \sqrt{2\omega_{k}} b^{*}_{k} \phi_{f}(-{\bf
k})
	   \e^{i\omega_{k}T_{f}}
\]
and the limit $T_{i} \to -\infty$, $T_{f} \to +\infty$ is assumed.
All
bosonic fields are denoted generically by $\phi$. Summation over
states is
represented by the integration over the coherent state variables with
exponential weight
\begin{equation}
  \sum_{i}~ \to~ \int~Da_{k} Da^{*}_{k}~
              \exp\left(- \int~d^{3}k~a^{*}_{k} a_{k}\right).
\label{E29**}
\end{equation}

The microcanonical probability can be written in the following form:
\begin{equation}
	 \sigma(E,n_{i}) = \sum_{i,f}
	 |\bra{b} \hat{S} \hat{P}_{P_{\mu}}
\hat{P}_{n_{i}}\ket{a}|^{2},
\label{E29+}
\end{equation}
where $\hat{P}_{P_{\mu}}$ and $\hat{P}_{n_{i}}$ are projectors onto
the
subspace of a fixed number of incoming particles and a fixed total
four-momentum $P_{\mu}=(E,0)$. Summation in eq. (\ref{E29+}) runs
over all
initial and final states. It can be  shown \cite{KRTPeriodic,RTMany}
that the
matrix elements of the projection operators are
\[
   \bra{\alpha} \hat{P}_{P_{\mu}} \ket{a} =
       \int~d^{4}\xi~\exp\left( -iP_{\mu}\xi^{\mu}
       + \int~d^{3}k~ \alpha^{*}_{k} a_{k} \e^{ik\xi}\right)
\]
\begin{equation}
   \bra{\alpha} \hat{P}_{n_{i}} \ket{a} =
       \int_{0}^{2\pi}~d\eta~\exp\left( -in_{i}\eta
       + \int~d^{3}k~ \alpha^{*}_{k} a_{k} \e^{i\eta}\right).
\label{E30+}
\end{equation}
Combining eqs. (\ref{E29*}), (\ref{E29**}) and (\ref{E30+}) and
performing
trivial integrations and changes of variables, one obtains the double
functional integral representation for the probability,
\[
  \sigma^{inst}(E,n_{i})
  =\int~d^{4}\xi d^{4}\xi'~d\eta d\eta'~
  Da_{k} Da^{*}_{k}~
  Db_{k} Db^{*}_{k}~
  D\phi(x) D\phi'(x')
\]
\[
   \exp\left[ -i P_{\mu}(\xi^{\mu} - \xi'^{\mu})
	      - in_{i}(\eta- \eta')
	      -\int~d^{3}k~ a^{*}_{k}a_{k} \e^{-iP(\xi - \xi') -
	      i n_{i}(\eta - \eta')}
	      -\int~d^{3}k~ b^{*}_{k}b_{k}\right]
\]
\begin{equation}
              \exp \left[ B_{i}(\phi_{i},a)
	      + B_{f}(\phi_{f},b^{*})
	      + B_{i}^{*}(\phi'_{i},a^{*})
	      + B_{f}^{*}(\phi'_{f},b)
	      + iS(\phi)  - iS(\phi') \right]
\label{E30*}
\end{equation}
The integrand here does not depend on $(\xi + \xi')$ due to
translational
invariance; the integration over $d(\xi + \xi')$ produces the usual
volume
factor. Similarly, the integration over $d(\eta + \eta')$ produces an
irrelevant pre-exponential factor.

The remaining integrations are of the saddle point character in the
regime
(\ref{E25*}): upon introducing the variables $\tilde{\phi} = g\phi$
and
$\tilde{a},\tilde{b} = g a, gb$, all terms in the exponent become
explicitly proportional to $1/g^{2}$. So, the general form of the
probability, eq. (\ref{E25**}), is immediate. Furthermore, the
exponent in
eq. (\ref{E25**}), $(4\pi/\alpha_{W})F(E/E_{0},\nu)$, is equal to the
extremum value of the exponent in eq. (\ref{E30*}).

\begin{figure}[p]
  \begin{center}
  \mbox{\epsffile{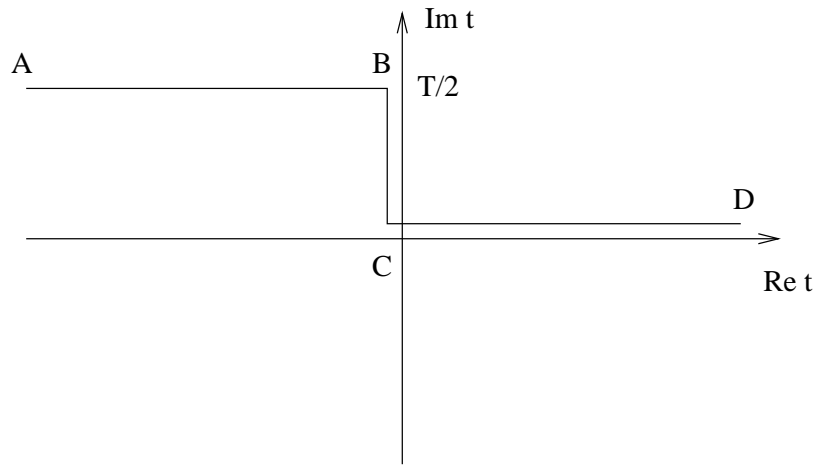}}
  \end{center}
  \caption{The contour in complex time plane appropriate for the
       formulation of the boundary value problem for $many \to many$
       transitions.
    \label{fig6?}}
\end{figure}

Thus, one has to extremize the functional
\[
  \Phi(\phi,\phi';a,a^{*};b,b^{*};T;\theta) =
\]
\[
              ET + n_{i}\theta
	      -\int~d^{3}k~ a^{*}_{k}a_{k} \e^{\omega_k T + \theta}
	      -\int~d^{3}k~ b^{*}_{k}b_{k}
\]
\begin{equation}
               + B_{i}(\phi_{i},a)
	      + B_{f}(\phi_{f},b^{*})
	      + B_{i}^{*}(\phi'_{i},a^{*})
	      + B_{f}^{*}(\phi'_{f},b)
	      + iS(\phi)  - iS(\phi')
\label{E31*}
\end{equation}
with respect to all its variables. Here we introduced the notation
\[
      \xi_{0} - \xi'_{0} = iT
\]
\[
      \eta - \eta' = i\theta
\]
Without loss of generality one considers only real values of $T$: the
imaginary part of $T$ may be removed by time translation.
Perturbative
calculations \cite{TinyakovMany} suggest also that the saddle point
value of $\theta$ is real. Extremization of the functional
(\ref{E31*}) is
conveniently performed by moving the contour in the complex time
plane from
the real axis to the contour ABCD shown in fig. \ref{fig6?}.
At this contour one obtains the following boundary value problem
\cite{RST}:

i) The field $\phi$ obeys the usual field equations,
\[
	    \frac{\delta S}{\delta \phi} =0.
\]

ii) In the future asymptotics, region D, the field is real, i.e. it
is
real on the line CD,
\[
         \phi|_{CD} = {\mbox{real}}.
\]

iii)  In the past asymptotics, $t = iT/2 + \tilde{t}$,
$\tilde{t}=\mbox{real} \to -\infty$
(region A), the field is a collection of
linear waves whose positive and negative frequency parts are related
to
each other
\begin{equation}
   \phi({\bf k},\tilde{t}) =
       f_{k}\e^{i\omega_{k}\tilde{t}} +
       \e^{-\theta}f^{*}_{k}\e^{-i\omega_{k}\tilde{t}}.
\label{E33+}
\end{equation}
The values of $T$ and $\theta$ are related to $E$ and $n_{i}$ through
\begin{equation}
   E = \int~d^{3}k~\omega_{k} f_{k}f^{*}_{k} \e^{-\theta}
\label{E33*}
\end{equation}
\begin{equation}
   n_{i} = \int~d^{3}k~ f_{k}f^{*}_{k} \e^{-\theta}.
\label{E33**}
\end{equation}
Equation (\ref{E33*}) is the natural result of the fact
 that the energy of the classical
solution is equal to the energy of the scattering process, while
eq. (\ref{E33**}) is the analogue of the usual relation between the
Fourier
components of linear classical fields and the corresponding number of
particles. The exponent $F(E/E_{0},\nu)$ is equal to the extremum
value of
the functional (\ref{E31*}) evaluated along the contour ABCD (in
fact, only
the part ABC is relevant). Of course, one has to make sure that the
solution to this boundary value problem indeed describes the
instanton-like
transition, i.e. that the topological numbers of initial and final
vacua
differ by one. In fact, an appropriate quantity in the case of a
finite
number of incoming and outgoing particles is the winding number of
the
Higgs field \cite{FarhiTransitions}.

As expected, requirements (i)--(iii) represent a purely classical
field
theory problem. However, the fields are necessarily complex on the
contour
ABC: eq. (\ref{E33+}) is the spatial Fourier transform of a complex
field.
Also, the field must have singularities somewhere between the real
axis and
the line AB, otherwise the conditions (ii) and (iii) would contradict
each
other. It is natural that when $F(E/E_{0},\nu)<0$ (we assume
implicitly that this is indeed the case), the
classical problem is formulated on the contour in the time plane that
contains both Minkowskian and Euclidean parts: we are dealing with a
kind
of tunnelling (``Euclidean'') process, but incoming and outgoing
particles
live in Minkowskian time.

A special case of the above boundary value problem emerges at
\[
       \theta = 0
\]
when the field both at final and initial asymptotics is real. In this
case
the classical solution is real on the
entire contour ABCD of fig. \ref{fig6?}, and
hence has turning points at $t=0$ and $t=iT/2$ (points B and C)
\cite{KRTPeriodic},
\[
 \partial_{t} \phi ({\bf x}, t=0) =
 \partial_{t} \phi ({\bf x}, t=i\frac{T}{2}) = 0~~~
 {\mbox{for~~all}}~~x.
\]
In other words, the solution is real in Euclidean space-time and
periodic
in Euclidean time with period $T$. This periodic instanton describes
the
instanton-like transition with {\em maximum probability} at a given
energy
$E$, and the corresponding number of  incoming particles is the
optimum
number at this energy \cite{KRTPeriodic}. The maximum probability is
determined by the truncated Euclidean action of the periodic
instanton,
\[
     \sigma^{inst}_{max}(E) \propto
	      \exp[ET - S_{per}(0,T)].
\]
At low energies, the periodic instanton is represented by the
instanton--anti-instanton chain, while at $E$ close to $E_{sph}$ its
fields
are those of the sphaleron plus small oscillations in the sphaleron
Minkowskian negative mode (which is the only positive mode in
Euclidean
time). At $E=E_{sph}$ the maximum probability becomes unsuppressed.
The number of incoming particles at which this occurs is
equal to that produced in the decay of the sphaleron,
\[
          n_{sph} \sim \frac{1}{\alpha_{W}}.
\]
At $E>E_{sph}$ the instanton-like transitions are unsuppressed at
some
$n_{i}$, and the periodic instanton does not exist. Periodic
instantons in
various models have been obtained numerically for all energies in
refs.\cite{VMatveev,Khabib,KuznetsovTinyakov} and analytically in
ref.\cite{SonR}.

The approach outlined above is of a fairly general nature.
It enables one to
study, at least in principle, various processes similar to
instanton-like
transitions induced by particle collisions. In fact, most extensive
results
obtained within this approach up to now refer to false vacuum decay
induced
by collisions of highly energetic particles in scalar theories. The
reason
is that solving classical field equations is easier in scalar
theories.
Also, the form of the scalar potential may be suitably adjusted.
Since the
false vacuum decay is in many respects analogous to instanton-like
processes in gauge theories, it serves as a theoretical laboratory to
probe
various sets of ideas
\cite{Hsu,VolFal,RSTFal,KiselevFal,VolFal2,SonR,RSon,KuznetsovTinyakov}.

\begin{figure}[p]
  \begin{center}
    \mbox{\epsffile{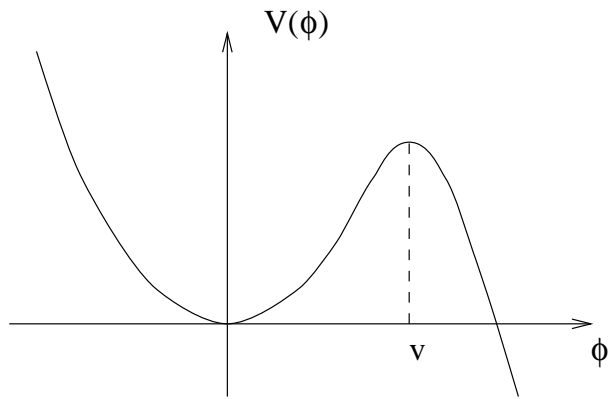}}
  \end{center}
  \caption{The scalar potential with unstable vacuum at $\phi=0$.
    \label{fig7?}}
\end{figure}

A particular model where the boundary value problem formulated above
can be
solved analytically in a wide range of the parameters $E/E_{0}$ and
$\nu$
is provided  by the theory of one scalar field in $(1+1)$-dimensional
space-time described by the scalar potential \cite{SonR}:
\begin{equation}
       V(\phi) = \frac{m^{2}}{2}\phi^{2}
	-\frac{m^{2}v^{2}}{2} \exp\left[2\Lambda \left(\frac{\phi}{v} -
		      1\right)\right]
\label{E40*}
\end{equation}
where $v^{2}\gg 1$ is the parameter that plays the role of the
inverse
coupling constant. $\Lambda$ is another free parameter which is
chosen
to be
large, $\Lambda \gg 1$ (but $\Lambda \ll v $). The potential has the
form
shown in fig.\ref{fig7?}. At large $\Lambda$ it is quadratic
almost up to $\phi = v$
and has a steep cliff at $\phi > v$. The problem that is
addressed is the decay
of the metastable vacuum $\phi = 0$ induced by collisions of
energetic
particles in this vacuum.

At low energies this process is described by the
bounce \cite{CBounce}, which is the analogue of the instanton. There
also exists an analogue of the sphaleron, which is a critical bubble
\cite{VolKobOkun}. Its energy equals  the height of the barrier
separating the two phases (the ``true vacuum'' in this model is
$\phi = \infty$, but this pathological property is irrelevant for the
problem of the false vacuum decay). The sphaleron energy in this
model is
\[
	    E_{sph} = mv^{2}
\]
and the characteristic number of incoming particles is
\[
	 n_{sph} = \frac{2}{\pi} v^{2}.
\]
At $E\geq E_{sph}$ and $n_{i}\geq n_{sph}$ the induced false vacuum
decay
occurs without exponential suppression. The parameters appearing in
the
classical boundary value problem are $E/E_{sph}$ and $n_{i}/n_{sph}$,
and
the probability of the induced vacuum decay has the form
\[
	\sigma(E,n_{i}) \propto
	 \exp\left[ S_{B} F\left( \frac{E}{E_{sph}}, \nu;
\Lambda\right)
	 \right]
\]
where $S_{B} = \const\cdot  v^{2}$  is the action of the bounce and
$F(E/E_{sph} \to 0) = -1$. The  question is whether the exponent $F$
becomes zero at some energy for $\nu<1$ and, most importantly,
whether it
becomes zero at high enough energies or remains always negative in
the
limit  $\nu \to 0$.

\begin{figure}[p]
  \begin{center}
    \mbox{\epsffile[130 170 500 450]{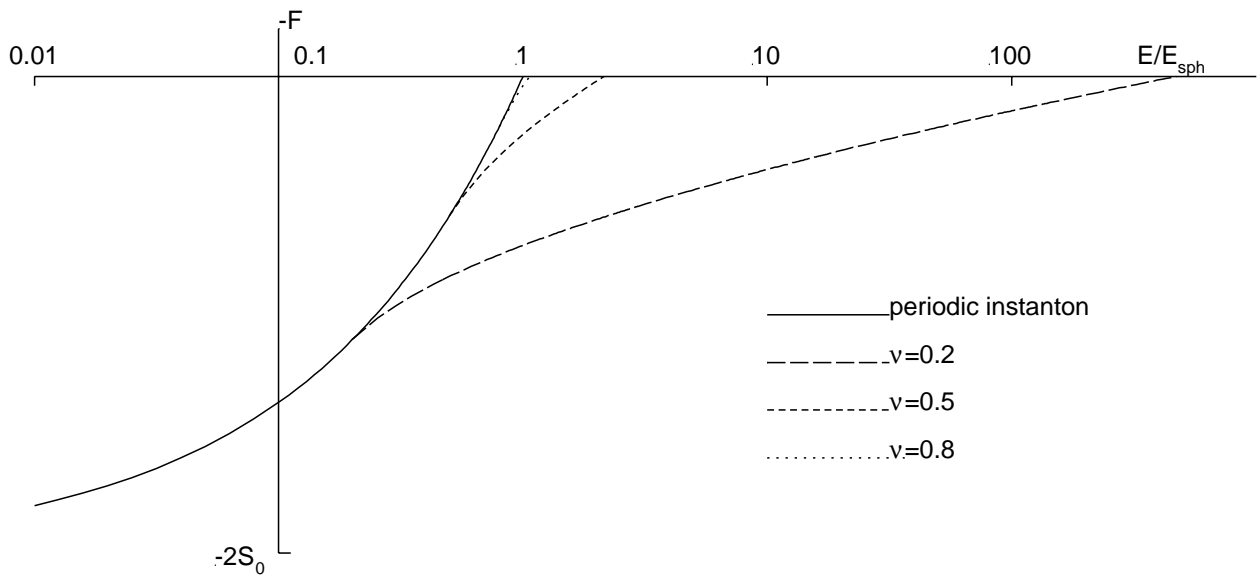}}
  \end{center}
  \caption[a]{\protect
 Exponent of the total probability $F(E)$ at different
     $\nu$ in the exponential model. From ref. \cite{SonR}.}
   \label{fig8?}
\end{figure}

The way to solve the classical boundary value problem in this model
is to
solve the {\em free massive} field equation in the region of the
space-time
(on  the contour of fig. \ref{fig6?})
where $\phi < v$, solve the {\em massless
Liouville} equation in the region where $\phi>v$, and then match the
solutions \cite{SonR}. In this way it is possible to analyse the
initial
multiplicities which are not too low,
\begin{equation}
	 \nu = \frac{n_{i}}{n_{sph}}
		\gg \Lambda^{-1}
\label{E42*}
\end{equation}
The result for the function $F$ as function of energy at various
$\nu$ is
shown in fig. \ref{fig8?}.
%

Figure \ref{fig8?}  shows that there exists, at $n_{i}$
obeying eq. (\ref{E42*}), some
critical energy $E_{crit}(n_{i})$ at which $F$ becomes equal to zero,
and
the exponential suppression disappears. At
$1\gg n_{i}/n_{sph} \gg \Lambda^{-1}$ the expression for this
critical
energy is fairly simple:
\[
   E_{crit}(n_{i}) =
     \frac{4}{\pi} \exp \left( \frac{\pi^{2}}{4}\frac{n_{sph}}{n_{i}}
     -1\right) \left(\frac{n_{i}}{n_{sph}}\right)^{2}\cdot E_{sph}.
\]
Clearly, $E_{crit}$ rapidly grows as $n_{i}$ becomes small. These
results
show that at $n_{i}<n_{sph}$ one can still have unsuppressed
instanton-like
transitions at the expense of increasing the energy. Furthermore, at
sufficiently large $n_{i}$, namely at $n_{i}$ obeying eq.
(\ref{E42*}),
there exist {\it real classical solutions} to the field equation {\it
in
Minkowski space-time} \cite{RSon} that describe the false vacuum
decay
induced by $n_{i}$ incoming particles in purely classical (hence
unsuppressed) manner at $E>E_{crit}(n_{i})$.

Unfortunately, analytical solutions to the boundary value problem are
lacking at $n_{i}/n_{sph}\lsim \Lambda^{-1}$. In particular, the
limit
$\nu\to 0$ cannot be studied even in the specially designed model
(\ref{E40*}). However, it has been shown \cite{RSon} that there are
no
classical Minkowskian solutions describing unsuppressed induced
false vacuum decay at $n_{i}/n_{sph}<\pi^{2}/\Lambda$ and any
energies.
This means that the false vacuum decay induced by collisions of {\it
two}
particles is exponentially suppressed at all energies, in accord with
the
general unitarity argument.

\begin{figure}[p]
  \begin{center}
    \mbox{\epsffile{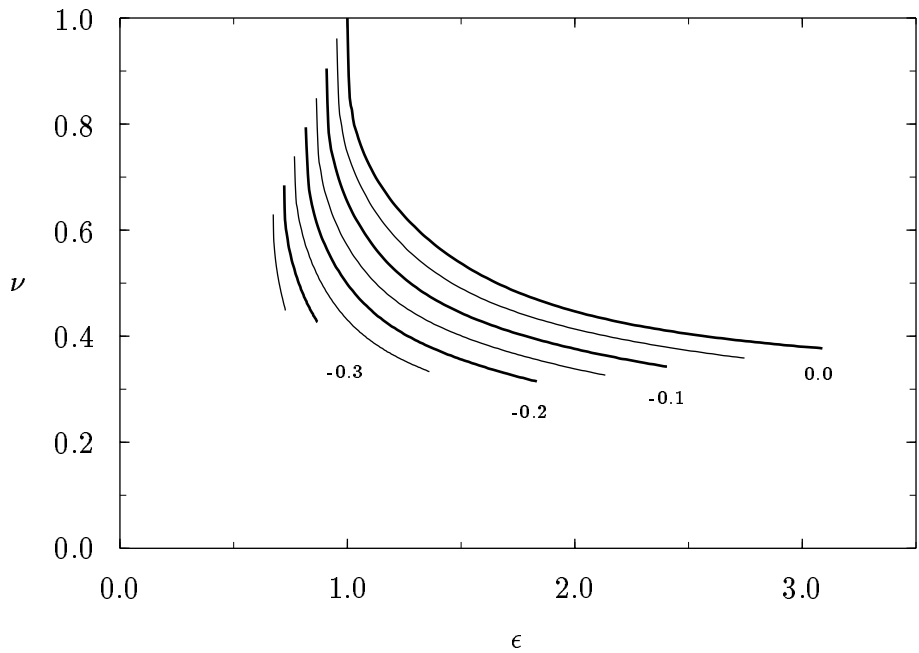}}
  \end{center}
  \caption[a]{\protect
     Lines of constant $F$  in the plane
     $(\epsilon=E/E_{sph}, \nu=n_{i}/n_{sph})$ in
     4d scalar model. From ref.
     \cite{KuznetsovTinyakov}.}
   \label{fig9}
\end{figure}

The above boundary value problem is suitable, at least in principle,
for
numerical calculations: after all, one has to solve classical field
equations with specified boundary conditions. The numerical study of
induced false vacuum decay has been undertaken in
ref.\cite{KuznetsovTinyakov} in the context of $(3+1)$-dimensional
scalar
theory with the potential
\[
       V(\phi) = \frac{m^{2}}{2}\phi^{2}
		- \frac{\lambda}{4} \phi^{4}
\]
which is similar to fig. \ref{fig7?}. The main result of this
study is summarized in
the plot of the lines of constant $F(\epsilon,\nu)$ in the
$(\epsilon,\nu)$
plane, where $\epsilon = E/E_{sph}$ and $\nu = n_{i}/n_{sph}$ (here
$E_{sph} \sim m/\lambda$ and $n_{sph} \sim 1/\lambda$ are again the
energy
of the critical bubble and the characteristic number of incoming
particles). This plot is shown in fig.\ref{fig9}.
%

The most interesting region, $\nu \ll 1$, was not accessible to
numerical
calculations for technical reasons. However, because of inequality
(\ref{E26+}) the results presented in fig. \ref{fig9} show
that the false vacuum
decay induced by {\it two}-particle collisions is exponentially
suppressed
at least at $E< 3E_{sph}$. Indeed, the line of unsuppressed induced
vacuum
decay, $F=0$, is above $\nu \sim 0.4$ at these energies (i.e. $n_{i}$
should be larger than $0.4 n_{sph}$  for the transitions to be
unsuppressed).  Further analysis \cite{KuznetsovTinyakov} reveals
that the
region of this suppression extends at least up to $10E_{sph}$, and,
most
likely, to infinity.  This is again in accordance with the
unitarity argument.

Another approach closely related to one discussed above is
to consider real classical solutions to
Minkowskian field equations, i.e. scattering of classical waves. To
every
classical solution that disperses into free waves at $t \to \pm
\infty$
one can assign the number of incoming and outgoing particles, both of
which
are naturally of order $1/\alpha_{W}$. The probability of the
scattering of
these multiparticle states is not suppressed. At given energy one
tries to
minimize the number of incoming particles under the condition that
the
topological number changes by 1 (instanton-like transitions) or that
the
phase of the system changes (false vacuum decay). If
the minimum number of incoming particles tends to zero (in units
$1/\alpha_{W}$) as the total c.m. energy approaches some $E_{cr}$,
then
{\it few} $\to$ {\it many} processes are not suppressed exponentially
at
$E>E_{cr}$ (this includes the more likely possibility that
$E_{cr}=\infty$,
in which case the exponential suppression of {\it few} $\to$ {\it
many}
cross sections disappears at asymptotically high energies). In the
opposite
case when the minimum number of incoming particles  needed to induce
the
classical transition remains larger than
$\const \cdot 1/\alpha_{W}$, the exponential
suppression of {\it few} $\to$ {\it many} persists at all energies,
but the
actual exponent cannot be calculated by studying classical
scattering.

We have pointed out already that the two approaches (complex time and
real
time) nicely match in the $(1+1)$-dimensional model (\ref{E40*}) and
that,
in this model, the number of incoming particles required for the
instanton-like transitions to occur classically is finite in units
$n_{sph}$. The classical real time approach is also fairly suitable
for
numerical analysis; considerable progress in this direction has been
reported in ref. \cite{Rebbi}, where it has also been found that the
instanton-like transitions may proceed classicaly at $n_{i}$ somewhat
below $n_{sph}$ at sufficiently high energies.

Let us stress that the idea of using $many \to many$ transitions as
an
intermediate step to $few \to many$ is not the only way (and, maybe,
not
the best way) to study the instanton-like processes semiclassically.
A completely different --- and promising --- approach
is the generalization of
the Landau technique for calculation of semiclassical matrix elements
\cite{landau} to quantum field theory
\cite{VoloshinPHI4,KhlebLa,DiakLa,GorskyLa,CornLa,SonLa}. It remains
to be
seen whether this approach will be able to provide new insight into
electroweak $B$ and $L$ violation at high energies.

To conclude, the existing calculations suggest the following overall
picture of the instanton-like transition in high energy collisions.
When the number of incoming particles is of order $1/\alpha_{W}$, the
instanton-like processes occur at unsuppressed rates provided that
the
energy is sufficiently high. This is perfectly consistent with the
calculations of the sphaleron rate at high temperatures:
multiparticle
collisions are possible in the high temperature plasma, and they are
responsible for the high rate of electroweak $B$ and $L$ violation.
On the
other hand, if the number of colliding particles is small (say, two),
the
instanton-like processes occur at exponentially low rates at all
energies,
which is in accord with unitarity argument. The actual suppression
factor
is, unfortunately, still unknown for the most interesting
range of energies.
While this suppression factor may be of limited interest for the
electroweak
theory (the suppression by $\exp (-\const /\alpha_{W})$ with
 almost any constant in the exponent will make the
transitions unobservably rare), it may become important for the study
of
similar processes in QCD (for the discussion of instanton-induced
processes in QCD see
refs. \cite{qcd1,qcd2,qcd3,qcd4}).
\section{Conclusion}

The non-conservation of baryon number in the early Universe, giving
rise
to the  baryon asymmetry, was proposed by Sakharov almost thirty
years ago. Still, its particle physics
origin is not established. Certainly,
there were baryon number violating interactions operating at
temperatures
well above 100 GeV --- these were anomalous electroweak reactions ---
but whether the baryon asymmetry came entirely from this source or
was a
combined effect of the electroweak processes and  grand unified
or/and
intermediate scale interactions is unclear at the moment. In any
case, the
explanation of the existing baryon asymmetry requires at least a mild
extension of the Minimal Standard Model.

One expects substantial further progress, in coming years, in the
understanding of the electroweak baryon number non-conservation and
its role in the early Universe. On the theoretical side, quantitative
estimates of the regions in the parameter space where the observed
amount of baryon asymmetry is produced, are to be obtained in various
extensions of the Minimal Standard Model. This requires further
development of the kinetics of the electroweak phase transition,
B-violating reactions and non-equilibrium description of
baryogenesis. The calculations of the suppression factor for the
anomalous reactions in high energy collisions are to be done in MSM
and its extensions, and the relevance of similar reactions for hard
processes in QCD is to be understood in detail. Most remarkable
progress is expected, however, on the experimental side. Uncovering
the physics at the energy scale of a few hundred GeV to a few TeV
(Minimal Standard Model? Supersymmetry? Extended Higgs sector?
Technicolor-type symmetry breaking? ...) will be crucial for
establishing or ruling out the electroweak origin of the baryon
asymmetry of the Universe. Finding out the mechanism of CP
non-conservation in Nature ($B$-meson physics, electric dipole
moments of neutron and electron) will become another important step.
Possible experimental discovery of lepton number violation (Majorana
neutrino masses, neutrino ocsillations, muon number non-conservation
in $\mu$-decays) and/or proton decay would be a strong indication of
the early origin ($T\gg 1$ TeV) of the baryon asymmetry. Having
understood, to a considerable extent, highly non-trivial aspects of
the electroweak physics in the early Universe, one needs strong
experimental input to solve one of the most challenging problems in
cosmology.

The authors are indebted to their colleagues at INR (Moscow), CERN
and elsewhere for numerous helpful discussions. We are grateful to
J. Ambj\o rn, A.I. Bochkarev, K. Farakos, G. Farrar, G. Giudice,
D.Yu. Grigoriev, K. Kajantie, S.Yu. Khlebnikov, N.V. Krasnikov, 
S.V. Kuzmin, V.A. Kuzmin,
M. Laine, M.L. Laursen,
M.V. Libanov, V.A. Matveev, L. McLerran, E. Mottola, K. Rummukainen, D.T.
Son, A.N. Tavkhelidze,
P.G. Tinyakov, I.I. Tkachev,  V.F. Tokarev, S.V. Troitsky, N. Turok and
M. Voloshin, for collaboration on different
issues related to this review. We thank F.L. Bezrukov, S.L. Dubovsky
and D.S. Gorbunov for their assistance in preparing the manuscript.
Careful reading of the manuscript and helpful comments by 
P. Huet,
M. Joyce, K. Kajantie, S. Khlebnikov,
M. Laine, and K. Rummukainen are greatly acknowledged. We are indebted
to New High Energy Theory Center, Rutgers University, where this
work has been completed, for hospitality.
The work of V.R. was supported in part by INTAS grant 94-2352 and
Russian Foundation for Basic Research grant 96-02-17449a.


\newpage


\end{document}